\documentclass[12pt,preprint]{aastex}

\usepackage{emulateapj5}

\slugcomment{to appear in the Astrophysical Journal}

\shorttitle{Recurrent Novae}
\shortauthors{Hachisu and Kato}

\begin{document}

\title{RECURRENT NOVAE AS A PROGENITOR SYSTEM OF TYPE 
I\lowercase{a} SUPERNOVAE. I.\\
RS OPHIUCHI SUBCLASS --- SYSTEMS WITH A RED GIANT COMPANION}


\author{Izumi Hachisu}
\affil{Department of Earth Science and Astronomy, 
College of Arts and Sciences, University of Tokyo,
Komaba, Meguro-ku, Tokyo 153-8902, Japan} 
\email{hachisu@chianti.c.u-tokyo.ac.jp}

\and

\author{Mariko Kato}
\affil{Department of Astronomy, Keio University, 
Hiyoshi, Kouhoku-ku, Yokohama 223-8521, Japan} 
\email{mariko@educ.cc.keio.ac.jp}

%\author{K. Nomoto}
%\affil{Department of Astronomy, University of Tokyo, 
%Bunkyo-ku, Tokyo 113-0033, Japan \\ e-mail: 
%nomoto@astron.s.u-tokyo.ac.jp}
%


%\altaffiltext{1}{Visiting Astronomer, Cerro Tololo Inter-American Observatory.
%CTIO is operated by AURA, Inc.\ under contract to the National Science
%\altaffiltext{2}{Society of Fellows, Harvard University.}
%\altaffiltext{3}{present address: Center for Astrophysics,
%    60 Garden Street, Cambridge, MA 02138}
%\altaffiltext{4}{Visiting Programmer, Space Telescope Science Institute}
%\altaffiltext{5}{Patron, Alonso's Bar and Grill}


\begin{abstract}
     Theoretical light curves of four recurrent novae
in outburst are modeled to obtain various physical parameters.
The four objects studied here are those with a red giant companion, 
i.e., T Coronae Borealis, RS Ophiuchi, V745 Scorpii, and V3890 Sagittarii.
Our model consists of a very massive white dwarf (WD)
with an accretion disk and a red giant companion. 
Light curve calculation includes reflection effects of the companion star 
and the accretion disk together with a shadowing effect 
on the companion by the accretion disk. 
We also include a radiation-induced warping instability of 
the accretion disk to reproduce the second peak of T CrB 
outbursts.  The early visual light curves are well reproduced 
with a thermonuclear runaway model on a very massive white dwarf 
close to the Chandrasekhar mass limit, i.e., 
$M_{\rm WD}= 1.37 \pm 0.01 ~M_\sun$ for T CrB, 
$1.35 \pm 0.01 ~M_\sun$ for RS Oph with solar metallicity ($Z=0.02$) but
$1.377 \pm 0.01 ~M_\sun$ for RS Oph with low metallicity ($Z=0.004$),
$1.35 \pm 0.01 ~M_\sun$ for V745 Sco, and
$1.35 \pm 0.01 ~M_\sun$ for V3890 Sgr. 
Optically thick winds, which blow from the WDs 
during the outbursts, play a key role
in determining the nova duration and the speed of decline 
because the wind quickly reduces the envelope mass on the WD.
Each envelope mass at the optical maximum is also estimated to be 
$\Delta M \sim 3 \times 10^{-6} M_\sun$ (T CrB),
$2 \times 10^{-6} M_\sun$ (RS Oph),
$5 \times 10^{-6} M_\sun$ (V745 Sco),
$3 \times 10^{-6} M_\sun$ (V3890 Sgr), indicating 
an average mass accretion rate 
$\dot M_{\rm acc} \sim 0.4 \times 10^{-7} M_\sun$ yr$^{-1}$ 
(80 yrs, T CrB), 
$1.2 \times 10^{-7} M_\sun$ yr$^{-1}$ (18 yrs, RS Oph), 
$0.9 \times 10^{-7} M_\sun$ yr$^{-1}$ (52 yrs, V745 Sco), 
$1.1 \times 10^{-7} M_\sun$ yr$^{-1}$ (28 yrs, V3890 Sgr) 
during the quiescent phase before the last outburst.
Although a large part of the envelope mass is blown in the wind,
each WD can retain a substantial part of the envelope mass
after hydrogen burning ends.  Thus, we have obtained net 
mass-increasing rates of the WDs as 
$\dot M_{\rm He} \sim 0.1 \times 10^{-7} M_\sun$ yr$^{-1}$ (T CrB), 
$0.12 \times 10^{-7} M_\sun$ yr$^{-1}$ (RS Oph), 
$0.05 \times 10^{-7} M_\sun$ yr$^{-1}$ (V745 Sco), 
$0.11 \times 10^{-7} M_\sun$ yr$^{-1}$ (V3890 Sgr). 
These results strongly indicate that the WDs in the recurrent novae
have now grown up to near the Chandrasekhar mass limit
and will soon explode as a Type Ia supernova if the WDs consist 
of carbon and oxygen. 
We have also clarified the reason why only T CrB shows a secondary
maximum but the other three systems do not.
\end{abstract}




\section{INTRODUCTION}
     Type Ia supernovae (SNe Ia) are one of the most luminous 
explosive events of stars.  Recently, SNe Ia have been used 
as good distance indicators that provide a promising tool 
for determining cosmological parameters \citep{per99,rie98} 
because of their almost uniform maximum luminosities.
These both groups derive the maximum luminosities, $L_{\rm max}$, 
of SNe Ia completely empirically from the light curve shape 
(LCS) of nearby SNe Ia, and assume that the same 
$L_{\rm max}$-LCS relation holds in high red-shift ($z$) SNe Ia. 
In order that this method works, 
the nature of SNe Ia and their progenitors 
should be the same between high-$z$ and low-$z$ SNe Ia.  
Therefore, it is necessary to identify the progenitor binary systems, 
thus confirming whether or not SNe Ia ``evolve'' from high red-shift 
to low red-shift galaxies \citep[e.g.,][]{ume99a}.
\par
    It is widely accepted that the exploding star itself is 
an accreting white dwarf (WD) in a binary
\citep*[e.g.,][]{nom82, nom84}. 
However, the companion star (and thus the observed binary system)
is not known.  Several types of objects have ever been considered, 
which include merging double white dwarfs
\citep[e.g.,][]{ibe84, web84}, 
recurrent novae \citep*[e.g.,][]{sta85}, 
symbiotic stars \citep[e.g.,][]{mun92} etc 
\citep[see, e.g.,][for recent summary] {liv00}.  
\par
     One of the candidates, the recurrent nova U Scorpii
underwent sixth recorded outburst in February 25, 1999 \citep{smr99}.  
For the first time, a complete light curve of the outburst has been
obtained from the rising phase to the final decline phase 
through the mid-plateau phase including eclipses
\citep*[e.g.,][]{hkkm00, mat00, mun99}. 
By analyzing these new data on a basis of 
the theoretical light-curve model of outbursts,  
\citet{hkkm00} have elucidated
the outburst nature of U Sco 
much more accurately than the previous estimates
\citep*[e.g.,][]{sta88, kat90}.  We summarize
some critically important values they have obtained. 
(1) The white dwarf mass in U Sco is $1.37 \pm 0.01 M_\sun$.
(2) The companion is a slightly evolved main-sequence star 
of 1.4---1.6 $M_\sun$ that expands to fill the Roche lobe
after a large part of the central hydrogen has been consumed and
is now transferring mass to the white dwarf 
in a thermally unstable manner. 
(3) The white dwarf is now growing in mass at an average rate of 
$\sim 1 \times 10^{-7} M_\sun$ yr$^{-1}$. 
Therefore, the white dwarf will reach the critical mass 
($M_{\rm Ia}=1.378 M_\sun$, in W7 model of \citealp{nom84}) 
for the SN Ia explosion in the quite near future.
They concluded that U Sco is a very strong candidate 
for the immediate progenitor of SNe Ia.
Other recurrent novae are also expected to be a strong candidate
for SNe Ia.
\par
     Thus, we have started our light curve analysis of recurrent
novae in which the mass and the growth rate of white dwarfs 
are determined as accurately as possible and examine 
the possibility of an SN Ia explosion
in the quite near future.  In this series of papers, 
details of our results on recurrent novae are presented,
although a part of our results has already been published 
in several short papers 
(\citealt{hac99k} for T CrB; 
\citealt{hkkm00, hkkmn00} for U Sco; 
\citealt{hac00ka} for RS Oph; 
\citealt{hac00kb} for V394 CrA). 
\par
     Here, in the first paper of this series, we have examined
four recurrent novae which have a red giant companion,
T CrB, RS Oph, V745 Sco, V3890 Sgr.  Two of them
(T CrB and RS Oph) are reexamined with new observational
knowledge.  We first summarize the nature
of recurrent novae and explain the physical reason why they 
become so massive as near the Chandrasekhar mass limit in \S 2.
Then, we show details of our numerical methods for calculating
evolution of bloated white dwarf envelopes in \S 3 and
for obtaining theoretical light curves in \S4, 
because details of our numerical method have not been fully
described yet.
New numerical results on determination of white dwarf masses
for four recurrent novae are shown and discussed
in \S 5 for T CrB, in \S 6 for RS Oph, 
in \S 7 for V745 Sco, and in \S 8 for V3890 Sgr.
Conclusions follow in \S 9.

\section{RECURRENT NOVAE AS A PROGENITOR OF TYPE I\lowercase{a} SUPERNOVAE}
     Recurrent novae are characterized by nova eruptions 
with its recurrence time scale from a decade to a century, 
its brightness amplitude larger than 6 magnitudes, and its expansion 
velocity faster than 300 km s$^{-1}$ \citep[e.g.,][]{web87}.  
They are considered to be binary systems in which the secondary
companion star overflows the inner critical Roche lobe and
transfers matter onto the primary white dwarf (WD) component.
When the hydrogen-rich envelope on the WD reaches 
a critical mass, hydrogen ignites to trigger
a nova eruption (hydrogen shell-flash). 
We call this a {\it thermonuclear runaway (TNR) event}. 
\par
     Recurrent novae are morphologically divided into three groups
depending on its companion star:  dwarf companions, 
only one member, T Pyxidis with $P=0.076$ days, \citep{sch90, sch92, 
pat98}; slightly evolved main-sequence (MS) companions,
U Scorpii with $P=1.23$ days \citep{sch90, sch95},
V394 Coronae Austrinae with $P=0.757$ days \citep{sch90}, 
Large Magellanic Cloud Nova 1990 No.2, hereafter LMC RN, with
$P=$unknown but $P \sim 1.3$ days was suggested \citep{sek92},
CI Aquilae with $P=0.618$ days \citep{men95}, 
and red giant (RG) companions, 
T Coronae Borealis with $P=227.6$ days \citep*{lin88},
RS Ophiuchi with $P=460$ days, \citep{dob94}, recently revised to be 
$P=457$ days \citep{fek00}, V745 Scorpii with $P=$unknown, 
V3890 Sagittarii with $P=$unknown, where $P$ is the orbital period 
of the binary in units of day.   These are sometimes called
T Pyx subclass, U Sco subclass, and T CrB (or RS Oph) subclass,
respectively \citep[e.g.,][]{war95}.  We have summarized 
these features in Table \ref{recurrent_novae}.
\par
     Classical novae, sharing many aspects with the recurrent novae, 
are also binary systems consisting of a white dwarf and a less 
massive main-sequence star that fills the Roche lobe 
and transfers its mass to the white dwarf. 
The white dwarf accumulates hydrogen-rich matter and
ignites hydrogen atop the white dwarf when the hydrogen-rich 
envelope on the white dwarf reaches a critical mass.
%Hydrogen shell-burning is unstable to trigger an explosion 
%of the envelope.  
A part of the hydrogen diffuses into 
the white dwarf before the ignition so that a very surface layer 
of the white dwarf is dredged up into the hydrogen-rich envelope 
and blown off in the outburst wind \citep[e.g.,][]{pri86, kov94}.
As a result, nova ejecta contain 
white dwarf matter, i.e., carbon and oxygen for carbon-oxygen white dwarfs
or oxygen, neon, and magnesium for O-Ne-Mg white dwarfs 
\citep[see, e.g.,][ and references therein]{pri86, kov94, war95}.   
Thus, the white dwarf is eroded gradually after many cycles of nova
outbursts, recurrence periods of which are usually longer than
thousands of years.   
\par
     Recurrent novae show some characteristic differences:
(1) heavy elements such as carbon, oxygen and neon
are not enriched in ejecta but similar to the solar values
\citep[e.g.,][for summary]{liv92, war95},
indicating that the white dwarf is not eroded;
(2) very short recurrence periods from a decade to a century 
theoretically require very massive white dwarfs close to 
the Chandrasekhar mass limit \citep[e.g., Fig. 9 of][]{nom82}.
These two arguments indicate that the mass of the white dwarfs 
in recurrent novae increases toward the Chandrasekhar mass limit
\citep*[e.g.,][]{sta85, sta88, liv92, del96}. 
We expect that, if the white dwarf is made of carbon and oxygen,
the white dwarf ignites carbon at the center and explodes 
as a Type Ia supernova when it reaches the critical mass of 
$M_{\rm Ia}= 1.378 M_\sun$ \citep{nom82, nom84}.
However, the evolutionary path to recurrent novae has not long 
been elucidated as well as the reason why so massive white dwarfs
can exist in recurrent novae.

\subsection{Accretion Wind Evolution vs. Common Envelope Evolution}
     In this subsection, we briefly explain how white dwarfs 
in recurrent novae have grown its mass 
toward the Chandrasekhar mass limit.
The reason why the standard binary evolution theory (will be cited 
below) has failed to explain the evolutionary path to recurrent novae 
is mainly in the theoretical prediction of a common envelope
formation and the ensuing spiral-in process, because these processes 
inhibit the growth of white dwarfs in binary systems.  As discussed
in many previous papers on binary evolutions, 
it had been widely accepted that hydrogen-rich envelopes on 
mass-accreting white dwarfs expand to a red giant size 
when its mass accretion rate, 
$\dot M_{\rm acc}$, exceeds a critical limit, $\dot M_{\rm cr}$, 
i.e., $\dot M_{\rm acc} > \dot M_{\rm cr} \sim 1 \times 10^{-6}
M_\sun$ yr$^{-1}$ \citep[see, e.g., Fig. 9 of][]{nom82}, 
and easily forms a common envelope \citep*[e.g.,][]{nom79}.
Once a common envelope is formed, two stars begin to spiral-in 
each other due to viscous drag; thus, producing
a double degenerate system \citep[e.g.,][]{ibe84, web84}.
\par
     However, \citet*{hkn96} found that white dwarfs begin to blow 
strong winds ($v_{\rm wind} \sim 1000$ km s$^{-1}$ and
$\dot M_{\rm wind} \gtrsim 1 \times 10^{-6} M_\sun$ yr$^{-1}$)
when the mass accretion rate exceeds the critical rate, i.e.,
$\dot M_{\rm acc} > \dot M_{\rm cr}$,
as illustrated in Figure \ref{accwind}; thus, 
preventing the binary from collapsing.
The binary does not spiral-in but keeps its separation. 
Hydrogen burns steadily and therefore the helium layer of 
the white dwarf can grow at a rate of 
$\dot M_{\rm He} \approx \dot M_{\rm cr}$.
The other transferred matter is blown in the wind 
($\dot M_{\rm wind} \approx \dot M_{\rm acc} - \dot M_{\rm cr}$),
where $\dot M_{\rm wind}$ is the wind mass loss rate. 
We call this {\it accretion wind} because it begins to blow 
when the accretion rate exceeds the critical rate.
\par
     Therefore, as shown in Figure \ref{accmap_z02},  
we have to revise Figure 9 of \citet{nom82},
which has been widely used in binary evolution scenarios as basic
processes of mass-accreting white dwarfs \citep[e.g.,][]{kah97}. 
The most important difference of Figure \ref{accmap_z02} from 
old Nomoto's (1982) is that white dwarfs grow its mass 
in a much wider parameter region. 
In our new diagram, the status of a mass-accreting white dwarf 
is specified by the following three phases:
(1) accretion wind phase ($\dot M_{\rm acc} > \dot M_{\rm cr}$); 
(2) steady shell-burning phase 
($\dot M_{\rm std} < \dot M_{\rm acc} < \dot M_{\rm cr}$);
and (3) intermittent shell-flash phase
($\dot M_{\rm acc} < \dot M_{\rm std}$), where
$\dot M_{\rm std}$ means the lowest limit of mass accretion rate
for steady hydrogen shell burning.  
The new growing region of white dwarfs,
$\dot M_{\rm acc} > \dot M_{\rm std}$, 
is much wider than old Nomoto's (1982) narrow strip,
$\dot M_{\rm std} < \dot M_{\rm acc} < \dot M_{\rm cr}$.

\subsection{Evolutionary Paths to Recurrent Novae and Type Ia Supernovae}
\begin{figure}
%\epsscale{.5}
%\plotone{sn1afig.eps}
\plotone{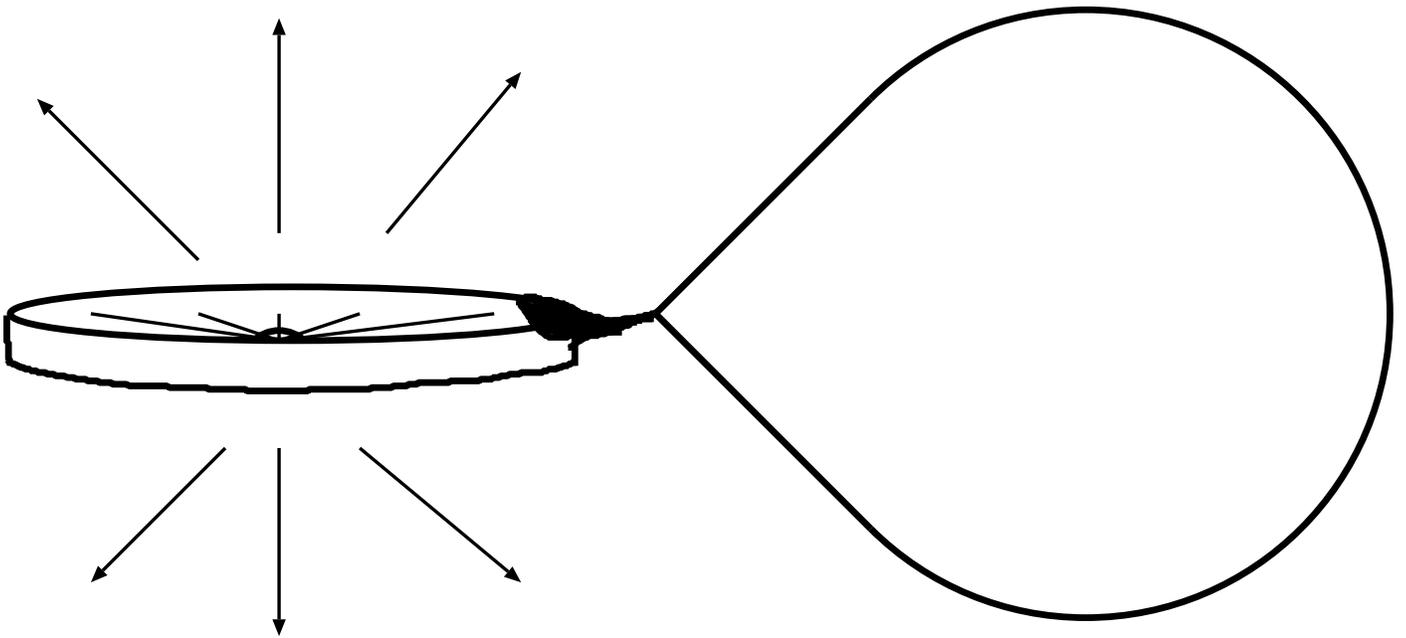}
%\plotfiddle{evolution1.ps}{5.0cm}{270}{0.4}{0.4}{-170}{220}
\caption{
Optically thick winds blow from mass-accreting white dwarfs
({\it left figure}) when the mass transfer rate from 
a lobe-filling companion ({\it right figure}) exceeds a critical 
rate, i.e., $\dot M_{\rm acc} > \dot M_{\rm cr}$.  Here, we assume 
that the white dwarf accretes mass from equatorial region and, 
at the same time, blows
winds from polar regions as illustrated in the figure.
\label{accwind}}
\end{figure}

\begin{figure}
%\epsscale{.5}
%\plotone{acc_map.eps}
\plotone{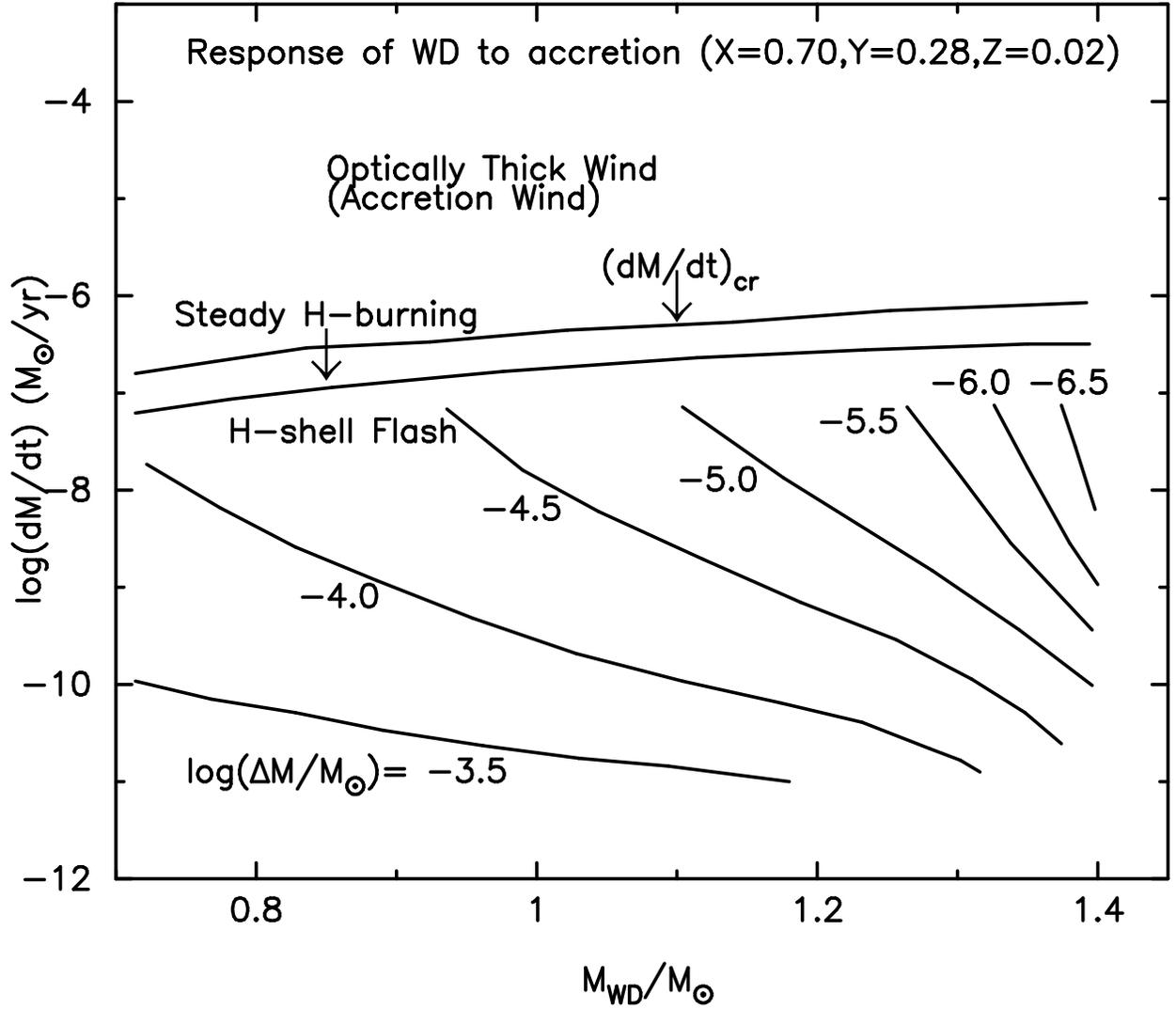}
%\plotfiddle{evolution1.ps}{5.0cm}{270}{0.4}{0.4}{-170}{220}
\caption{
Response of white dwarfs to mass accretion is illustrated
in the white dwarf mass and the mass accretion rate plane,
i.e., in $M_{\rm WD}$-$\dot M_{\rm acc}$ plane.
Strong optically thick winds blow above the line of
$\dot M_{\rm acc} > \dot M_{\rm cr}$.  The wind mass loss rate is 
$\dot M_{\rm wind} \approx \dot M_{\rm acc} - \dot M_{\rm cr}$.
Steady hydrogen shell burning with no optically thick winds occur 
between $\dot M_{\rm std} \le \dot M_{\rm acc} \le \dot M_{\rm cr}$.
There is no steady-state burning below 
$\dot M_{\rm acc} < \dot M_{\rm std}$.  Instead, intermittent
shell flashes occur.  The envelope mass at which hydrogen shell
flash occurs is also shown (taken 
from Fig. 9 of \citealp{nom82}).
\label{accmap_z02}}
\end{figure}

     Based on the new mechanism of Figures \ref{accwind} 
and \ref{accmap_z02}, \citet{hkn99, hknu99} have found two 
paths to SNe Ia \cite[see also,][]{lih97}.  In these two paths,
white dwarfs can accrete mass continuously from 
the companion and grow at a rate of $\dot M_{\rm cr} \sim 1 
\times 10^{-6} M_\sun$ yr$^{-1}$.
Therefore, candidates of SN Ia progenitors are systems at the final
stages on these paths; i.e., one is a supersoft
X-ray source (SSS) consisting of a white dwarf and a lobe-filling
more massive main-sequence star \citep[WD+MS system, e.g.,][]{lih97} 
and the other is a symbiotic star consisting of a white dwarf
and a mass-losing red giant \citep[WD+RG system, e.g.,][]{hkn96}. 
Both the systems contribute to main parts of the SN Ia birth rate
\citep{hkn99, hknu99} \citep[see also][for a recent review]{liv00}.
\par
\begin{figure}
%\epsscale{.5}
%\plotone{zregevl10.eps}
\plotone{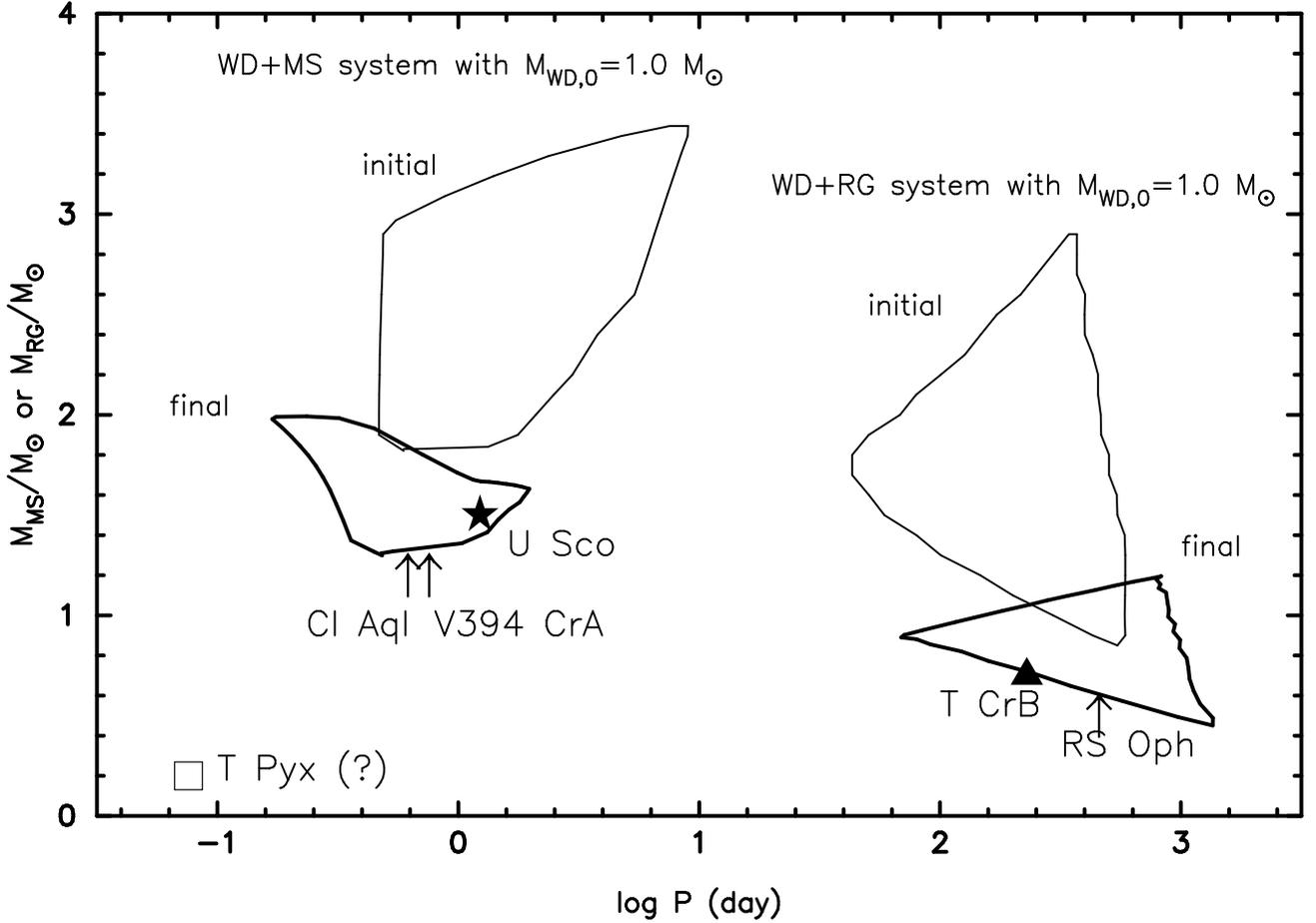}
%\plotfiddle{evolution1.ps}{5.0cm}{270}{0.4}{0.4}{-170}{220}
\caption{
Regions producing SNe Ia are plotted  
in the $\log P - M_{\rm d}$ (orbital period --- donor mass) plane 
for the WD+MS system (left in the figure) and the WD+RG system
(right in the figure).
Here we assume an initial white dwarf mass of $1.0 M_\sun$. 
The initial system inside the region encircled by a thin solid line
(labelled by "initial") is increasing its white dwarf mass 
up to the critical mass ($M_{\rm Ia}= 1.378 M_\sun$) 
for Type Ia supernova explosion, the regions of which are encircled 
by a thick solid line (labelled by "final").
Currently known positions of each recurrent nova are indicated 
by a star mark ($\star$) for U Sco \citep[e.g.,][]{hkkm00}, 
a filled triangle for T CrB \citep[e.g.,][]{bel98},
an open rectangle for T Pyx \citep[e.g.,][]{pat98},
but by arrows for the other three recurrent novae, V394 CrA,
CI Aql, and RS Oph,
because the mass of the companion is not yet available explicitly.
Two subclasses of the recurrent novae, U Sco subclass and
RS Oph (or T CrB) subclass correspond to the helium-rich supersoft 
X-ray source channel \citep{hknu99} and the symbiotic channel 
\citep{hkn99} of Type Ia supernovae, respectively.
\label{zregevl10}}
\end{figure}

     \citet{hkn99, hknu99} have followed many evolutionary paths
and obtained the initial parameter regions of binaries 
that finally produce an SN Ia.  Figure \ref{zregevl10} 
shows such two regions of the WD+MS systems (SSS channel) 
and the WD+RG systems (symbiotic channel).  We have added 
in this figure the final parameter regions just before an SN Ia 
explosion occurs.
Starting from the initial region
encircled by the thin solid lines, binary systems evolve and 
can explode as an SN Ia in the regions enclosed 
by the thick solid lines.  \citet{hkn99, hknu99}
showed that some recurrent novae lie on the boarder of their 
parameter region that can produce an SN Ia.
Among six recurrent novae with known orbital periods,
five fall in the final regions of SN Ia progenitors 
just before the explosion.  Only T Pyx is far out of the regions 
of SNe Ia.

\placefigure{accwind}
\placefigure{accmap_z02}
\placefigure{zregevl10}
\placetable{recurrent_novae}
\placetable{recurrent_novae_outburst}

\section{OPTICALLY THICK WINDS IN DECAY PHASE OF NOVAE}
     In the thermonuclear runaway (TNR) model of nova explosions, 
WD envelopes expand greatly as large as $\sim 100 ~R_\sun$ or 
more and then the photospheric radius gradually shrinks to the 
original size of the white dwarfs 
(e.g., $R_{\rm WD}= 0.0032 ~R_\sun$ for $M_{\rm WD}= 1.37 ~M_\sun$). 
Here, we call this {\it decay phase}, i.e., 
the phase after the maximum expansion of the photosphere.  
In the decay phase of novae,
the photospheric temperature increases from $T_{\rm ph} \sim 10^4$ K
to $\sim 10^{6}$ K with the bolometric luminosity 
being almost constant (near the Eddington limit).  Therefore, the main 
emitting region moves from optical to soft X-ray through UV.  
Correspondingly, the optical luminosity
reaches its maximum at the maximum expansion of the photosphere and
then decays toward the level in quiescent phase.
Since the WD envelope reaches a steady-state in the decay phase of novae
\citep[e.g.,][]{pri86, kat94h},
we are able to approximate the development of the envelope structure
with a unique sequence of steady-state wind solutions having
a different envelope mass ($\Delta M$) as shown by
Kato \& Hachisu (1994). 
\par 
     Optically thick winds, blowing from WDs in the decay
phase of novae, play a key role in determining the nova duration,
because the winds eject a large part of the envelope mass and
drastically reduce the fuel.  The winds blow when
the WD photosphere is larger than $\sim 0.05-0.1 R_\sun$, i.e.,
when the photospheric temperature is lower than
$\log T_{\rm ph} \lesssim 5.5$, because the wind is driven by 
a strong peak at $\log T \sim 5.2$ in OPAL opacity \cite[e.g.,][]{igl96},
and $\log T_{\rm ph} \sim 5.5$ corresponds to the shoulder of the peak
on the high temperature side.  This opacity peak strongly blocks 
photons and, as a reaction, matter is effectively accelerated 
to blow a wind deep inside the photosphere
\cite[see, e.g., Figs. 2, 3, and 4 in][]{kat94h}. 
\par
     Assuming the solar abundance of heavy elements
($Z=0.02$), we have calculated the steady-state envelope solutions
and obtained sequences for WDs with various masses 
of $M_{\rm WD}= 0.6$, 0.7, 0.8, 0.9, 1.0, 1.1, 1.2, 1.3, 1.34,  
1.35, 1.36, 1.37, and $1.377 M_\sun$.
Here, we have adopted the updated OPAL opacity \citep{igl96}
with lower hydrogen contents of $X=0.70$, 0.50, 0.35, 0.15, 0.10,
0.08, 0.07, 0.06, 0.05, and 0.04.
The numerical methods and various assumptions are 
essentially the same as those in Kato \& Hachisu (1994) except 
the opacities.  Some of them are plotted in Figure \ref{dmdtallsol}. 
\par
     For RS Oph and LMC RN, 
we have also obtained the envelope solutions
for lower metallicities, i.e., $Z=0.001$,  0.002, 0.004, and 0.01 with
$X=0.70$, 0.50, 0.35, 0.15, 0.10, 0.08, 0.07, 0.06, 0.05, and 0.04.
Some of them are also shown in Figure \ref{dmdtall_low_m}.

\subsection{Evolution of White Dwarf Envelope}
     Hydrogen shell-flashes occur when the mass transfer rate 
is below the lowest limit
for the steady hydrogen shell-burning, i.e., 
$\dot M_{\rm acc} < \dot M_{\rm std}$. 
In this subsection, we describe such a rapid development
of white dwarf envelopes.  As already mentioned above,
optically thick winds blow in the decay phase of 
hydrogen shell-flashes.  
Since each wind solution is a unique function of the envelope mass
$\Delta M$ for a given WD mass, we can follow 
the development of the envelope structure by calculating 
the envelope mass, which is decreasing due to the wind mass loss 
at a rate of $\dot M_{\rm wind}(\Delta M)$ 
and hydrogen shell burning 
at a rate of $\dot M_{\rm nuc}(\Delta M)$, i.e., 
\begin{equation}
{{d} \over {d t}} \Delta M = \dot M_{\rm acc} - 
\dot M_{\rm wind} - \dot M_{\rm nuc},
\label{dmdt_envelope_mass}
\end{equation}
where $\dot M_{\rm acc}$ is the mass accretion rate of the WD.
Directly integrating equation (\ref{dmdt_envelope_mass}), 
we obtain the time-development of the envelope, $\Delta M$,
and then the evolutionary changes of physical quantities such as 
the photospheric temperature $T_{\rm ph}(\Delta M)$, 
photospheric radius $R_{\rm ph}(\Delta M)$, 
photospheric velocity $v_{\rm ph}(\Delta M)$, 
wind mass loss rate $\dot M_{\rm wind}(\Delta M)$, 
and nuclear burning rate $\dot M_{\rm nuc}(\Delta M)$,
all of which are a unique function of the envelope mass, $\Delta M$.
\par
     We have shown, in Figures \ref{dmdtallsol} and \ref{dmdtall_low_m},
the envelope mass $\Delta M$, the photospheric radius $R_{\rm ph}$,
the photospheric temperature $T_{\rm ph}$, 
and the photospheric velocity $v_{\rm ph}$
against the mass decreasing rate of the envelope,
$\dot M_{\rm wind} + \dot M_{\rm nuc}$.
The wind mass loss rate becomes as large as 
$10^{-4}$---$10^{-3} M_\sun$ yr$^{-1}$ at the maximum expansion
of the photosphere $R_{\rm ph} \sim 100 R_\sun$,
and then decreases down to $1 \times 10^{-6} M_\sun$ yr$^{-1}$ 
just before the wind stops, i.e., when $R_{\rm ph} \sim 0.1 R_\sun$.
The photospheric temperature increases from $1 \times 10^4$ K 
to $3 \times 10^5$ K.  The photospheric velocity remains 
as high as $\sim 1000$ km s$^{-1}$ during the wind phase.
\par
     When the envelope mass decreases to below the critical mass,
the wind stops and the photosphere rapidly shrinks 
towards its original size before the outburst, for example,
from $R_{\rm ph}= 0.06 R_\sun$ to $0.005 R_\sun$ within
a week for $M_{\rm WD}= 1.37 ~M_\sun$. 
Shortly after the wind stops, 
steady-state shell burning on the WD ends but  
hydrogen burning itself is still continuing and supplying a part of 
the luminosity.  In this phase, the envelope mass is decreased only 
by nuclear burning.
The rest of the luminosity comes from the thermal energy of 
the hydrogen envelope and the hot ash (helium) below the hydrogen
layer.  The thermal energy amounts to several times $10^{43}$ ergs
for the case of $M_{\rm WD}= 1.37 ~M_\sun$, $X=0.05$, and $Z=0.02$
(for the 1999 outburst of U Sco). 
It can supply a bolometric luminosity of $10^{38}$ ergs s$^{-1}$ 
for about ten days or so.
\par
     The mass lost by the wind, $\Delta M_{\rm wind}$, and the mass
added to the helium layer of the WD, $\Delta M_{\rm He}$, are
calculated from
\begin{equation}
\Delta M_{\rm wind}= \int \dot M_{\rm wind} ~d t, 
\label{wind_lost_mass}
\end{equation}
and 
\begin{equation}
\Delta M_{\rm He}= \int \dot M_{\rm nuc} ~d t, 
\label{wind_accum_mass}
\end{equation}
respectively.
The efficiency of hydrogen shell-flashes (nova outbursts) is defined by
\begin{equation}
\eta_{\rm H} = {{\Delta M_{\rm He}} \over {\Delta M_{\rm max}}}
\approx {{\Delta M_{\rm He}} 
\over {\Delta M_{\rm wind} + \Delta M_{\rm He}}}, 
\label{efficiency_H_flashes}
\end{equation}
where $\Delta M_{\rm max}$ is the envelope mass at the optical
maximum.  
\par
     Assuming a high constant rate of 
$\dot M_{\rm acc} \sim 10^{-3} M_\sun$ yr$^{-1}$ 
in equation (\ref{dmdt_envelope_mass}), we start the calculation.
The envelope mass is quickly increased and the optical luminosity 
soon reaches the observational maximum in a day or so.  
Then, we drop the rate to
$\dot M_{\rm acc} \sim 1 \times 10^{-7} M_\sun$ yr$^{-1}$ or zero
and further follow the decay phase of nova. 
Therefore, our rising phase is not a realistic one but just a rough
sketch.  Once the envelope has reached the optical maximum, 
our theoretical light curve in the decay phase represents 
a realistic one because the flow in the envelope is
approximately in a steady-state as already mentioned above.

\subsection{White Dwarf Envelope in Steady Mass Accretion} 
\begin{figure}
%\epsscale{.5}
%\plotone{dmdtallsol.eps}
\plotone{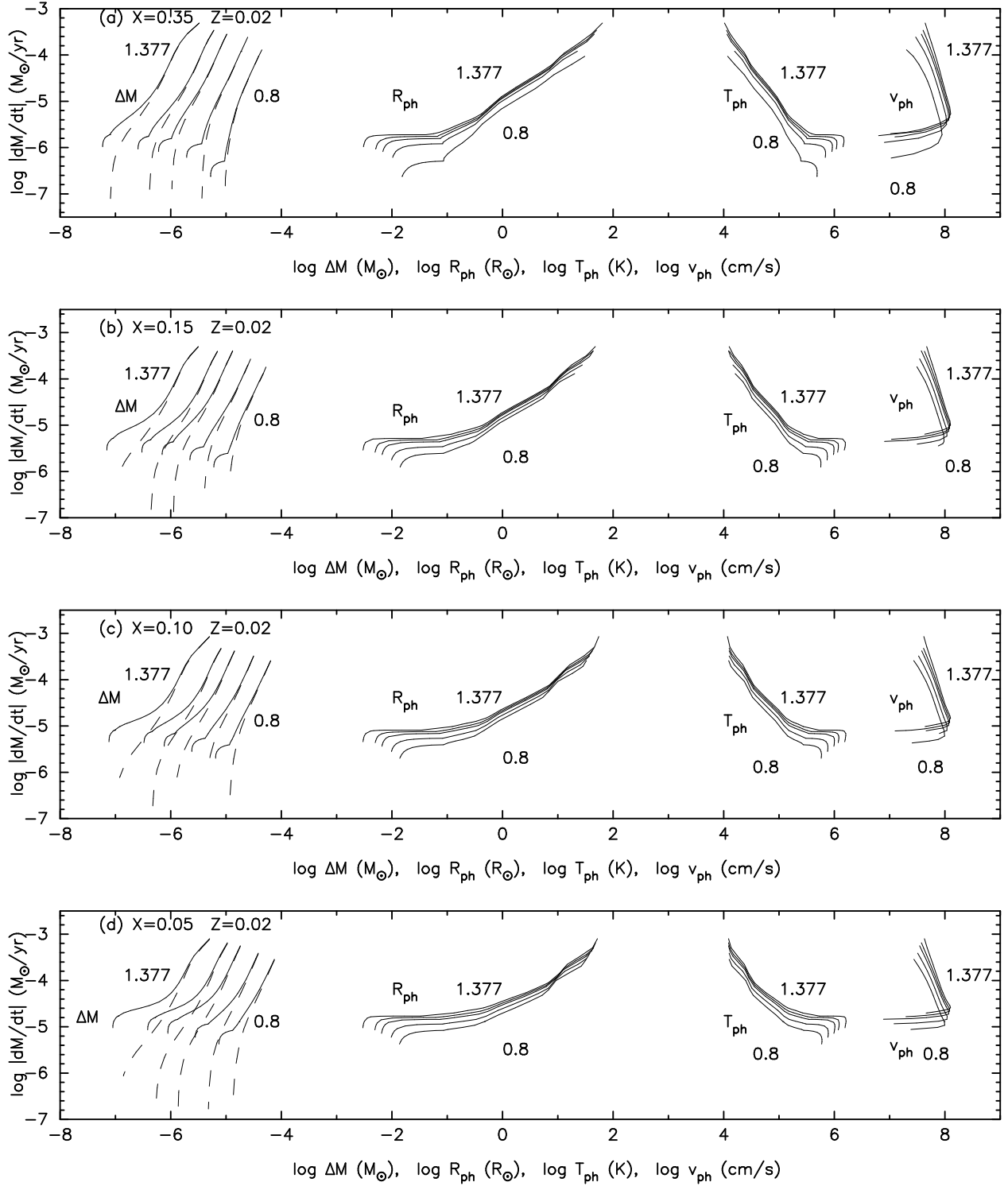}
%\plotfiddle{evolution1.ps}{5.0cm}{270}{0.4}{0.4}{-170}{220}
\caption{
Wind mass loss rate (dashed lines) and total mass decreasing
rate of the envelope (solid lines labeled by $\Delta M$), i.e., 
wind mass loss rate plus nuclear burning rate, 
$\dot M_{\rm wind} + \dot M_{\rm nuc}$,
are plotted against the envelope mass for WDs with 
masses of 0.8, 1.0, 1.2, 1.3 and $ 1.377 M_\sun$.  
There exist only wind solutions above the breaks on the solid lines 
while only static (no wind) solutions exist below the breaks.
Thus, an optically thick wind stops at this break.
The photospheric radius, $R_{\rm ph}$, the photospheric temperature,
$T_{\rm ph}$, and the photospheric velocity, $v_{\rm ph}$, are 
also plotted against the total decreasing rate of the envelope mass
in the figure.   We have shown four different cases of 
hydrogen content of the envelope material, i.e.,
(a) $X=0.35$, (b) 0.15, (c) 0.10, and (d) 0.05, from top to bottom panel, 
for the solar metallicity of $Z=0.02$.  Each sequence is plotted 
from the wind solution with $\log T_{\rm ph}= 4.0$ to the static solution
with $\dot M_{\rm nuc}= \dot M_{\rm std}$, i.e., the lower limit for
steady hydrogen burning. 
\label{dmdtallsol}}
\end{figure}

\begin{figure}
%\epsscale{.5}
%\plotone{dmdtall_low_m.eps}
\plotone{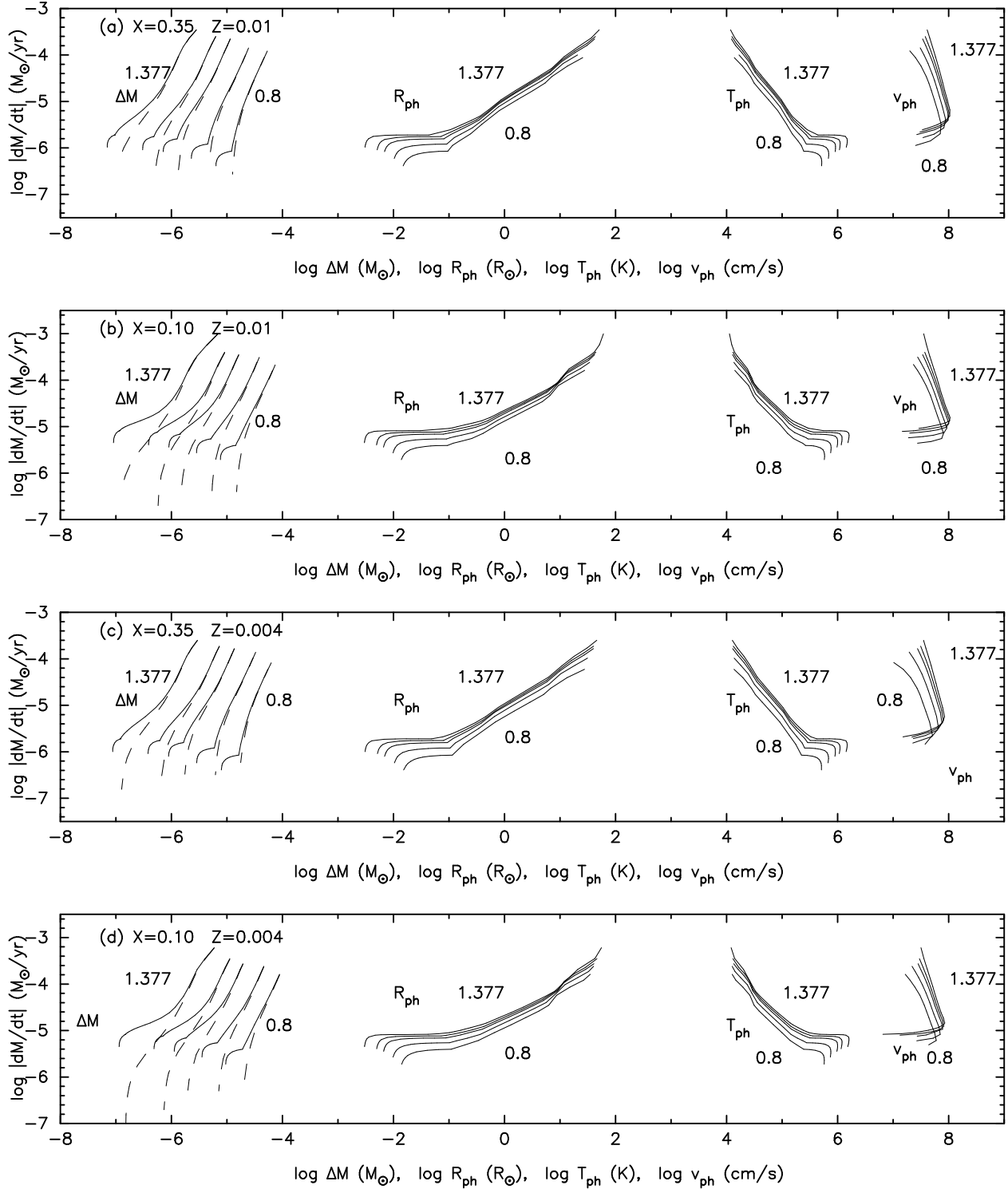}
%\plotfiddle{evolution1.ps}{5.0cm}{270}{0.4}{0.4}{-170}{220}
\caption{
Same as Fig. 4, but for low metallicities of
$Z=0.01$ (upper two panels) and $Z=0.004$ (lower two panels), i.e.,
(a) $X=0.35$, $Z=0.01$, (b) $X=0.10$, $Z=0.01$, 
(c) $X=0.35$, $Z=0.004$, and (d) $X=0.10$, $Z=0.004$.
\label{dmdtall_low_m}}
\end{figure}

     If the mass accretion rate ($\dot M_{\rm acc}$) does not vary,
the envelope solution is obtained from
$d(\Delta M)/dt=0$ in equation (\ref{dmdt_envelope_mass}).  
The mass accretion rate is in a balance with the mass decreasing rate 
of the envelope,
\begin{equation}
\dot M_{\rm acc} = \dot M_{\rm wind} + \dot M_{\rm nuc}.
\label{envelope_steady_state}
\end{equation}
If the steady mass-accretion rate is given,
we can directly (without integration) obtain the corresponding state
of the envelope from Figures \ref{dmdtallsol} and \ref{dmdtall_low_m}.
The critical envelope-mass decreasing rate, i.e.,
at the cease point of wind, exactly coincides with
the critical mass-accretion rate, i.e., at the beginning point of wind,
$\dot M_{\rm cr} \equiv (\dot M_{\rm acc} )_{\rm cr}
= (\dot M_{\rm nuc} )_{\rm cr}$,
which can be approximately expressed by
\begin{equation}
\dot M_{\rm cr} = 5.3 \times10^{-7}{{1.7-X} \over {X}} 
\left({M_{\rm WD} \over {M_\sun}}  -
0.40\right) M_\sun {\rm ~yr}^{-1},
\label{critical}
\end{equation}
for various WD masses ($M_{\rm WD}$) and hydrogen contents ($X$) 
with solar metallicity ($Z=0.02$).  
This relation is reduced from our wind solutions
for $X=0.70$, $X=0.50$, $X=0.35$, $X=0.15$, and $X=0.10$ ($Z=0.02$)
and has accuracy within to 3---5\%.
\par
     In the same steady mass-accretion mentioned above, 
we also obtain the lower limit of the mass accretion rate
for steady hydrogen shell-burning as
\begin{eqnarray}
\dot M_{\rm std} &=& 2.6 \times10^{-7}{{1.7-X} \over {X}} 
\left({M_{\rm WD} \over {M_\sun}}  -
0.40\right) M_\sun {\rm ~yr}^{-1} \cr
&\approx& 0.5 \dot M_{\rm cr},
\label{steady_hydrogen_burning}
\end{eqnarray}
for solar metallicity ($Z=0.02$).
This relation is reduced from our steady-state solutions
for $X=0.70$, $X=0.50$, $X=0.35$, $X=0.15$, and $X=0.10$ ($Z=0.02$)
and has accuracy better than 10\%.

\placefigure{dmdtallsol}
\placefigure{dmdtall_low_m}

\section{THEORETICAL LIGHT CURVES}
\begin{figure}
%\epsscale{.5}
%\plotone{m1370_tcrb_fig.eps}
\plotone{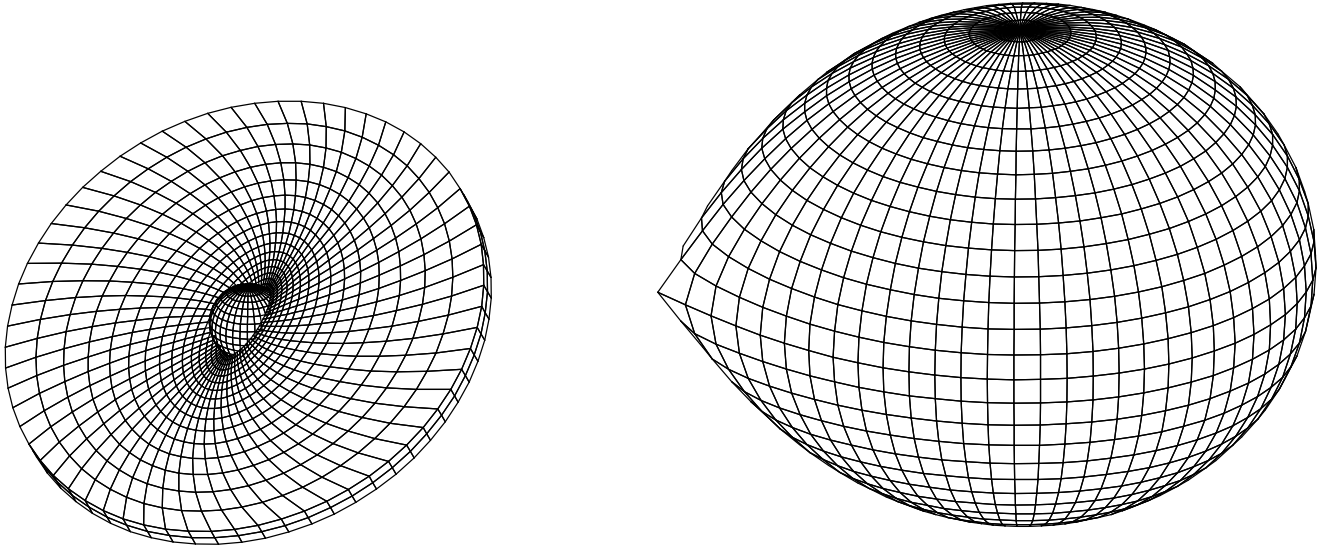}
%\plotfiddle{evolution1.ps}{5.0cm}{270}{0.4}{0.4}{-170}{220}
\caption{
Model configuration near the second peak of the recurrent 
nova T CrB.  The cool component ({\it right in the figure}) is 
a red giant filling up its inner critical Roche lobe.  
The hemisphere is heated up by the hot component 
($1.37 ~M_\sun$ white dwarf, {\it left in the figure}).  
The photospheric radius of the hot component 
near the second peak is as small as 
$\sim 0.003 ~R_\sun$, about $\sim 0.00003$ times the size 
of the cool component, but 
it is exaggerated in this figure to easily see it. 
We assume a warping accretion disk around the hot component, which is 
precessing at a period of about 140 days.  
The surface of the accretion disk is also heated up by the hot 
component.  The radius of the accretion disk near the second peak 
is as small as $\sim 6 ~R_\sun$, i.e., about $\sim 0.07$ times the 
inner critical Roche lobe size.  
The accretion disk is also exaggerated in this figure to easily see it. 
\label{m1370_tcrb_fig}}
\end{figure}

     One of our binary models is graphically shown 
in Figure \ref{m1370_tcrb_fig}.  A circular orbit is assumed. 
Our light curves include the contributions of three components: 
photosphere of WD, red giant (RG) companion, and
accretion disk (ACDK or DK).

\placefigure{m1370_tcrb_fig}

\subsection{White Dwarf Photosphere}
     We assume a black-body photosphere of the WD envelope
and calculate the $V$-magnitude 
with a window function given by Allen (1973).   
The photospheric surface is divided into 32 pieces 
in the latitudinal angle ($\Delta \theta = \pi/32$) 
and into 64 pieces in the longitudinal angle 
($\Delta \phi = 2 \pi/64$) as illustrated 
in Figure \ref{m1370_tcrb_fig}.   The contributions 
to the $V$ light are summed up from each piece.
We do not consider the limb-darkening effect 
because it is as small as $\Delta V \lesssim 0.2$ mag.
\par
     We have assumed the luminosity of the WD given by
\begin{equation}
L_{\rm WD} =  L_{\rm WD, 0} + 
{1 \over 2} {{G M_{\rm WD} \dot M_{\rm acc}} \over {R_{\rm WD}}},
\label{accretion-luminosity}
\end{equation}
where the first term is the intrinsic luminosity of the WD
calculated by our wind and steady hydrogen shell burning 
solutions and the second term is the accretion luminosity 
(e.g., Starrfield, Sparks, \& Shaviv 1988); 
$R_{\rm WD}$ is the radius of the WD (e.g.,  
$R_{\rm WD}= 0.0032 R_\sun$ for $1.37 M_\sun$ WD).
The accretion luminosity is as large as 
700 $L_\sun$ for a mass accretion rate of
$\dot M_{\rm acc} \sim 1 \times 10^{-7} M_\sun$ yr$^{-1}$
onto a $1.37 M_\sun$ WD.
Therefore, the accretion luminosity itself is negligibly small 
during the steady hydrogen shell burning phase 
($L_{\rm WD, 0} \gtrsim 10^4 L_\sun$) but may be important after 
the hydrogen burning diminishes.
In such a case, we recalculate the surface temperature
of the white dwarf, i.e.,
\begin{equation}
\sigma T_{\rm ph}^4 = {{L_{\rm WD}} \over {4 \pi R_{\rm WD}^2}},
\label{accretion-temperature}
\end{equation}
including the accretion luminosity.

\subsection{Companion's Irradiated Photosphere}
     We have to include the contribution of the companion star 
when it is strongly irradiated by the WD photosphere.  
The companion star is assumed to fill the inner 
critical Roche lobe as shown in Figure \ref{m1370_tcrb_fig}
for T CrB, 
%U Sco, V394 CrA, LMC RN, and T Pyx, 
but underfill for RS Oph, V745 Sco, and V3890 Sgr as described later.
Dividing the latitudinal angle into 
32 pieces ($\Delta \theta = \pi/32$) and the longitudinal angle into
64 pieces ($\Delta \phi = 2 \pi/64$), we have also summed up
the contribution from each area,
but we neglect both the limb-darkening effect and
the gravity-darkening effect of the companion star because
the both effects are negligibly small ($\Delta V \lesssim 0.2$ mag)
for the strongly irradiated companion.  
Here, we assume that 50\% of the absorbed energy 
is reemitted from the hemisphere of the companion with a black-body 
spectrum at a local temperature, that is, the efficiency of 
irradiation is $\eta_{\rm RG}=0.5$
unless otherwise specified.
\par
     The original (non-irradiated) photospheric temperature 
of the companion star is obtained from the light curve fitting,
for example, $T_{\rm RG, org} = 3100$ K for T CrB.
The irradiated surface temperature is roughly estimated by
\begin{equation}
\sigma T_{\rm RG}^4 \approx \eta_{\rm RG}
{{L_{\rm WD}} \over {4 \pi r^2}}\cos\theta 
+ \sigma T_{\rm RG, org}^4,
\label{irradiation_RG_angle}
\end{equation}
where $r$ is the distance from the WD 
and $\cos\theta$ is the inclination angle of the surface.
The irradiated temperature is as high as $T_{\rm RG} \sim 5000$ K 
for $\cos\theta=0.5$, $M_{\rm WD}= 1.37 ~M_\sun$ 
($L_{\rm WD}= 2 \times 10^{38}$ ergs~s$^{-1}$), 
and $r \sim 150 R_\sun$ (T CrB).  
Therefore, the effect of the irradiation
becomes very important in the late phase of outbursts 
when the irradiated hemisphere faces towards the Earth.
\par
     If the accretion disk around 
the WD blocks the light from the WD photosphere, it makes 
a shadow on the surface of the companion star.  Such an effect is also 
included in our calculation as explained below.

\subsection{Accretion Disk Surface}
     We have to include the luminosity coming from the accretion disk 
when it is irradiated by the WD photosphere.  Here, we assume 
a circular accretion disk, the edge of which is defined by 
\begin{equation}
R_{\rm disk} = \alpha R_1^*,
\label{accretion-disk-size}
\end{equation}
where $\alpha$ is a numerical factor indicating a size of 
the accretion disk, and $R_1^*$ the effective radius of 
the inner critical Roche lobe given by Eggleton's (1983) formula.
\par
     The surface of the accretion disk is strongly irradiated
and dragged outward by the strong wind from the WD 
during the nova outburst.  Its thickness may be 
increasing outward by both the ablation and
the Kelvin-Helmholtz instability.
We here approximate such a surface by 
\begin{equation}
h = \beta R_{\rm disk} \left({{\varpi} 
\over {R_{\rm disk}}} \right)^{\nu},
\label{flaring-up-disk}
\end{equation}
where $h$ is the height of the surface from the equatorial plane,
$\varpi$ the distance on the equatorial plane from the center of the WD, 
and $\beta$ is a numerical factor showing the degree of thickness.
For comparison, we have calculated the three cases of $\nu$, i.e., 
$\nu=9/8$ for the case of standard accretion disk model 
by Shakura \& Sunyaev (1973), $\nu=2$ and $\nu=3$ 
for the case of flaring-up disk \citep*{sch97}.
\par
     The surface of the accretion disk is divided into 32 pieces
logarithmically evenly in the radial direction and into 64 pieces 
evenly in the azimuthal angle as shown in Figure \ref{m1370_tcrb_fig}.
The outer edge of the accretion disk is also divided into 64 pieces
in the azimuthal direction and 2 (or 8) pieces in the vertical direction
by rectangles.  When the photosphere of the WD becomes very small, e.g.,
$R_{\rm disk}/R_{\rm ph} > 10$, we attribute the first 16 meshes 
to the outer region (from $\varpi=R_{\rm disk}$ to 
$\varpi=R_{\rm disk}/\sqrt{10}$) to avoid coarse meshes in the outer part, 
and then 16 meshes to the inner region (from 
$\varpi=R_{\rm disk}/\sqrt{10}$ to $\varpi=R_{\rm ph}$), 
each region of which is divided logarithmically evenly.  
\par
     The surface of the accretion disk 
absorbs photons and reemits in the same way as the companion does.
Each area of the heated disk surface emits blackbody photons at its
local temperature, $T_{\rm disk}$.
We assume that 50\% of the absorbed energy is 
emitted from the surface 
while the other is carried into interior of the accretion disk 
and eventually brought into the WD, that is, the efficiency of
irradiation is $\eta_{\rm DK}=0.5$.  
\par
     The non-irradiated temperature of the disk surface is assumed 
to be determined by the viscous heating of the standard accretion
disk model \citep{shak73}.  
Then, the original disk surface temperature is given by 
\begin{equation}
\sigma T_{\rm disk, org}^4 = {{3 G M_{\rm WD} \dot M_{\rm acc}} 
\over {8 \pi \varpi^3}}.
% + \eta_{\rm ir,DK} 
%{{L_{\rm WD}} \over {4 \pi r^2}} \cos\theta,
%\left[ 1 - \left({{R_{\rm WD}} \over {\varpi}}\right)^{1/2}
%\right]
\end{equation}
%where $\sigma$ is the Stefan-Boltzmann constant, 
%$r$ the distance from the WD center, and $\cos\theta$ 
%the incident angle of the surface \citep*[e.g.,][]{sch97}.  
The viscous temperature is estimated to be 
$T_{\rm disk,org}$ $\sim$ 9,000 K at
$\varpi= 1 R_\sun$ for $M_{\rm WD}= 1.37 M_\sun$ and 
$\dot M_{\rm acc}= 1 \times 10^{-7} M_\sun$ yr$^{-1}$. 
However, the viscous heating is negligibly
small when the disk surface is strongly irradiated
because of
$T_{\rm disk}$ $\sim$ 40,000---50,000 K at $\varpi= 1 R_\sun$ 
for the luminosity of the WD, $L_{\rm WD}= 50,000 L_\sun$, 
and the average inclination of the surface, $\cos\theta= 0.1$, 
which is estimated from \citep[e.g.,][]{sch97}
\begin{equation}
\sigma T_{\rm disk}^4 \approx \eta_{\rm DK}
{{L_{\rm WD}} \over {4 \pi \varpi^2}}\cos\theta
+ {{3 G M_{\rm WD} \dot M_{\rm acc}} 
\over {8 \pi \varpi^3}}.
% + \eta_{\rm ir,DK} 
%{{L_{\rm WD}} \over {4 \pi r^2}} \cos\theta,
%\left[ 1 - \left({{R_{\rm WD}} \over {\varpi}}\right)^{1/2}
%\right]
\label{irradiation_disk_angle}
\end{equation}
The outer rim of the accretion disk is not irradiated 
by the WD photosphere so that the temperature of the disk rim 
is simply assumed, for example, to be 2000 K for T CrB.   

\subsection{Radiation-induced Warping}
     Radiation-induced warping of accretion disks have been 
suggested by Pringle (1996).  Here, we have introduced 
a warping surface calculated by \citet*{mal96}.
In their formulation, when the tilt vector of the accretion disk surface
is defined by
\begin{equation}
\mbox{\boldmath{$l$}}(\varpi,t) = (\sin\tilde{\beta}\cos\tilde{\gamma},
\sin\tilde{\beta}\sin\tilde{\gamma},
\cos\tilde{\beta}),
\label{tilt-vector}
\end{equation}
one of their stationary solutions,
\begin{equation}
W= \tilde{\beta} e^{i \tilde{\gamma}},
\end{equation}
can be analytically solved as
\begin{equation}
M(1,2,2 i x) = e^{i x} {{\sin(x)} \over {x}},
\label{warping-solution-original}
\end{equation}
where $x \propto \sqrt{\varpi}$.  So, we have
\begin{equation}
x = c_1 \pi \sqrt{\varpi},
\end{equation}
and 
\begin{equation}
\tilde{\beta} = c_2 {{\sin(c_1 \pi \sqrt{\varpi})} 
\over {c_1 \pi \sqrt{\varpi}}},
\label{beta-angle}
\end{equation}
\begin{equation}
\tilde{\gamma} = c_1 \pi \sqrt{\varpi},
\label{gamma-angle}
\end{equation}
where $c_1$ and $c_2$ are numerical factors indicating a degree of
warping.  Here, we adopt $c_1=0.5$ and $c_2=-1.0$, unless otherwise 
specified.  Then, we have transformed the coordinates of the disk,
i.e., equation (\ref{flaring-up-disk}), by a rotation matrix of
\begin{eqnarray}
\mbox{\boldmath{$T$}} = 
\left( 
\begin{array}{ccc}
\cos\tilde{\beta}\cos\tilde{\gamma} & -\sin\tilde{\gamma} 
& -\sin\tilde{\beta}\cos\tilde{\gamma} \\
\cos\tilde{\beta}\sin\tilde{\gamma} & \cos\tilde{\gamma} 
& -\sin\tilde{\beta}\sin\tilde{\gamma} \\
\sin\tilde{\beta} & 0 & \cos\tilde{\beta}
\end{array}
\right).
\label{rotation-matrix}
\end{eqnarray}
Such an example is shown in Figure \ref{m1370_tcrb_fig} for T CrB.
As will be discussed later, the condition of warping instability 
is satisfied only in T CrB.

\placefigure{patch}
\placefigure{shadow}

\subsection{Numerical Method}
\begin{figure}
%\epsscale{.5}
%\plotone{patch.eps}
\plotone{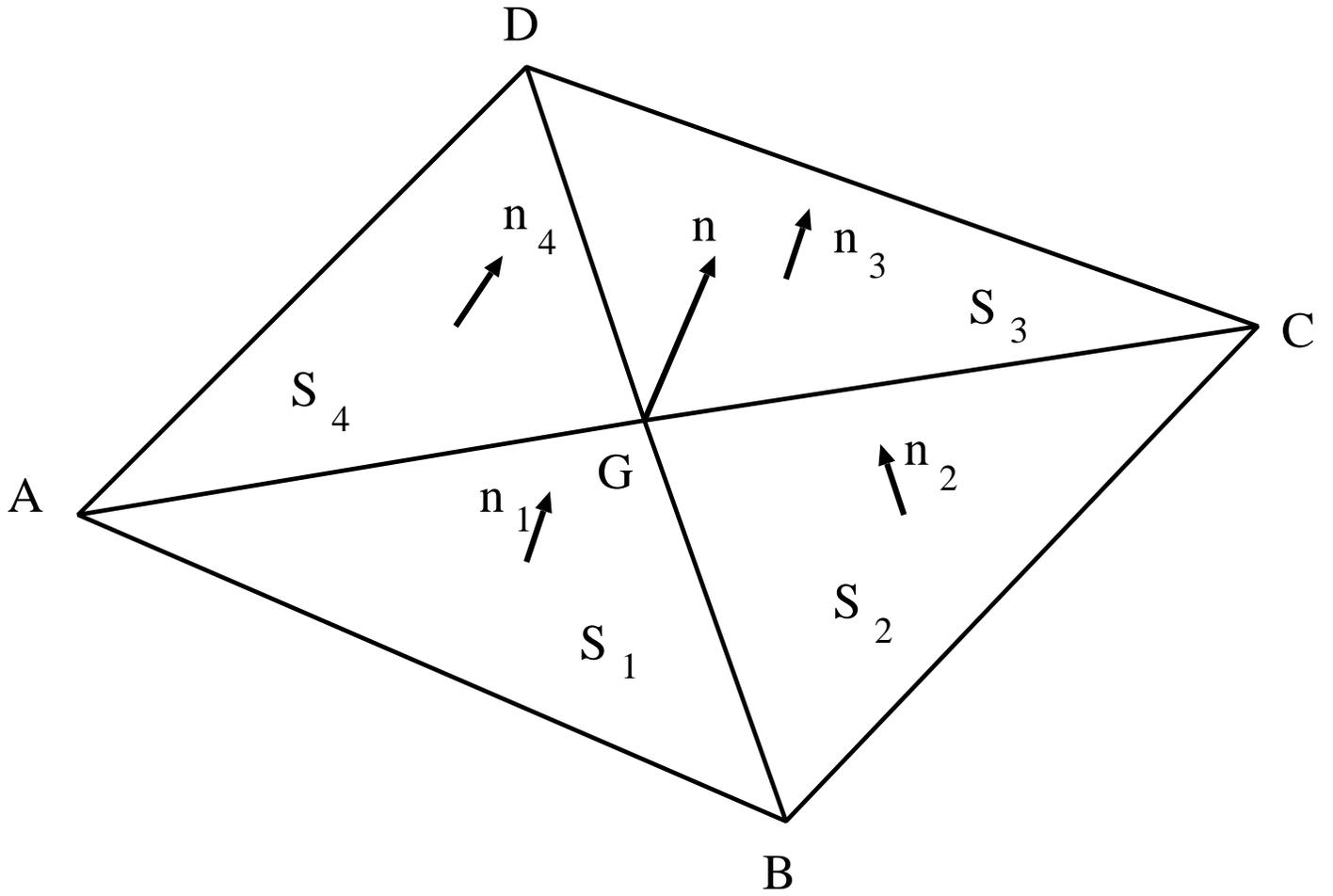}
%\plotfiddle{evolution1.ps}{5.0cm}{270}{0.4}{0.4}{-170}{220}
\caption{
Each patch is divided into four triangle parts as shown in the figure,
because each patch is almost a rectangular, 
but strictly speaking, not a rectangular for the Roche geometry of 
the companion star and for the warping disk surface. So that 
the normal unit vector to each patch, 
$\mbox{\boldmath $n$}$,
is approximately calculated from the four triangle parts 
by equation (\ref{center_of_gravity_patch}).
\label{patch}}
\end{figure}

     Total $V$ light of the system is calculated from 
the contribution of each patch with a window (response) 
function of the $V$-filter \citep[e.g.,][]{all73}.
Each patch is almost quadrangular but, strictly speaking, 
not quadrangular for the Roche geometry of the companion star 
and for the warping surface of the accretion disk, because 
the four vertices are not on a plane.  Therefore, we define
approximately the normal unit vector to each patch as follows:
the center of gravity of each patch can be defined by
\begin{equation}
\mbox{\boldmath $r$}_{\rm G} = {1 \over 4} \left(
\mbox{\boldmath $r$}_{\rm A} + \mbox{\boldmath $r$}_{\rm B} +
\mbox{\boldmath $r$}_{\rm C} + \mbox{\boldmath $r$}_{\rm D}
\right), 
\label{center_of_gravity_patch}
\end{equation}
where G denotes the center of gravity of the patch, 
and A, B, C, and D are
the apexes of each quadrangle as illustrated in Figure 
\ref{patch}.  Defining 
\mbox{\boldmath $n$}$_1$ as the normal unit vector to $\triangle$ABG,
\mbox{\boldmath $n$}$_2$ to $\triangle$BCG,
\mbox{\boldmath $n$}$_3$ to $\triangle$CDG, and
\mbox{\boldmath $n$}$_4$ to $\triangle$DAG,
we obtain an approximate normal unit vector of \mbox{\boldmath $n$} 
to patch $\Box$ABCD by
\begin{equation}
\mbox{\boldmath $n$} = {{
S_1 \mbox{\boldmath $n$}_1 + S_2 \mbox{\boldmath $n$}_2 +
S_3 \mbox{\boldmath $n$}_3 + S_4 \mbox{\boldmath $n$}_4}
\over {S_1+S_2+S_3+S_4}},
\end{equation}
where each area of the triangles is corresponding to
\begin{eqnarray}
S_1 &=& \triangle {\rm ABG},~~~
S_2 = \triangle {\rm BCG}, \cr
S_3 &=& \triangle {\rm CDG},~~~
S_4 = \triangle {\rm DAG},
\end{eqnarray}
as shown in Figure \ref{patch}, and the total area of this patch 
$\Box$ABCD is given by
\begin{equation}
S = S_1 + S_2 + S_3 + S_4.
\end{equation}
\par
\begin{figure}
%\epsscale{.5}
%\plotone{shadow.eps}
\plotone{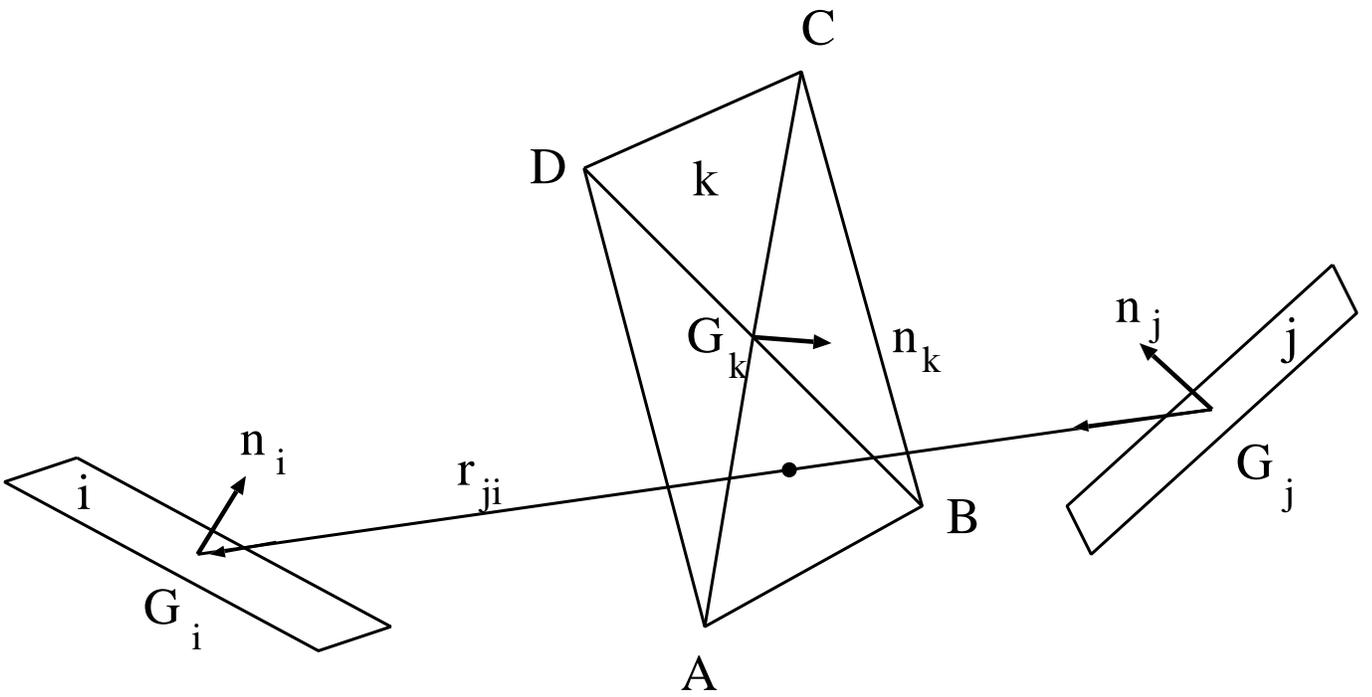}
%\plotfiddle{evolution1.ps}{5.0cm}{270}{0.4}{0.4}{-170}{220}
\caption{
When patch $k$ blocks the light from patch $j$ to 
patch $i$, we do not sum up the contribution 
from $j$ to $i$ as shown in equations (\ref{sum_condition_ij_angle}),
(\ref{sum_up_ij}), and (\ref{sum_up_ij_block}).  
\label{shadow}}
\end{figure}

     Then, patch $i$ receives flux from patch $j$ 
as much as
\begin{equation}
\Delta L_{ij} = 
{{\sigma T_j^4 S_j 
(\mbox{\boldmath $n$}_j \cdot \mbox{\boldmath $r$}_{ji})
S_i (\mbox{\boldmath $n$}_i \cdot \mbox{\boldmath $r$}_{ij})
}
\over {2 \pi |\mbox{\boldmath $r$}_{ji}|^4}},
\end{equation}
where 
\begin{equation}
\mbox{\boldmath $r$}_{ij}= 
\mbox{\boldmath $r$}_{{\rm G},j} - \mbox{\boldmath $r$}_{{\rm G},i},
\end{equation}
\begin{equation}
\mbox{\boldmath $r$}_{ji}= 
\mbox{\boldmath $r$}_{{\rm G},i} - \mbox{\boldmath $r$}_{{\rm G},j},
\end{equation}
as illustrated in Figure \ref{shadow}.  
We sum up all combinations of pair patches on the condition
of
\begin{equation}
(\mbox{\boldmath $n$}_j \cdot \mbox{\boldmath $r$}_{ji}) \ge 0,
\mbox{~~and~~}
(\mbox{\boldmath $n$}_i \cdot \mbox{\boldmath $r$}_{ij}) \ge 0,
\label{sum_condition_ij_angle}
\end{equation}
and obtain the total flux from all patches, i.e.,
\begin{equation}
L_i = \sum_j \sum_k \xi_{i,j,k} 
{{\sigma T_j^4 S_j (
\mbox{\boldmath $n$}_j \cdot \mbox{\boldmath $r$}_{ji})
S_i (\mbox{\boldmath $n$}_i \cdot \mbox{\boldmath $r$}_{ij})
}
\over {2 \pi |\mbox{\boldmath $r$}_{ji}|^4}}.
\label{sum_up_ij}
\end{equation}
Here, we set 
\begin{eqnarray}
\xi_{i,j,k}=\cases{1, & \mbox{~~for ~~not-blocked} \cr
0, & \mbox{~~for ~~blocked} \cr
}
\label{sum_up_ij_block}
\end{eqnarray}
depending whether the line connecting the centers of gravity
is blocked by patch $k$ as shown in Figure \ref{shadow}.
\par
     The surface (patch $i$) absorbs photon flux and reemits
by a black-body spectrum with a local temperature of $T_i$, which is
determined by
\begin{equation}
\sigma T_i^4 S_i = \eta_{\rm BB} L_i + \sigma T_{i,{\rm org}}^4 S_i, 
\end{equation}
where $\eta_{\rm BB}$ is the efficiency of the irradiation,
$T_{i,{\rm org}}$ the unheated temperature of the surface.
We adopt 50\% efficiency, i.e., $\eta_{\rm BB}= 0.5$ unless
otherwise specified.
\par
     Using the window functions, 
$W_V(\lambda)$ and $W_B(\lambda)$, 
for $V$- and $B$-filters \citep{all73}, respectively,
we have calculated $V$-, $B$-, and bolometric magnitudes 
of the patches, i.e.,
\begin{equation}
\Delta E_i({\rm V}) = 
(\mbox{\boldmath $n$}_i \cdot \mbox{\boldmath $e$})
S_i \int_0^{\infty} W_V (\lambda) 
B_{\lambda}(T_i) d \lambda,
\end{equation}
\begin{equation}
\Delta E_i({\rm B}) = 
(\mbox{\boldmath $n$}_i \cdot \mbox{\boldmath $e$})
S_i \int_0^{\infty} W_B (\lambda) 
B_{\lambda}(T_i) d \lambda,  
\end{equation}
\begin{equation}
\Delta E_i({\rm bol}) = 
(\mbox{\boldmath $n$}_i \cdot \mbox{\boldmath $e$})
S_i \sigma T_i^4,
\end{equation}
respectively.   
Here, the unit vector \mbox{\boldmath $e$} denotes the direction
from the patch to the Earth, $B_{\lambda}(T_i)$ the Planck function
for the temperature of $T_i$. 
We have summed up the patches having 
$(\mbox{\boldmath $n$}_i \cdot \mbox{\boldmath $e$}) > 0$, i.e.,
\begin{equation}
E(V) = \sum_i \Delta E_i(V),
\mbox{~~for~~} 
(\mbox{\boldmath $n$}_i \cdot \mbox{\boldmath $e$}) > 0
\end{equation}
\begin{equation}
E(B) = \sum_i \Delta E_i(B),
\mbox{~~for~~} 
(\mbox{\boldmath $n$}_i \cdot \mbox{\boldmath $e$}) > 0
\end{equation}
\begin{equation}
E({\rm bol})= \sum_i \Delta E_i({\rm bol}),
\mbox{~~for~~} 
(\mbox{\boldmath $n$}_i \cdot \mbox{\boldmath $e$}) > 0.
\end{equation}
We have finally calculated the $V$-magnitude, $m_V$, 
$B$-magnitude, $m_B$, and the bolometric magnitude, $m_{\rm bol}$, as
\begin{equation}
m_V = {5 \over 2}\left(- \log E(V) + 33.582 \right)
+ m_{V,0} + 5 ~ \log({{d} \over {10}}),
\end{equation}
\begin{equation}
m_B = {5 \over 2}\left(- \log E(B) + 33.582 \right)
+ m_{B,0} + 5 ~ \log({{d} \over {10}}),
\end{equation}
\begin{equation}
m_{\rm bol} = {5 \over 2}\left(- \log E({\rm bol}) + 33.582 \right)
+ 4.75 + 5 ~ \log({{d} \over {10}}),
\end{equation}
where $d$ is the distance to the object in units of pc, 
$m_{V,0}$ is a constant determined to satisfy
\begin{equation}
m_V = m_{\rm bol}, \mbox{~~at~~} T=6500 \mbox{ K},
\end{equation} 
and $m_{B,0}$ is also a constant determined to satisfy
\begin{equation}
m_B = m_V, \mbox{~~at~~} T=9600 \mbox{ K (A0)},
\end{equation}
so that we use 
\begin{equation}
m_{V,0} = 4.52,
%%% 4.75-0.2777+0.04788
\end{equation} 
and
\begin{equation}
m_{B,0} = 4.91.
%%% 4.75+0.0343+0.2706-0.1423
\end{equation}
In addition, in order to compare our brightness with
the recent CCD photometry, we also calculate the Cousins 
$R_{\rm c}$- and $I_{\rm c}$-magnitudes with the window functions
given by Bessell (1990) along the same way as those described above
for $V$- and $B$-magnitudes.  Then, we adopt
\begin{equation}
m_{R,0} = 4.91,
%%% 4.75+0.161
\end{equation} 
and
\begin{equation}
m_{I,0} = 4.50.
%%% 4.75-0.251
\end{equation} 

%         DMB = -2.5*DLL+33.582*2.5+4.75+5.D0*LOG10(D110)
%         DMV = DMB+BOC
%
%   Bolometric correction = 0  at T=6500 K
%
%      BOCV=-LOG10(BOCV1)/0.4D0-0.2777D0
%      DMB=-2.5D0*DLL+33.582*2.5+4.75+5.*LOG10(D110)
%      MVI=-2.5*DLL+33.582*2.5+4.75+BOCV
%      DMVI=DMB+BOCV
%      BOLF=LOG10(LUM/4./3.141/D/D)
%      VIFLUX=LOG10(LUM/4./3.141/D/D)-BOCV/2.5

\par
     The most time-consuming part of the calculation is 
proportional to cubic of the patch number $N$, that is, 
$\propto N^3$.  Therefore, it is hard to increase the patch 
number substantially more than the present maximum case, i.e.,
\begin{equation}
N= 32 \times 64 + 32 \times 64 + (32 \times 64 \times 2 
+ 8 \times 64)=8704,
\end{equation}
because it takes about a half day on the Compaq (old DEC) 
Alpha 21264 (500 MHz) machine but about two days 
on the Intel Pentium III (500 MHz) machine to compute one time
sequence for the 1999 U Sco outburst, about 3000 steps for 
the $m_V$ and $m_B$ light curve calculations \citep{hkkm00}. 

%%%%% \placefigure{m_giants_cc}

\section{T CORONAE BOREALIS}
\subsection{Nature of T CrB Outbursts}
      T Coronae Borealis (T CrB) is one of the well observed recurrent 
novae and characterized by a secondary maximum occurring 
$\sim 150$ days after the primary peak.  
Historically, T CrB bursted twice, in 1866 and 1946, 
with the light curves very similar each other 
\citep[e.g.,][b, c, d]{pet46a}.
Large ellipsoidal variations of optical and infrared light curves 
in quiescence suggest that an M3 giant 
fills its Roche lobe \cite[e.g.,][]{bai75, lin88, 
yud93, lei97, shahb97, bel98}.
From the ellipsoidal variations, the orbital period and ephemeris were
determined to be HJD 2,431,931.05 $+$ 227.67$E$ at the epoch of
spectroscopic conjunction with the M-giant in front \citep{lin88}.
\par
     There have been debates on the nature of the hot component 
of the binary system. 
Radial velocity curves of $K_1 \sim 31$ km s$^{-1}$ ({\it hot} component) 
and $K_2 \sim 23$ km s$^{-1}$ ({\it cool} component) 
indicated the mass ratio of $q=M_2/M_1 \sim 1.4$
\citep{kra58, pac65, ken86}.  
Adopting the inclination angle of
$i = 68 \arcdeg$, the resulting masses were 
$M_1= M_{\rm hot} \sim 1.9 M_\sun$ and 
$M_2= M_{\rm cool} \sim 2.6 M_\sun$, thus placing the hot component
above the Chandrasekhar limit.  Therefore,
\citet{web76} and \citet{web87} proposed 
an outburst mechanism of T CrB 
based on their main-sequence accretor model:
the outburst was caused by the transfer of a chunk of matter 
ejected by the giant and, then, by the subsequent shock 
dissipation of the kinetic energy.  
This accretion event model was investigated 
further by 3D numerical simulations \citep*{can92, ruf93}.
\par
     However, \citet*{sel92} 
have been opposed to the main-sequence accretor model from their
analysis of {\it IUE} data in quiescence, which 
indicates the existence of a mass-accreting white dwarf (WD)
as follows:
(1) the bulk of the luminosity is emitted in the UV with little 
contribution to the optical; 
(2) the presence of strong \ion{He}{2} and \ion{N}{5} emission
lines suggests temperatures of the order of $10^{5}$ K;
(3) the rotation broadening of the high-excitation lines is larger
than 1300 km s$^{-1}$ without inclination correction.
Moreover, the detection of X-rays \citep*{cor81} 
and the presence of flickering in
the optical light curves \citep[e.g.,][]{ian64, zam98}
are also naturally explained in terms 
of accretion onto a WD.
\par
     The mass accretion rate estimated 
by \citet{sel92} in quiescence is very high 
($\dot M_{\rm acc} \sim 2.5 \times 10^{-8} M_\sun$ yr$^{-1}$) 
and is exactly required by the thermonuclear runaway (TNR) theory 
to produce a TNR event every 80 yr on a massive ($\gtrsim 1.3 M_\sun$) 
WD.  They also pointed out that the measurement of the emission-line
radial velocities \citep{kra58} was intrinsically difficult
and uncertain.  
Thus, the 1866 and 1946 outbursts can be interpreted 
in terms of a TNR event on a very massive WD \citep[e.g.,][]{sta85}.

\subsection{Very Massive White Dwarf in T CrB}
     Rapid decline rates of the light curves indicate 
a very massive WD close to the Chandrasekhar limit. 
Assuming the solar metallicity ($Z=0.02$) of the WD envelope,
\citet{kat95, kat99} calculated nova light curves 
for the WD mass of $M_{\rm WD}= 1.2$, 1.3, 1.35 and 1.377 $M_\sun$,
and found that the light curve of the $1.377 M_\sun$ model 
is in better agreement with the observational light curve 
of T CrB than the other lower mass models.  
\par
     Recently, other observational supports of 
a massive WD in T CrB have been reported.
\citet{bel98} derived a
permitted range of binary parameters, 
$M_{\rm WD}= 1.2 \pm 0.2 M_\sun$ 
and $q= M_{\rm RG}/ M_{\rm WD} \sim 0.6$, i.e.,  
$M_{\rm RG} \sim 0.6$---$0.8 M_\sun$, 
from amplitude of the ellipsoidal
variability and constraints from the orbital solution of M-giants. 
In \citet{shahb97}, a massive WD of 
$M_{\rm WD}= 1.3-2.5 M_\sun$ is suggested
from the infrared light curve fitting.
Moreover, the measurement of the radial velocities \citep{kra58}
has been revised including the effect of gas-streams around 
the hot component \citep{hri98}.  
\citet{hri98} estimated the masses,
$M_{\rm WD}= 1.2 \pm 0.2 ~M_\sun$ for the hot component and
$M_{\rm RG}= 1.38 \pm 0.2 ~M_\sun$ for the M-giant, assuming
the inclination of $i=68\arcdeg$.
Combining these permitted ranges of the WD mass in T CrB, 
we may conclude that a mass of the WD is between 
$M_{\rm WD}= 1.3-1.4 M_\sun$, 
which is very consistent with the light curve analysis 
$M_{\rm WD} \sim 1.37-1.38 M_\sun$ by Kato (1999).

\subsection{Secondary Maximum of Outbursts}
     The secondary maximum in outbursts is rarely
observed in fast novae and in other recurrent novae.  
Selvelli et al. (1992) suggested a possibility of an irradiated
stationary shell around the binary system, although the presence
of such a shell is just speculation.  Recently, Hachisu \& Kato (1999)
proposed another mechanism of the second peak, i.e.,
radiation-induced tilting disk around a massive WD.  
The main results of their analysis are:
(1) the first peak is naturally reproduced by a fast developing
photosphere of the WD envelope based on the TNR model of
a very massive WD, $M_{\rm WD} \sim 1.35 M_\sun$;
(2) the second peak is not fully reproduced by an irradiated M-giant 
model as simply estimated by \citet{web87};
(3) instead, the second peak can be well reproduced if an irradiated 
tilting accretion disk around the WD is introduced together with
the partly irradiated M-giant companion.  
\par
     Such a radiation-induced, tilting instability of an accretion 
disk has been suggested by \citet{pri96} when a central star 
is as luminous as the Eddington limit.
This radiation-induced instability sets in if the condition
\begin{eqnarray}
{{\dot M_{\rm acc}} \over {10^{-7}~M_\sun \mbox{~yr}^{-1}}}
& \lesssim &
\left( {{R_{\rm disk}} \over {10~R_\sun}} \right)^{1/2}
\left( {{L_{\rm bol}} \over {2 \times 10^{38} 
\mbox{~ergs~s}^{-1}}} \right) \cr
&\times & \left( {{R_{\rm WD}} \over {0.003 ~R_\sun}} \right)^{1/2}
\left( {{M_{\rm WD}} \over {1.37 ~M_\sun}}\right)^{-1/2}
\label{radiation_condition_T_CrB}
\end{eqnarray}
is satisfied \citep*{sou97}, where $R_{\rm disk}$ is the radius 
at the edge of the optically thick accretion disk.
\citet{sel92} estimated the accretion rate of T CrB as low as 
$\dot M_{\rm acc} \sim 0.25 \times 10^{-7} M_\sun$ yr$^{-1}$, 
which meets condition (\ref{radiation_condition_T_CrB}) above
even if the optically thick radius is as small as 
$R_{\rm disk} \sim 1 ~R_\sun$.
Therefore, it is likely that the radiation induced instability 
grows during the outburst in T CrB.  The growth timescale of warping
is estimated by Pringle (1996) and Livio \& Pringle (1996) as 
\begin{eqnarray}
\tau_{\rm prec} & \simeq & 30
\left( {{M_{\rm disk}} \over {10^{-8} ~M_\sun}} \right)
\left( {{M_{\rm WD}} \over { 1.37 ~M_\sun}} \right) 
\left( {{R_{\rm disk}} \over { R_\sun}} \right)^{-1/2} \cr 
&\times & 
\left( {{L_{\rm bol}} \over {2 \times 10^{38} \mbox{~ergs~s}^{-1}}}
\right)^{-1} \mbox{~day~},
\label{precession_time_T_CrB}
\end{eqnarray}
where we assume that the $\alpha$-parameter in the Shakura-Sunyaev
(1973) standard accretion disk is $\alpha \sim 0.1$. 
Thus, the growth timescale is short enough to excite warping 
of the accretion disk.

\subsection{Revised Model of Warping Accretion Disk}
      We have revised \citet{hac99k} model to be consistent with
the observations described below.  The main difference is in 
the accretion disk, which plays an essential role in the mid/late
phase of the $V$ light curve.
\citet{hac99k} assumed that the size of the accretion disk 
is as large as 0.7 times the Roche lobe size, 
%$R_{\rm disk} = 0.7~R_1^*$ in equation (\ref{accretion-disk-size}), 
i.e., $R_{\rm disk} \sim 50 ~R_\sun$.
This is because the mass transfer is not a wind-fed type
but a Roche lobe overflow type \citep[e.g.,][]{sch97}.
However, \citet{sel92} estimated the optically thick region 
of the accretion disk in quiescence is as small as 
$R_{\rm disk} \lesssim 1 ~R_\sun$ because the disk luminosity
contributes mostly to the satellite UV.  
\citet{bel98} pointed out
that the optical contribution of the accretion disk is clearly 
visible in the $U$, $B$, and $V$ bands in 1981--85 
(optically high state) while it is practically absent in 1990--94
(optically low state) as first shown by \citet{pay46}.  
Thus, Belczy{\'n}ski \& Miko{\l}ajewska derived the disk radius
(hot spot) of $R_{\rm disk} \sim 0.1 a ~(20 ~R_\sun)$ 
in the optically high state.
\par
     Therefore, we here adopt a small size of the optically thick disk,
i.e., $R_{\rm disk} \sim 1~R_\sun$ just before the outburst. 
A large accretion disk of $\sim 50~R_\sun$ requires that
(1) the irradiation efficiency is too small and that (2)
the disk is inclined from too early phase of the outburst.
If we introduce a small size of the accretion disk,
we do not need to assume such unnatural assumptions.
This accretion disk satisfies the warping condition 
(\ref{radiation_condition_T_CrB}) so that we
adopt an analytic form proposed by \citet{mal96}.

\placefigure{vmag1370va1_tcrb1946}

\subsection{Light Curve of Early Phase}
     The very fast decay during the first 10 days can be well 
reproduced only by a WD photosphere of a very massive WD,
$M_{\rm WD} \sim 1.35$---$1.37 ~M_\sun$, as already shown by
\citet{kat99} and \citet{hac99k}.  
We have found that the decay rate of 
the $1.37 ~M_\sun$ light curve is in better agreement with 
the observational points than the others.  
The difference from \citet{hac99k} comes from
the difference of the disk size which affects the light curve 
10---30 days after maximum.
\citet{hac99k} adopted a large accretion disk which 
in general contributes more to the luminosity.  
Therefore, they reduced the WD luminosity to be consistent 
with the observational points by occulting the WD photosphere
with the accretion disk.  
In the present model, on the other hand, the small disk
cannot occult a large part of the WD photosphere 
during the day 10---30.  Thus, we need a more massive 
WD to generate a faster decline.
Therefore, we have determined the WD mass of 
$M_{\rm WD} =1.37 \pm 0.01 ~M_\sun$ in T CrB, giving
a much better fit with the observational points even
in the day 0---10.
\par
     \citet{deu48} suggested from the observations of 
\citet{ash46}, \citet{wri46}, and \citet{pet46a} that
the color index of T CrB was about zero February 12 
(4 days after maximum).  Our model indicates the photospheric color
of $(\bv)_0= -0.2$ 
($T_{\rm ph}\sim 18,500$ K, $R_{\rm ph} \sim 18 ~R_\sun$),
which is consistent with the observational color
because the reddening is about $E(\bv)=0.15$.
\par
     The mass of the companion is estimated as follows:
Adopting $M_{\rm WD}= 1.37 ~M_\sun$ and the radial velocity 
of the cool component $K_2 = 23.32$ km s$^{-1}$ \citep{ken86}, i.e.,
\begin{equation}
f(M)= {{M_{\rm WD}^3 \sin^3 i} \over {(M_{\rm WD}+M_{\rm RG})^2}}
= 0.30 \pm 0.01 M_\sun,
\label{mass_function_TCrB}
\end{equation}
we have $M_{\rm RG}= 1.30$, 1.25, 1.16, and $1.0 ~M_\sun$, 
for the inclination angle of $i=70\arcdeg$, $68\arcdeg$, 
$65\arcdeg$, and $60\arcdeg$, respectively. 
Following \citet{bel98},
$q \equiv M_{\rm RG}/ M_{\rm WD} \sim 0.6$ and an inclination
angle of $i \sim 60\arcdeg$,
we have assumed the red giant mass of $M_{\rm RG}= 1.0~M_\sun$
($i= 60\arcdeg$).
So we show here only the results for $M_{\rm RG}= 1.0 M_\sun$.
In this case, the separation is $a= 209.2 R_\sun$, the effective 
radius of the inner critical Roche lobe for the WD component is 
$R_1^*= 85.0 R_\sun$, the effective radius for the red giant 
companion is $R_2= R_2^*= 73.6 R_\sun$.
We have examined the other two cases of the red giant mass,
i.e., $M_{\rm RG}= 1.25 ~M_\sun$ ($i=68\arcdeg$) 
and $M_{\rm RG}= 1.16 ~M_\sun$ ($i=65\arcdeg$), 
and obtained similar light curves to that of
$M_{\rm RG}= 1.0 ~M_\sun$ ($i=60\arcdeg$).

% AV. DM2/DT=  0.4356E-05  WIND=  0.8856   M1= 1.3700  M2= 1.0000
% AA=209.173  RL1= 85.017  RL2= 73.632  RDISK=  4.251  FLUPD=  0.010

%------ v2=23.32 km/s (Kenyon and Garcia, AJ, 91, 125, 1986)
% WD =   1.370     RG =  0.9930     I= 60 DEG
% WD =   1.370     RG =   1.029     I= 61 DEG
% WD =   1.370     RG =   1.063     I= 62 DEG
% WD =   1.370     RG =   1.096     I= 63 DEG
% WD =   1.370     RG =   1.129     I= 64 DEG
% WD =   1.370     RG =   1.160     I= 65 DEG
% WD =   1.370     RG =   1.191     I= 66 DEG
% WD =   1.370     RG =   1.220     I= 67 DEG
% WD =   1.370     RG =   1.248     I= 68 DEG
% WD =   1.370     RG =   1.275     I= 69 DEG
% WD =   1.370     RG =   1.301     I= 70 DEG

\par
\begin{figure}
%\epsscale{.5}
%\plotone{vmag1370va1_tcrb1946.eps}
\plotone{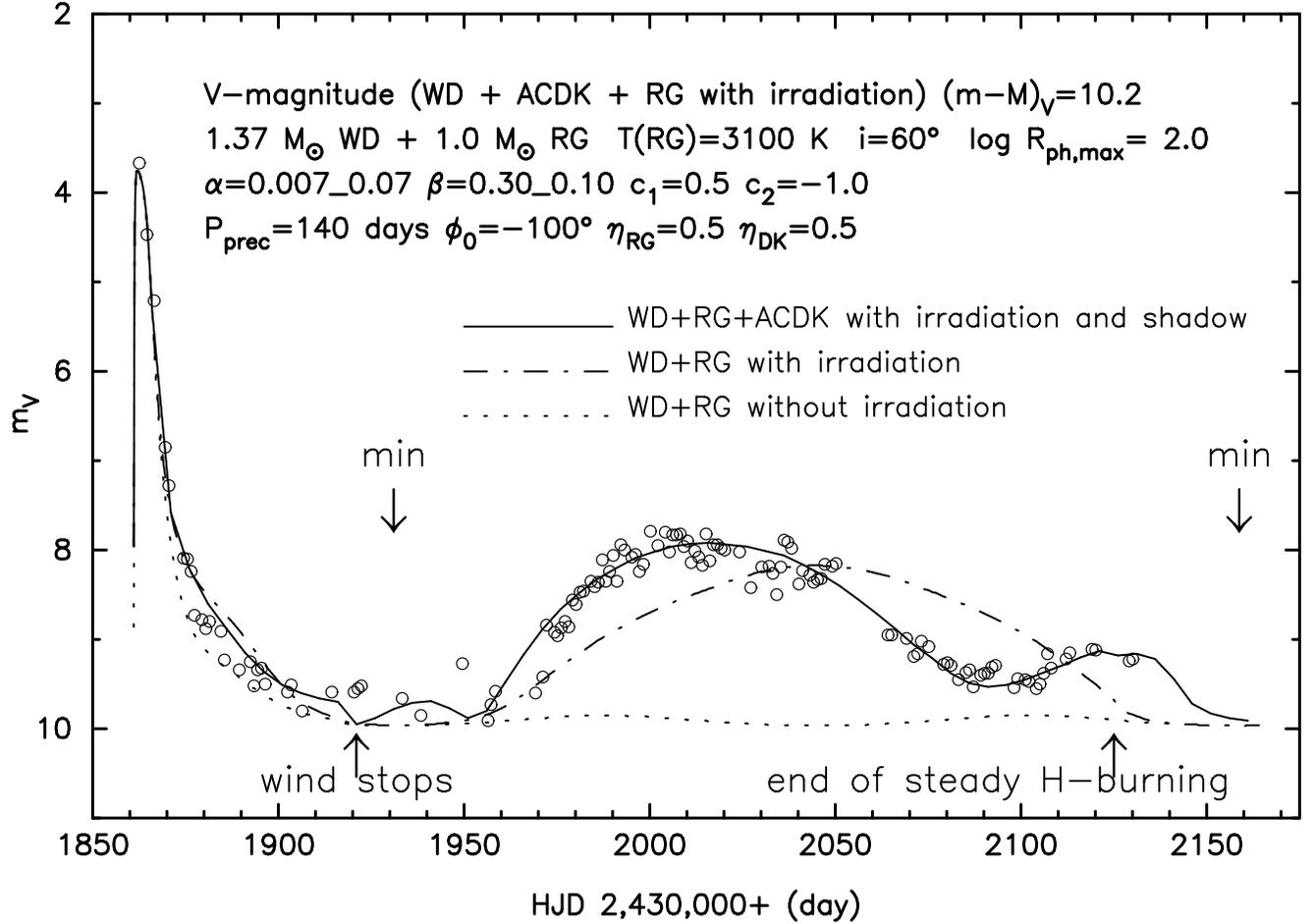}
%\plotfiddle{evolution1.ps}{5.0cm}{270}{0.4}{0.4}{-170}{220}
\caption{
Model light curves are plotted against time (HJD 2,430,000+) together
with the observational points 
\citep[{\it open circles: \rm}][]{pet46a, pet46b, pet46c, pet46d}.
{\it Dotted}: the total $V$ light of the white dwarf (WD) photosphere and
the red giant (RG) photosphere without irradiation.  Large arrows 
attached by "min" indicate epochs at the spectroscopic conjunction
with the M-giant in front.  An ellipsoidal variation can be seen
in the late phase.
{\it Dash-dotted}: the total $V$ light of the WD photosphere and 
the RG photosphere irradiated by the WD.  
The observed second peak cannot be reproduced only by the irradiation
of the M-giant.
{\it Solid}: the total $V$ light of the WD photosphere, the RG photosphere
with irradiation, and the accretion disk (ACDK) surface heated-up 
by the hot component.  
The parameters specifying the light curves are shown in the figure.
\label{vmag1370va1_tcrb1946}}
\end{figure}

     Figure \ref{vmag1370va1_tcrb1946} shows the best fitted 
model to the 1946 outburst together with the observational
points.  Here, we assume the solar metallicity ($Z=0.02$) 
and hydrogen content of $X=0.70$ for the WD envelope.  We also assume 
that the $V$ magnitude reached its maximum on HJD 2,431,861.
The dotted curve denotes the total $V$ light from the WD photosphere 
and the non-irradiated red giant at $T_{\rm RG, org}= 3100$ K.
Ellipsoidal light variations can be seen in the late stage (dotted).
The effective temperature of the M-giant is estimated to be
$T_{\rm RG, org}= 3100$ K from our light-curve fitting, which  
is a little bit lower than $T_{\rm eff}= 3560$ K by \citet{bel98},
but roughly consistent with $T_{\rm eff}= 3200$ K by \citet{shahb99}.  
\par
     If we include irradiation of the companion, the light curve
changes to the dash-dotted curve.   This irradiation effect
can be well seen both 10---40 days after the optical maximum
and 120---260 days after the optical maximum.
The irradiation effect was once examined and discarded 
by \citet{web87} as a main mechanism of the secondary maximum,
mainly because its maximum largely deviates from the observational 
secondary maximum as seen in the figure.
\par
     The solid curve includes further the irradiated accretion disk.
Here, we have started the calculation by assuming
the disk size of $\alpha= 0.007$ in equation (\ref{accretion-disk-size}),
i.e., $R_{\rm disk}= 0.6 ~R_\sun$.
\par
     The photospheric radius expands to $R_{\rm ph} \sim 100 ~R_\sun$ 
in order to fit the peak luminosity of the $V$ light.  We have
derived the distance modulus of $(m-M)_V= 10.2$. 
The light curve during the first 10 days of the 1946 
outburst can be well reproduced only with the fast-developing WD
photosphere of $M_{\rm WD}= 1.37 ~M_\sun$ as already shown by
\citet{kat99}.
The envelope mass of the WD at the optical maximum is 
$\Delta M= 3.3 \times 10^{-6} M_\sun$ from our wind solutions.
Then, the mass accretion rate was estimated to be
$\dot M_{\rm acc}= 4 \times 10^{-8} M_\sun$ yr$^{-1}$ 
during the quiescent phase between 1866 and 1946.  
This value is roughly consistent with the 
estimation of $2.5 \times 10^{-8} M_\sun$ yr$^{-1}$ by 
\citet{sel92}.  
\par
     After the WD photosphere shrinks to smaller than 
$R_{\rm disk} \sim 0.6 ~R_\sun$, the accretion disk reappears
but its contribution to the $V$ light is small as shown 
in Figure \ref{vmag1370va1_tcrb1946}, because
radiation emits mainly in UV.
We assume a flaring-up rim of the accretion disk,
i.e., $\beta= 0.30$ in equation (\ref{flaring-up-disk}), during
the strong wind phase for the 1946 outburst of T CrB, 
since \citet{hkkm00} suggested that the rim of the accretion disk is 
flaring-up during the 1999 outburst of U Sco,
Then, the $V$ light is reduced 20---30 days after maximum because
the disk rim makes shadow on the red giant component. 
The warping of the accretion disk does not occur yet, because
the strong wind suppresses the warping.  The spherically
symmetric wind momentum is much larger than the radiation
momentum as shown by \citet{kat94h}.

\subsection{Light Curve Model of the Second Peak}
     Warping (or wobbling) instability of the accretion disk begins to 
grow after the strong wind stops $\sim 60$ days after maximum 
(on HJD 2,431,920) and, as a result,
the $V$ light attains another peak.
We expect that the radiation induced
warping of the accretion disk 
grows in a time scale of a few tens of days
as derived in equation (\ref{precession_time_T_CrB})
and further that the mass transfer from the red giant to the WD 
increases because the red giant surface is strongly irradiated and
the Roche lobe overflow is enhanced.  Such enhancement of the mass
transfer certainly increases the radius of the optically 
thick region of the accretion disk up to $\sim 10$---$20 ~R_\sun$
as suggested by \citet{bel98}.
Thus, we conclude that the radius of the optically thick
accretion disk increased from $\alpha= 0.007$ 
($R_{\rm disk} \sim 0.6 ~R_\sun$) to $\alpha= 0.07$ 
($R_{\rm disk} \sim 6 ~R_\sun$) during the day $\sim$60---150,
by fitting with the observational points as shown in Figure
\ref{vmag1370va1_tcrb1946}.  
\par
     The second peaks of the 1866 and 1946 outbursts are
very similar.  This is explained as
follows:  (1) The first peak ends up when the wind stops
about 60 days after maximum.
This period ($\sim 60$ days), between the optical peak and 
the termination of the strong wind, is almost uniquely 
determined by the WD mass.
(2) The second peak begins to rise
when the warping of the accretion disk grows, which is about 120 days 
after maximum.  The growth time is estimated to be 
about a few times 30 days 
as in equation (\ref{precession_time_T_CrB}).
This growth time depends on the mass ($M_{\rm disk}$)
and the optically thick radius ($R_{\rm disk}$) of 
disk, which are determined mainly by the mass transfer
rate from the red giant companion. We can naturally expect
the similar time scales between the 1866 and 1946 outbursts, 
because the irradiation of the red giant by the WD photosphere
is almost the same.
(3) Thus, we are able to obtain quite similar light curves 
if the orbital phases of the 1866 and 1946 outbursts 
happen to be the same 
\citep[see the light curves of the 1866 and 1946 outbursts,][]{pet46d}.  
Accidentally, the two orbital phases were almost similar to
each other.
\par
     The precessing period of the warping disk is similar
to the growth time of warping \citep[e.g.,][]{liv96}.
About 60\% faster precessing angular velocity 
than the orbital motion is required from 
the phase relation between the rising shoulder of the second peak 
near HJD 2,431,960, a small dip near HJD 2,432,090, and then a small 
bump near HJD 2,432,120 as shown in Figure \ref{vmag1370va1_tcrb1946}.
These dips are caused by a large shadow on the companion cast 
by the accretion disk.  
In other words, the accretion disk blocks 
the light from the WD photosphere and makes its shadow 
on the companion surface.  Finally, we have determined 
$P_{\rm prec}= 145$ days by fitting.  
This period is similar to the 164 days precessing period of SS 433.
\par
     \citet{web87} summarized the observational development
of the 1946 outburst, showing the color index at the second peak 
of $\bv \sim 0.4$ after correcting the reddening of 
$E(\bv)= 0.15$.  They further pointed out that the reflection 
by the M-giant is about $\bv \sim 1.35$, therefore, 
rejected the possibility of reflection by the M-giant.  
Instead, they suggested that the much bluer color index 
of $\sim 0.4$ is typical of the accretion disks seen 
in outbursting accretion-powered symbiotic stars 
\citep[e.g.,][]{ken84}.  In our warping-disk model,
the color index is ranging from $\bv= 0.15$ 
(on HJD 2,432,000) to $\bv= 0.35$ (on HJD 2,432,050), which
is reasonably consistent with the observational color at
the second peak because the reflection effect by the warping
disk dominates during the second peak rather than the 
irradiated M-giant.
\par
     Ever since the emission lines of \ion{O}{3},
\ion{N}{3}, \ion{He}{2}, [\ion{O}{3}],
[\ion{Ne}{3}], and [\ion{Ne}{5}] first appeared,
they have been always present until two years 
after the outburst, except in early June, at the most, 
from May 15 to July 17, 1946 
\citep[$\sim 105$---135 days after maximum,][]{san49}.  
In our modeling, the warping-up disk occulted both
the WD and the innermost region of the disk during these period,
being consistent with the disappearance of the emission lines.

\subsection{The Distance, Envelope Mass of the White Dwarf, 
and Possibility of SN Ia Explosion}
     The distance to T CrB is estimated to be $d= 0.94$ kpc
with the absorption of $A_V= 0.35$ \citep*{har93} 
because the distance modulus is 
$(m-M)_0= (m-M)_V - A_V = 10.2 - 0.35 = 9.85$.
This is consistent with both Harrison et al.'s
estimation of $d= 1100$ pc, Belczy\'nski  \&  Miko{\l}ajewska's 
(1998) estimation of $d= 960\pm 150$ pc, and Bailey's (1981) 
estimation of $d= 1180$ pc.
Harrison et al. determined the infrared spectral
type of the cool companion as M3 III.
%%%%% (see Fig. \ref{m_giants_cc}).
Assuming the absolute magnitude of $M_K = -5.3$ (M3 III), 
they obtain the distance modulus of 
$(m-M)_0 = m_K - M_K - A_K = 4.79 - (-5.3) - 0.04 =10.05$,
where they used a relation between the absorptions of $A_K$ and
$A_V$ given by \citet{rie85}, i.e.,
$A_K / A_V = 0.112$.
\par
     The envelope mass of the WD at the first peak is
$\Delta M= 3.3 \times 10^{-6} M_\sun$ from our wind solutions.
Therefore, the mass accretion rate can be estimated to be
$\dot M_{\rm acc}= 4 \times 10^{-8} M_\sun$ yr$^{-1}$ 
if the envelope mass had been accumulated during the quiescent phase
between 1866 and 1946.  This value is roughly consistent with the 
estimation of $2.5 \times 10^{-8} M_\sun$ yr$^{-1}$ by 
\citet{sel92}.  
The wind mass loss rate at the first peak is 
$\dot M_{\rm wind}= 3.7 \times 10^{-4} M_\sun$ yr$^{-1}$ while
the nuclear burning rate is as small as 
$\dot M_{\rm nuc}= 9.3 \times 10^{-7} M_\sun$ yr$^{-1}$.
The wind mass loss rate is quickly decreasing to 
$\dot M_{\rm wind} \sim 1 \times 10^{-6} M_\sun$ yr$^{-1}$ 
about 30 days after the first peak.  Then, 90\% of the envelope
mass has been blown in the wind.  The residual 10\% is left 
on the WD as a part of the helium layer.
Therefore, the net growth rate of the WD is 
$\dot M_{\rm He}= 4 \times 10^{-9} M_\sun$ yr$^{-1}$.
\par
      At the end of this section, we discuss whether or not 
T CrB will explode as a Type Ia supernova in the near
future.  The conditions for SN Ia explosions are summarized
as follows \citep[see, e.g.,][]{nom91}:  (1) the accretion rate
of the C+O WD should be higher than the critical accretion rate
under which helium detonation occurs, i.e.,
\begin{equation}
\dot M_{\rm acc} > \dot M_{\rm He-det} \sim 
1 \times 10^{-8} M_\sun \mbox{~yr}^{-1},
\label{low_accretion_condition_for_SNIa}
\end{equation}
for Population I stars; 
(2) the initial mass of the C+O WD at the beginning 
of the mass transfer is less massive than
\begin{equation}
M_{\rm C+O,0} < 1.2 ~M_\sun.
\label{initial_condition_for_SNIa}
\end{equation}
T CrB satisfies the first condition as has already been
suggested to be $\dot M_{\rm acc} \sim 4 \times 10^{-8} M_\sun$ 
yr$^{-1}$ (even if $\dot M_{\rm acc} \sim 2.5 \times 10^{-8} M_\sun$ 
yr$^{-1}$ by \citealt{sel92}).
For the second condition, we cannot estimate the initial WD mass of T CrB.
However, evolutionary scenario strongly suggests 
that the mass of the WD in T CrB has been
increased up to the present mass from a much lower initial mass
\citep[see e.g., the evolutionary path to SNe Ia proposed by][]{hkn99}.
Moreover, it is theoretically predicted that an upper limit of
degenerate carbon-oxygen core is about $1.07 ~M_\sun$, i.e.,
$M_{\rm C+O,0} < 1.07 ~M_\sun$ \citep{ume99b}. 
Thus, we may conclude that T CrB will explode as a 
Type Ia supernova in quite a near future if the hot component is
a carbon-oxygen white dwarf.

\subsection{Summary of T CrB Outburst}
     The main features of the T CrB outbursts can be well understood
with a thermonuclear runaway model on a very massive white dwarf
having a warping accretion disk.
We have summarized the following important points 
from the fitting of the modeled light curve with observation: \\
(1) The early phase ($\sim 0$---10 days) visual light curve 
of T CrB outbursts
can be well reproduced only by a bloated white dwarf photosphere
of $M_{\rm WD}= 1.37 \pm 0.01 ~M_\sun$ for solar metallicity of $Z=0.02$.
This is because decline rates of optical light curves 
depends sensitively on white dwarf mass especially 
for white dwarf masses near the Chandrasekhar mass limit.
Thus, we are able to specify the mass of white dwarf with
a great accuracy (three digits). 
(2) The ensuing early phase ($\sim 10$---40 days) visual light
curve cannot be reproduced by any combination of a white dwarf 
photosphere and a non-irradiated red giant.  
It requires an irradiation effect of the red giant component.
This heated-up red giant surface is consistent with the very 
early appearance of M-type spectrum (TiO bands) shortly after
the outburst, indicating the luminosity of M-type giant 
much brighter than 10th magnitude \citep{deu48}.  \\
(3) Through the mid and late phase ($\sim 40$---300 days),
the secondary maximum can be well modeled 
if we introduce the radiation-induced warping instability 
of the accretion disk, which sets in after the wind stops 
$\sim 60$ days after maximum.
Introducing an analytic form of warping \citep{mal96},
we obtain the size of the accretion disk as large as 
$R_{\rm disk} \sim 6 ~R_\sun$ and
the precessing period of the warping disk 
as fast as $\sim 140$ days to reproduce the late phase light curve. \\
(4) The envelope mass of the bloated white dwarf is
$3.3 \times 10^{-6} M_\sun$, suggesting the mass accretion
rate of $0.4 \times 10^{-7} M_\sun$ yr$^{-1}$ between the 1866 and
1946 outbursts. \\
(5) The wind carries away about 90\% of the envelope mass,
so that the residual 10\% can be accumulated to the helium
layer on the white dwarf.  As a result, the white dwarf grows 
at a rate of $0.4 \times 10^{-8} M_\sun$ yr$^{-1}$. \\
(6) T CrB will certainly explode as a Type Ia supernova
in quite a near future if the white dwarf consists of carbon and oxygen.

\section{RS OPHIUCHI}
      RS Oph is also one of the well-observed recurrent novae.
It is characterized by a long orbital period of 460 days
(\citealt{dob94}, but recently 457 days by \citealt{fek00}),
a relatively short recurrence period of $\sim 10$---20 yrs
compared with 80 yrs of T CrB.  It has been suggested
that a companion (M-giant) star is underfilling the Roche lobe
and losing its mass by massive stellar winds \citep[e.g.,][]{dob96}.
RS Oph underwent five recorded outbursts (in 1898, 1933, 1958, 
1967, and 1985), with the light curves being very similar to each other 
\citep[e.g.,][]{ros87}.   The latest (1985) outburst has been observed
at all wave lengths from radio to X-rays (see papers in \citealt{bod87}).
\par 
      Although there had been intense debates on the mechanism
of RS Oph outbursts \citep*[e.g.,][]{liv86b, web87}, 
various observational aspects
favor TNR models on a very massive WD
\citep[see, e.g.,][for a recent summary]{anu99}.  
Rapid decline rates of the $V$ light curves
also indicate a very massive WD close to the Chandrasekhar limit. 
\citet{kat91} has first calculated RS Oph outburst light curves 
for the WD masses of 1.33, 1.35, 1.36 and 1.37 $M_\sun$ and  
found that the light curve of the $1.36 M_\sun$ model 
is in better agreement with the observational light curve 
of RS Oph than the other mass models.  
\par
      Since \citet{kat91} adopted an old opacity in her calculation,
\citet{hac00ka} recalculated the light curves of
RS Oph outbursts with the new opacity \citep[e.g.,][]{igl96}, including 
irradiation effects of both an M-giant companion and
an accretion disk around the WD.  
Their main results are summarized as: 
(1) the first 4 days light curve of the outbursts can be well 
reproduced with a TNR model based 
on a $M_{\rm WD}= 1.35 \pm 0.01 M_\sun$ and
an apparent distance modulus of $(m-M)_V= 11.09$. 
(2) To reproduce the $V$ and UV light curves in the late stage
($t \sim 4$---100 days after the optical maximum), 
the contributions of the irradiated M-giant and the irradiated
accretion disk must be introduced.
(3) The distance to RS Oph is estimated to be 0.6 kpc both from 
the $V$ and UV light curve fittings with the absorption 
of $A_V=2.3$.
(4) About 90\% envelope mass of the WD has been blown off in the wind 
but the residual 10\% has been left on the WD.  Thus, \citet{hac00ka} 
has concluded that RS Oph will explode as an SN Ia 
in the near future if the hot component is a carbon-oxygen white
dwarf.
\par
      In Hachisu \& Kato's (2000a) calculations, they assumed
the solar metallicity ($Z=0.02$) of the M-giant envelope.
However, it has recently been suggested that the cool component
of RS Oph is a metal-poor giant.  \citet{sco94} obtained
the carbon abundance of [C/H]$=-3$ for RS Oph 143 days after 
the 1985 outburst and of [C/H]$=-4$ in 1992.  
Similar depletions of carbon, oxygen, neon, sulphur, and iron
with respect to solar values are suggested by \citet*{con95}.
They obtained one or two tenths 
of the solar values for C, O, Ne, S, and Fe, in order to
reproduce the line ratios based on their shock models of nova shells.  
\citet{smi96} estimated metal abundance of AG Dra as
[Fe/H]$= -1.3$ and suggested that all yellow symbiotic stars
including RS Oph are metal-poor.  They attributed K6 I--II 
to the cool component of RS Oph.  
%The position of RS Oph 
%in the infrared two color ($J-H$)-($H-K$) diagram deviates 
%much from the locus of the normal giants 
%(\citet{har93}; see also Fig. \ref{m_giants_cc}),
%indicating that the cool component of RS Oph is not a normal giant.  
%\citet{sek90} suggested that RS Oph is a member of the
%thick disk, which has a similar color to those found in the bulge 
%M-giants.
\par
     If the envelope of the white dwarf is also metal-poor, 
the mass determination of the white dwarf from the light curve
fitting may not be correct because the optically thick wind is
driven by iron lines and the decay timescale is much longer 
in low metallicity envelopes.  Therefore, we must reexamine 
the light curves of the 1985 outburst by introducing lower 
metallicities and study the possible range of the metallicity 
and the WD mass.

\subsection{The Model of the 1985 Outburst}
\begin{figure}
%\epsscale{.9}
%\plotone{metal_mix_rsoph1985.eps}
\plotone{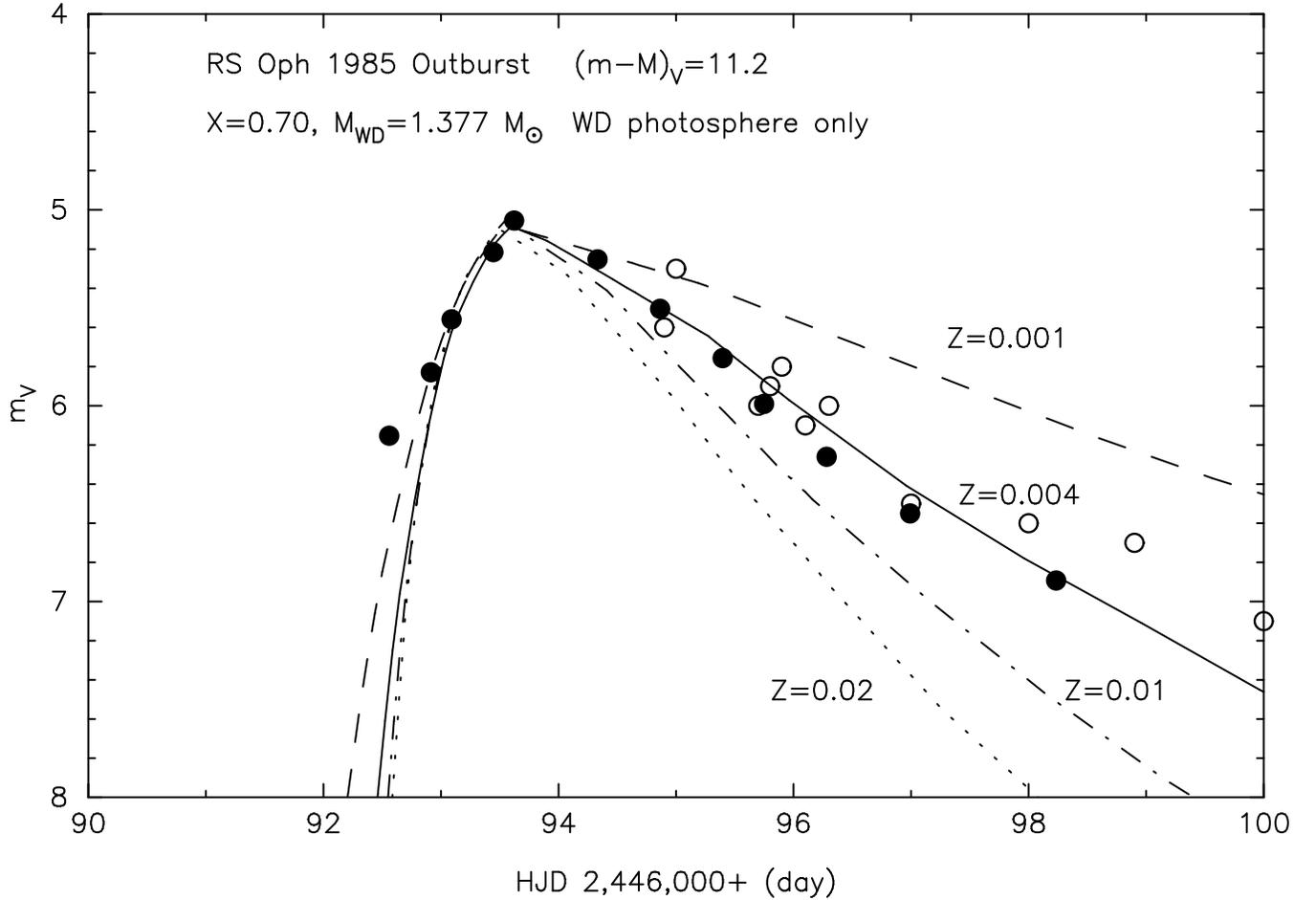}
%\plotfiddle{sn1a96let/evolution1.ps}{5.0cm}{270}{0.4}{0.4}{-170}{220}
\caption{
Model $V$ light curves for various metallicities are plotted 
against time (HJD 2,446,000+) 
together with the observational points of RS Oph outbursts.
Filled circles indicate observational points with the previous
outbursts superposed 
(taken from \citealt{ros87}, see also \citealt{hac00ka}). 
Open circles correspond to the observational points of the 1985 
outburst, which are taken from IAU Circulars
\citep{sco85, low85, alc85, hur85a, hur85b, 
dan85, duc85a, duc85b, duc85c,  
mcn85a, mcn85b, mcn85c, mcn85d,
lub85a, lub85b, kin85, wei85, rig85, 
coo85a, coo85b, coo85c, med85, 
bor85a, bor85b, bor85c,
ell85, ver85, sve85, 
alb85},
i.e., No.4031, 4036, 4048, 4049, 4066, 4074, and 4091.
The WD mass is assumed to be the extreme case of 
$M_{\rm WD}= 1.377 ~M_\sun$ just before the SN Ia explosion.
The metallicity is attached to each line (for $Z=0.001$, $Z=0.004$,
$Z=0.01$, $Z=0.02$).  Each line denotes the $V$ light of the WD
photosphere only.
The hydrogen content of the WD envelope is assumed to be $X=0.70$
for all models.  
The apparent distance modulus is $(m-M)_V= 11.2$.
The lowest metallicity case of $Z=0.001$ cannot reproduce 
the rapid decline of the $V$ light curve of RS Oph outburst.
Since the hydrogen content, $X$, hardly affect the decline rate of 
the $V$ light curve \citep[e.g.,][]{kat99}, we may conclude that
$Z \gtrsim 0.004$ in RS Oph outbursts. 
\label{metal_mix_rsoph1985}}
\end{figure}

     Assuming an extremely
massive white dwarf of $M_{\rm WD}= 1.377 ~M_\sun$ just prior 
to the SN Ia explosion, we have calculated light curves for
various metallicities of $Z= 0.001$, 0.004, 0.01, and 0.02
as shown in Figure \ref{metal_mix_rsoph1985}.
The $V$ light of the first five days (HJD 2,446,092 -- 2,446,096)
can be reproduced only by the bloated WD photosphere as already
shown in \citet{hac00ka}.  The case of $Z=0.001$ is too
slow to be compatible with the observations.  Thus, we may conclude
that the metallicity is $Z \gtrsim 0.004$.
Here, we assume $Z=0.004$ for RS Oph as a lower limit of metallicity.
For the set of $M_{\rm WD}= 1.377 ~M_\sun$ and $Z=0.004$,
we have obtained the envelope mass at the optical maximum,
$\Delta M= 2.2 \times 10^{-6} M_\sun$, which indicates a mass 
accretion rate of $1.2 \times 10^{-7} M_\sun$ yr$^{-1}$
during the quiescent phase between the 1968 and the 1985 outbursts.

\placefigure{metal_mix_rsoph1985}

\par
      The irradiation effects of the accretion disk and the red giant
(RG) are necessary to reproduce the optical and UV light curves of 
the outburst as already shown by \citet{hac00ka}.  
\citet{dob94} obtained the radial velocities of
the M-type feature and the A-type feature but concluded that 
the A-type absorption feature does not associate with any of
the components.  Thus, RS Oph is a single line binary with
\begin{equation} 
{{M_{\rm WD}^3 \sin^3 i} \over {\left(M_{\rm WD}+M_{\rm RG} \right)^2}}
\approx 0.1 ~M_\sun,
\end{equation} 
from the semiamplitude of the radial velocity $K_{\rm RG}= 12.8 \pm 2.0$
km s$^{-1}$ and the orbital period of $P= 460$ days.
% WD MASS=   1.377     RG MASS=   .5000     INCL= 30 DEGREE
% WD MASS=   1.377     RG MASS=   .5130     INCL= 31 DEGREE
% WD MASS=   1.377     RG MASS=   .5950     INCL= 32 DEGREE
% WD MASS=   1.377     RG MASS=   .6780     INCL= 33 DEGREE
% WD MASS=   1.377     RG MASS=   .7610     INCL= 34 DEGREE
% WD MASS=   1.377     RG MASS=   .8440     INCL= 35 DEGREE
% WD MASS=   1.377     RG MASS=   .9270     INCL= 36 DEGREE
% WD MASS=   1.377     RG MASS=   1.010     INCL= 37 DEGREE
% WD MASS=   1.377     RG MASS=   1.093     INCL= 38 DEGREE
% WD MASS=   1.377     RG MASS=   1.175     INCL= 39 DEGREE
% WD MASS=   1.377     RG MASS=   1.258     INCL= 40 DEGREE
They suggested a relatively low inclination angle of 
$i \sim 30$---$40\arcdeg$.  If $M_{\rm WD}= 1.377 ~M_\sun$, 
then we obtain $M_{\rm RG}= 0.5 ~M_\sun$ for $i=30\arcdeg$,
$M_{\rm RG}= 0.7 ~M_\sun$ for $i=33\arcdeg$,
$M_{\rm RG}= 0.9 ~M_\sun$ for $i=36\arcdeg$,
$M_{\rm RG}= 1.0 ~M_\sun$ for $i=37\arcdeg$,
or $M_{\rm RG}= 1.2 ~M_\sun$ for $i=39\arcdeg$.
We assume $M_{\rm RG}= 1.0 ~M_\sun$ and $i= 37\arcdeg$ 
in the present paper, although we have already confirmed that
the other set of $M_{\rm RG}= 0.7 ~M_\sun$ and $i= 33\arcdeg$
reproduces almost the same light curves.
\par
     For the cool component of the RS Oph system, 
as suggested by \citet{dob96}, the RG lies well 
within the inner critical Roche lobe, i.e.,
\begin{equation}
R_{\rm RG} = \gamma R_2^* ~~~ (\gamma < 1),
\label{RG-size}
\end{equation}
where $R_2^*$ is the effective radius of the inner critical
Roche lobe for the RG component and $\gamma$ is a numerical factor.
\citet{dob96} adopted $\gamma= 0.4$ for the distance
of $d=1.5$ kpc to RS Oph.  On the other hand, \citet{hac00ka}
estimated the distance of $d= 0.6$ kpc to RS Oph much shorter than
Dobrzycka et al.'s value, so that we adopt $\gamma= 0.25$ for 
$d=0.6$ kpc in this paper.   Even if we use $\gamma= 0.4$,
the light curve itself is not so different from that for $\gamma= 0.25$
as shown in \citet{hac00ka}.  Only the difference is in 
the irradiation effect of the M-giant, but the irradiation itself 
is relatively small compared with the case of T CrB ($\gamma=1$).  
\par
     Recent studies on the radii of M-giants in symbiotic stars 
indicate $\gamma \sim 0.5$ for the normal S-type symbiotic stars
\citep[e.g.,][]{mur99}.  The cool component of the RS Oph system 
is not a normal giant.  It is suggested to be metal-poor and 
sometimes classified to K-giant \citep[e.g.,][]{smi96} rather than M-giant.
Therefore, we may adopt the relatively small radius of $\gamma= 0.25$. 
\par
     As for the orbit of the binary system,
the ephemeris for the inferior conjunction of the red giant in front 
is 2,444,999.9$+ 460 \times E$ \citep{dob94}.
The separation is $a= 325.0 R_\sun$, the effective radii 
of the inner critical Roche lobes for the WD component and
the RG component are $R_1^*= 138.6 R_\sun$ 
and $R_2^*= 108.2 R_\sun$.  For $\gamma= 0.25$, 
we have $R_{\rm RG} = 0.25 R_2^* \sim 27 R_\sun$.
% AV. DM2/DT=  0.5759E-05  WIND=  0.8959  M1= 1.3770  M2= 0.8000
% AA=324.970  RL1=138.649  RL2=108.219  RDISK=  1.109  FLUPD=  0.050
\par
\begin{figure}
%\epsscale{.9}
%\plotone{vmag1377va1_rsoph1985fixdisk.eps}
\plotone{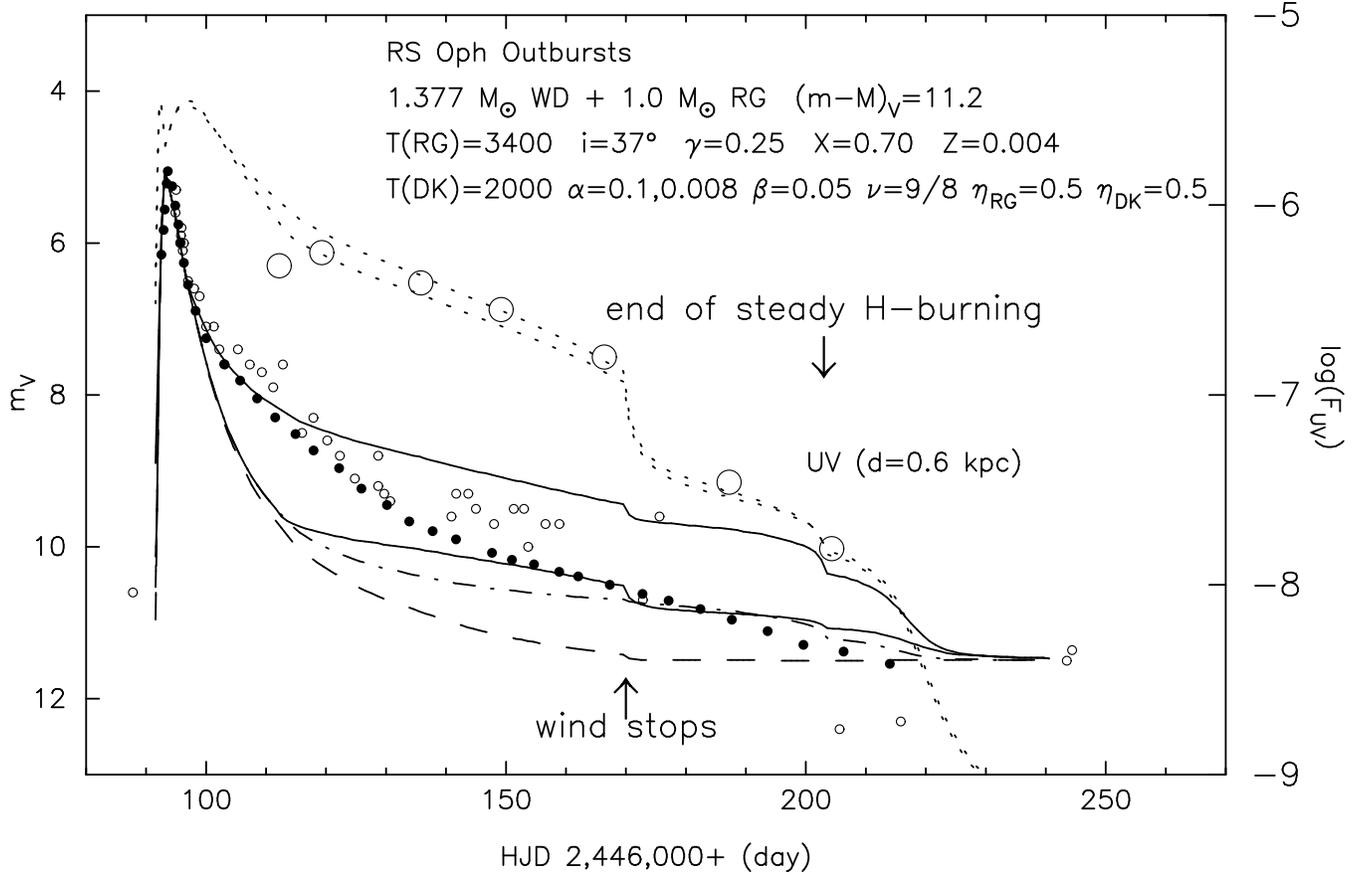}
%\plotfiddle{sn1a96let/evolution1.ps}{5.0cm}{270}{0.4}{0.4}{-170}{220}
\caption{
Same as Fig. 10 but for the entire outburst phase for 
$M_{\rm WD}= 1.377 M_\sun$, $X=0.7$, and $Z=0.004$.
{\it Dashed line}: Sum of the contributions 
from the white dwarf (WD) photosphere
and the non-irradiated red giant (RG) photosphere with the surface
temperature of $T_{\rm ph}= 3400$ K.  The RG lies well within 
the inner critical Roche lobe, i.e., its radius is 0.25 times
the Roche lobe size ($\gamma = 0.25$).
{\it Dashed-doted}: Sum of the WD photosphere 
and the irradiated RG photosphere.
{\it Solid}: the total $V$ light of the WD photosphere,  
the irradiated RG photosphere and the accretion disk (ACDK) surface.
{\it Dotted}: ultra-violet (UV) radiation (911--3250 \AA) 
from the WD photosphere, the irradiated RG photosphere, 
and the irradiated ACDK surface.  The UV data (large open circles)
are taken from \citet{sni87a}.
Here both the disk shape and size are assumed to be constant 
during the outburst, i.e., $\alpha=0.01$
($R_{\rm disk} \sim 14~ R_\sun)$ and $\beta=0.05$
for the upper solid (and upper dotted) line, 
and $\alpha=0.008$ ($R_{\rm disk} \sim 1 ~R_\sun)$ and $\beta=0.05$ 
for the lower solid (and lower dotted) line, respectively. 
Two epochs with "wind stops" and "end of steady H burning"
are indicated by arrows.   
The optically thick wind blows during the period from the first
phase of the outburst (HJD 2,446,092) to 79 days after 
maximum (HJD 2,446,170).  The steady hydrogen shell burning
ends at 112 days after maximum (HJD 2,446,203).  
\label{vmag1377va1_rsoph1985fixdisk}}
\end{figure}

     We assume 50\% efficiency of the irradiation of the red giant (RG) 
component ($\eta_{\rm RG}=0.5$).  The nonirradiated photospheric 
temperature of the RG is a parameter which is determined to be fitted
with observation.  Here, we adopt $T_{\rm RG} = 3400$ K 
against $\gamma=0.25$ as shown 
in Figure \ref{vmag1377va1_rsoph1985fixdisk}.
\par
     We have also included the contribution of irradiated accretion
disk (ACDK) luminosity.
Here, we also assume 50\% efficiency of the ACDK irradiation
($\eta_{\rm DK}= 0.5$).
The viscous heating is neglected because it is much smaller 
than that of the irradiation effects.
The temperature of the unheated surface of the ACDK including
the rim is assumed to be $T_{\rm disk}= 2000$ K.  
We have checked two other cases of $T_{\rm disk}= 1000$ and 0 K
and have found no significant differences in the light curves. 
The power of $\nu$ is assumed to be $\nu=9/8$ from the standard disk model. 
We have checked the dependency of the light curves on the parameter 
$\nu$ by changing from $\nu=9/8$ to $\nu=2$ but cannot find
any significant differences as far as the disk rim is small
($\beta=0.01$---0.05). 

\placefigure{vmag1377va1_rsoph1985fixdisk}

\begin{figure}
%\epsscale{.9}
%\plotone{vmag1377va1_rsoph1985vardisk.eps}
\plotone{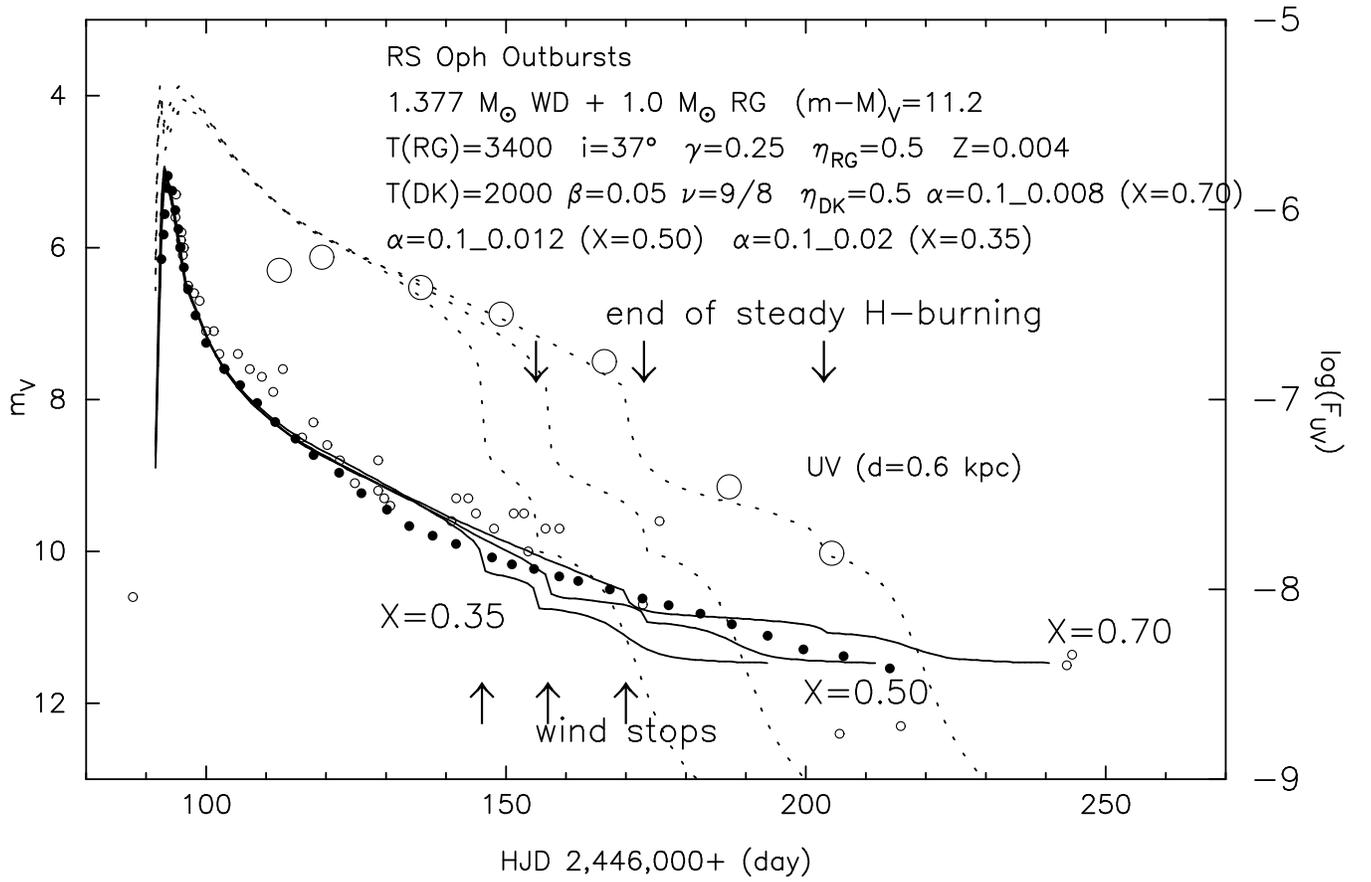}
%\plotfiddle{sn1a96let/evolution1.ps}{5.0cm}{270}{0.4}{0.4}{-170}{220}
\caption{
Same as Fig. 11 but for a case of gradually evaporating accretion disk.
Here, we assume that the radius of the accretion disk 
is gradually shrinking from $\sim 14 R_\sun ~(\alpha= 0.1)$ 
to $\sim 1 R_\sun ~(\alpha=$ 0.008) for $X=0.7$, 
to $\sim 2 R_\sun ~(\alpha=$ 0.012) for $X=0.5$, 
to $\sim 3 R_\sun ~(\alpha=$ 0.02) for $X=0.35$, 
during the wind phase.  Its time variation is exponential
as given by equation (\ref{disk_vanishing_time}).
\label{vmag1377va1_rsoph1985vardisk}}
\end{figure}

     The late-phase $V$ light curve ($t \sim 5$---100 days after maximum) 
indicates strong irradiation of the ACDK. 
We have examined the case of a lobe-filling RG companion ($\gamma=1$),
but the irradiation of the RG is too luminous to
be compatible with the observational light curves.
On the other hand, a smaller size of the RG such as $\gamma=0.25$
($\sim 27 R_\sun$) gives a reasonable fit with the observations 
as shown in Figures \ref{vmag1377va1_rsoph1985fixdisk}
and \ref{vmag1377va1_rsoph1985vardisk}.  
\par
     To fit the late-phase light curve, we have calculated the two
cases of the disk parameters, i.e., $(\alpha, \beta)=$ (0.1, 0.05) 
and (0.008, 0.05), as shown in Figure \ref{vmag1377va1_rsoph1985fixdisk}.  
Here, two solid  lines denote the cases of $\alpha=0.1$ 
($R_{\rm disk}= 14 ~R_\sun$) and $\beta=0.05$
and $\alpha=0.008$ ($R_{\rm disk}= 1 ~R_\sun$) 
and $\beta=0.05$, respectively.  Something between 
these two can roughly reproduce the light curve of the 1985 outburst 
but the gradual transition from $\alpha=0.1$ to $\alpha=0.008$ is 
much better for reproducing the observations as shown in Figure
\ref{vmag1377va1_rsoph1985vardisk}.  
It is very likely that the surface of the accretion disk has been 
dragged by the strong wind and been gradually blown 
in the strong wind.  As a result, the accretion disk has become 
smaller during the strong wind phase ($\sim 70$ days
after maximum).  Thus, we adopt 
\begin{equation}
\alpha = \alpha_0 
\left( {{\alpha_1} \over {\alpha_0}}\right)^{(t-t_0)/72},
\mbox{~for~} t_0 < t < t_0+72 \mbox{~day}  
\label{disk_vanishing_time}
\end{equation} 
where $\alpha_0=0.1$, $\alpha_1= 0.008$, $t$ is the days
after maximum, $t_0=7$ days is the reappearance day 
of the accretion disk, and 72 days is roughly the wind period.

\placefigure{vmag1377va1_rsoph1985vardisk}

\par
     We have also calculated UV light curves with a response
function of 911---3250\AA ~to fit the reddening corrected UV data 
\citep{sni87a}.  Figure \ref{vmag1377va1_rsoph1985fixdisk}
shows two cases of UV light curves 
with the parameters of ($\alpha=0.1$, $\beta=0.05$) and
($\alpha=0.008$, $\beta=0.05$).
The calculated UV light curves can 
reproduce well the UV observations both for $\alpha=0.1$ 
and $\alpha=0.008$ with the distance of $d=0.57$ kpc
because the innermost region ($\lesssim 0.5 ~R_\sun$) 
of the accretion disk radiates strong UV but the outer region
of $\gtrsim 1 ~R_\sun$ mainly contributes to optical light.

\subsection{Development of the 1985 Outburst}
      In this subsection, we summarize the observational developments
of the RS Oph 1985 outburst and discuss them
in relation to our model calculation.
\par
     \citet{ros87} and \citet{ros87i} divided the outburst
into several phases along with the development of the optical spectrum:
[[Phase 0---2 days]] after the optical maximum, broad emission lines of
\ion{H}{1}, \ion{He}{1}, and \ion{Fe}{2} are
observed with narrow absorption component (40---60 km s$^{-1}$)
and broad P Cygni troughs ($-2700$ to $-3700$ km s$^{-1}$).
[[Phase 2---14 days]], \ion{He}{2} 4686~\AA~, 
[\ion{N}{2}], [\ion{O}{1}], [\ion{O}{3}],
[\ion{Ne}{3}], [\ion{S}{2}], 
and [\ion{Fe}{2}] 6374~\AA~ appear.  
[[Phase 14---29 days]], \ion{He}{2} 4686~\AA~ and 
\ion{N}{3} 4640~\AA~ become strong.  
[[Phase 29---51 days]], coronal lines appear and become strong.
[[Phase 51---72 days]], the coronal line phase lasts.
[[Phase 72---119 days]], the coronal lines disappear. 
Most of high excitation lines, including \ion{He}{2} 4686
\AA, are fading.
\par
      \citet{sni87a} presented an overall development of 
the UV light curve during the 1985 outburst of RS Oph
as already shown in Figures \ref{vmag1377va1_rsoph1985fixdisk} and
\ref{vmag1377va1_rsoph1985vardisk}.  
\par
     Since RS Oph exploded in a circumstellar envelope that is formed
by the slow wind of the red giant component \citep{wal58},
the rapidly expanding ejecta \citep[$\sim 3,000$---4,000 km s$^{-1}$,
e.g.,][]{ros87, sni87a} of the outburst is decelerated 
by the slowly expanding ($\lesssim 20$ km s$^{-1}$) 
circumstellar matter to form a strong shock
as noticed by \citet{pot67}.  The secular variation of 
coronal lines was interpreted by \citet{gor72} in terms of
a decelerating shock wave \citep[see also][for the 1985 outburst]{wal86}.
This shock-heated layer is also
responsible for soft X-ray \citep{mas87} and radio 
\citep{hje86} emission as will be mentioned below.  
\par
     In our numerical modeling, the photospheric temperature
of the WD envelope has once decreased down to
$T_{\rm ph}= 12,200$ K at the optical maximum (HJD 2,446,093).
Then, it increases gradually to 
[[Phase 0---2]] $T_{\rm ph}= 12,200$ --- 17,700 K 
for the WD photospheric radius of
$R_{\rm ph}= 45$ --- 21 $R_\sun$,
[[Phase 2---14]] $T_{\rm ph}= 17,700$ --- 67,100 K for 
$R_{\rm ph}= 21$ --- 1.6 $R_\sun$,
[[Phase 14---29]] $T_{\rm ph}= 67,100$ --- 114,000 K for
$R_{\rm ph}= 1.6$ --- 0.56 $R_\sun$,
[[Phase 29---51]] $T_{\rm ph}= 114,000$ --- 181,000 K for
$R_{\rm ph}= 0.56$ --- 0.23 $R_\sun$,
[[Phase 51---72]] $T_{\rm ph}= 181,000$ --- 302,000K for
$R_{\rm ph}= 0.23$ --- 0.083 $R_\sun$,
[[Phase 72---119]] $T_{\rm ph}= 302,000$ --- 1,020,000 K for
$R_{\rm ph}= 0.083$ --- 0.0037 $R_\sun$.
The accretion disk reappears 4 days
after the optical maximum, i.e., the photosphere shrinks to 
$R_{\rm ph} \sim 8~R_\sun$, which is smaller than the 
size of the accretion disk $R_{\rm disk} \sim 10~R_\sun$
that we assume.
The optically thick strong wind, which is blowing from the WD,
stops 77 days after the optical maximum (HJD 2,446,170).  
The photospheric radius shrinks from $R_{\rm ph}= 0.06 ~R_\sun$ to
$R_{\rm ph}= 0.01 ~R_\sun$ within a few days.  Correspondingly,
the photospheric temperature increases from $T_{\rm ph}= 343,000$ K
to $T_{\rm ph}= 1 \times 10^6$ K.  This sharp decrease in the photospheric
radius causes the sharp drop in the UV light curve
in Figures \ref{vmag1377va1_rsoph1985fixdisk} and
\ref{vmag1377va1_rsoph1985vardisk}, because the effective angle
of $\cos\theta$ in equation (\ref{irradiation_disk_angle}) 
against the disk surface
drops by a factor of three.  The steady hydrogen
shell-burning also ends 109 days after maximum (HJD 2,446,202).
\par
      The appearance of \ion{He}{2} 4686~\AA~, 
[\ion{N}{2}], [\ion{O}{1}], [\ion{O}{3}],
[\ion{Ne}{3}], [\ion{S}{2}], 
and [\ion{Fe}{2}] 6374~\AA~ in [[Phase 2---14]] and 
[[Phase 14---29]] is 
consistent with our model because the WD photospheric temperature
increases to high enough to excite these elements.
The appearance of the coronal lines are related with
the appearance of X-rays heated by the blast shock as shown below 
\citep{mas87} because they need high temperature photons
of $T \gtrsim 1 \times 10^6$ K as in the solar corona. 
The disappearance of the coronal lines is usually understood 
by the shock breakout at the day $\sim 70$.
However, the quick disappearance of the coronal lines 
in [[Phase 72---119]] might be caused by the extinction 
of the hydrogen shell burning, as suggested
by \citet{sho96}, because it is consistent with the extinction time
of the steady hydrogen shell burning in our model.
\par
      \citet{eva88} presented infrared observation 
($J$, $H$, $K$, and $L$ bands) of the RS Oph 1985 outburst.
The decline rates of $J$, $H$, $K$, and $L$ are almost the same
as $t_2 \sim 30$ days ($t_2$ is the time taken to drop 2 mag 
from maximum).  At each wave length it is noticeable that
the IR flux of RS Oph had returned to its pre-outburst value
by the day $\sim 85$ (after the optical maximum, but
Evans et al. took the optical maximum on HJD 2,466,095.5 as a
origin of time) and that the decline continued beyond this level.
However, the IR fluxes had reverted to their pre-outburst levels
until the day $\sim$400.  The post-outburst minimum lasted
during about 300 days, i.e., from the day $\sim$100 to the day 
$\sim$400.  This post-outburst minimum is discussed below
in relation to the super-soft X-ray detection on the day 251
and the distance to RS Oph.
\par
     \citet{eva88} also estimated the IR flux from 
the circumstellar material in the RS Oph system by removing
the contribution from the M-giant companion.  
Assuming that the \ion{He}{2} 1640 \AA~ line
and the IR continuum do in fact arise in the same region,
they fitted the IR continuum on HJD 2,466,119 by a nebular continuum 
having electron temperature $T_{\rm e} \sim 1.6 \times 10^5$ K.
In our model calculation, we have the photospheric temperature
of the bloated WD envelope $T_{\rm ph} \sim 1.1 \times 10^5$ K 
on the same day, being roughly consistent with Evans et al.'s 
estimation, though Evans et al. attributed it to the cooled shocked 
layer.  Evans et al. also estimated that the spectral type of RS Oph 
cool component is K8 III ($\pm 2$ sub-divisions) 
from $[2.35]-[2.2]$ color.
\par
      \citet{sho96} also divided the outburst into
two phases along with the development of the UV spectrum:
The first phase, [[Phase 0---60 days]] after the optical maximum,
is the broad line phase.  The high velocity material 
dominated the emission.  This is roughly consistent with 
the strong wind phase of our numerical modeling, i.e., until
HJD 2,446,170.  In the second phase, i.e., [[Phase 60---130 days]],
the newly ionized red giant wind dominated the spectrum.  
This period is also consistent with the steady hydrogen burning phase
of our numerical modeling, i.e., until HJD 2,446,202.
\par
     Shore et al. also discussed that 
the line ratio of \ion{Si}{3}] 1892\AA~ to \ion{C}{3}]
1910 \AA, which measures the electron density \citep{nuss89},
steadily decreased from an initial value of 1.0 on HJD 2,446,106 to 
a minimum of 0.3 on HJD 2,446,164.  Then, the ratio began to rise 
thereafter, reaching 0.6 by HJD 2,446,202.
This trend can be explained by our wind model:
(1) The wind mass loss rate is gradually decreasing from
$\sim 10^{-4}$ to $\sim 10^{-7} M_\sun$ yr$^{-1}$ and 
this causes the decrease in the electron density because of
$\rho \sim \dot M_{\rm wind}/ 4 \pi r^2 v_{\rm wind}$.
(2) After the hot wind from the WD stops, 
the cool red giant wind resumes to expand slowly
and the newly ionized red giant wind dominates the spectrum. 
\par
      RS Oph became a strong X-ray source with characteristic
temperature of a few times $10^6$ K \citep{mas87}.
Mason et al. observed X-rays from 55 days after the
optical maximum through 251 days with the medium-energy proportional 
counter array (ME) and the low-energy imaging telescope (LE) 
onboard {\it EXOSAT}.  They made six separate observations,
55, 62, 74, 83, 93, and 251 days after the optical maximum.
The source became very weak in the ME from day 83 and only upper
limits were derived.  However, there was still some residual X-ray 
emission detected even more than 250 days after maximum
only with the LE band.  
The authors interpreted the early X-ray emission in terms of
emissions by the pre-outburst M-giant's wind that had been shock-heated
by the outburst ejecta \citep*[see also][]{bod85, obr87, ito90, obr92}.  
On the other hand, the late X-ray emissions have been 
interpreted in terms of the continued hydrogen shell burning 
on top of the white dwarf.  This interpretation is not consistent
with our model calculation because the steady hydrogen shell-burning
ended long before the day 251, i.e., at the day $\sim 110$.
We will discuss this point again in the next subsection in relation
to the distance to RS Oph.

\subsection{The Distance to RS Ophiuchi}
     The distance to RS Oph is determined to be 0.57 kpc
from the dereddened UV flux \citep{sni87a} as shown 
in Figures \ref{vmag1377va1_rsoph1985fixdisk}
and \ref{vmag1377va1_rsoph1985vardisk}.
We assume the accretion disk exists during the outburst.
We cannot reproduce the UV light curve if the disk does not exist or 
its size is smaller than $\alpha < 0.003$, i.e., $\lesssim 0.5 R_\sun$,
as already shown in \citet{hac00ka}.
The UV light curves depend hardly on the disk parameters of
$\alpha$, $\beta$, or $\nu$ when $\alpha \gtrsim 0.01$, mainly
because the UV light is coming from the innermost part of
the accretion disk.  The distance depends weakly on 
the irradiation efficiency ($\eta_{\rm DK}$), for example,
$d=0.70$ kpc for $\eta_{\rm DK}=1.0$ or
$d=0.50$ kpc for $\eta_{\rm DK}=0.25$.
The red giant does not contribute to the UV because its irradiated
temperature is as low as $\sim 4000$ K.
Thus, the distance estimation is rather robust.
\par
     We have calculated the absorption 
of $A_V= 11.09 - 5 \log(570/10) = 2.3$ for $d=0.57$ kpc
from our light curve fitting.
This absorption of $A_V=2.3$ is consistent with the color excess
of $E(\bv)=0.73$ \citep{sni87b} because of $A_V = 3.1~E(\bv)$. 
This distance of $\sim 0.6$ kpc is not consistent 
with other estimations of $d=1.6$ kpc from the hydrogen column 
density \citep{hje86}, $d=1290$ pc \citep{har93}, and 
$d=1$ kpc \citep{sek90}, both of which are determined 
from the K-band luminosity.  In the following, therefore, 
we discuss extensively other distance estimations supporting
our short distance of 0.6 kpc to RS Oph.
\par
     The VLBI map 77 days after the optical maximum
\citep{tay89} shows an elongated structure with a long
axis of $0 \farcs 22$ and a short axis of $0 \farcs 08$.
If we assume the distance of $d=1.6$ kpc to RS Oph,
this fastest average expansion velocity should be 4,000 km s$^{-1}$.
This is not consistent with 
the initial velocity of ejecta $\sim 3,000$---$4,000$ km s$^{-1}$
because the ejecta is strongly decelerated by the circumbinary
matter that is formed by the red giant wind in quiescence.  
If the distance is $d= 0.6$ kpc, on the other hand, the average expansion 
velocity is much slower ($\sim 1,400$ km s$^{-1}$), which is
consistent with the initial velocity of the ejecta.  
Then, the slowest average expansion 
velocity is estimated to be $\sim 500$ km s$^{-1}$, which is also 
consistent with the optical line width (e.g., $\sim 490$ km s$^{-1}$
110 days after optical maximum, \citealt{sni87a}).
\par  
     The outer edge of the circumbinary matter (CBM) that 
is formed by the red giant wind is about 
$R_{\rm CBM} \lesssim 1 \times 10^{15}$ cm, since 
an upper limit of 20 km s$^{-1}$ to the velocity of the red giant
wind has been given by \citet{gor72} based on spectra 
by \citet{duf64}.  Here, we assume that the ejecta of the 
previous 1967 outburst swept up the previously existing CBM far away.
The soft X-ray rapidly fell down from 60---70
days after the optical maximum so that the shock had broken out
of the circumbinary matter on that day \citep{mas87, ito90, llo93}. 
If the outer edge of the circumbinary matter corresponds to the elongated 
radius of the VLBI map on the day 77 ($0 \farcs 11$ of half axis), 
we obtain the distance to RS Oph,
$d \lesssim R_{\rm CBM} / (\Delta \theta) = 2 \times 10^{21}$ cm 
= 0.65 kpc.
\par  
     Very soft X-rays were observed 251 days after the 
optical maximum in the 1985 outburst of RS Oph \citep{mas87}.
Mason et al. estimated a blackbody temperature of 
$3.5 \times 10^5$ K and a total energy flux of 
\begin{eqnarray}
L_{\rm X,obs} = 1 \times 10^{37} \left( 
{{d} \over {1.6 \mbox{~kpc}}} \right)^2 \mbox{~ergs~s}^{-1},
\label{x-ray_obs_rsoph}
\end{eqnarray}
for a hydrogen column density of 
$N_{\rm H}= 3 \times 10^{21}$ cm$^{-1}$,
and concluded that these soft X-rays come 
from a WD photosphere with steady hydrogen shell burning.
However, steady hydrogen shell burning has stopped 
109 days after the optical maximum (HJD 2,446,202) in our model of 
$X=0.70$, $Z=0.004$, and $M_{\rm WD}= 1.377 M_\sun$ 
(Figs. \ref{vmag1377va1_rsoph1985fixdisk} and 
\ref{vmag1377va1_rsoph1985vardisk}), that is, this soft X-ray
detection is indicating accretion luminosity 
rather than hydrogen shell burning luminosity.
Using a blackbody temperature of $T_{\rm BB} \sim 3.5 \times 10^5$ K
and a WD radius of 0.003 $R_\sun$ for $1.377 ~M_\sun$, 
we estimate a total luminosity of 
\begin{eqnarray}
L_{\rm X, WD}= 4 \pi R_{\rm WD}^2 \sigma T_{\rm BB}^4 \sim 120 L_\sun .
\label{x-ray_acc_rsoph}
\end{eqnarray}
Thus, a rather short distance of 0.4 kpc to RS Oph can be 
derived by equating equation (\ref{x-ray_obs_rsoph}) 
and (\ref{x-ray_acc_rsoph}), i.e., $L_{\rm X,obs} = L_{\rm X, WD}$.
Thus, our short distance of 0.6 kpc is roughly consistent with
the soft X-ray observation by \citet{mas87} about eight
months after the optical maximum.
This low luminosity corresponds to the accretion luminosity of 
\begin{eqnarray}
\dot M_{\rm acc}= {{2 L_{\rm X,WD} R_{\rm WD}} \over {G M_{\rm WD}}} 
\sim 0.2 \times 10^{-7} M_\sun \mbox{~yr}^{-1}.
\end{eqnarray}
This accretion rate is six times lower than our estimated
mass accretion rate of 
$\dot M_{\rm acc}= 1.2 \times 10^{-7} M_\sun$ yr$^{-1}$ averaged
between 1967 and 1985.  
\par
     \citet{dob96} pointed out that the line fluxes 
in H {\footnotesize I} and He {\footnotesize I} before the 1985 
outburst are about four times larger than the fluxes after the
1985 outburst, thus indicating a decrease in the mass accretion rate
after the outburst (see also the low luminosity during the post-outburst
minimum $\sim$100---400 days after maximum pointed by \citealt{eva88}).
If our value of $\dot M_{\rm acc}= 1.2 \times 10^{-7} M_\sun$ yr$^{-1}$ 
is adopted as the mass accretion rate before the 1985 outburst, 
these line fluxes before/after
the outburst are roughly consistent with the ratio 
of the estimated mass accretion rates 
before and after the outburst $(1.2/0.2)$ if the line fluxes 
depend proportionally on the gravitational energy release. 
The day 251 was just in the period of 
the post-outburst minimum ($m_V \sim 11.5$) of
the day $\sim$100---400, after which the visual luminosity
increased to $m_V \sim 10.5$, by about one magnitude
\citep[e.g., Fig. 1 of][]{eva88}.
\par
     \citet{boh89} estimated the H$\beta$ intensity photoionized
by X-ray photons from the central hot WD 
and compared it with the observation 201 days after maximum.  
Their theoretically estimated value of
\begin{eqnarray}
I(\mbox{H}\beta) &=& 1.24 \times 10^{-10} \mbox{ergs~cm}^{-1} 
\mbox{s}^{-1} \cr
& \times &\left({{R_{\rm WD}} \over {1 \times 10^9 \mbox{~cm}}} \right)^2
\left({{d} \over {1.6 \mbox{~kpc}}} \right)^{-2}
\end{eqnarray}
is 3.5 times larger than the observed value of 
$I(\mbox{H}\beta) = 3.55 \times 10^{-11}$ ergs cm$^{-1}$ s$^{-1}$
when we adopt $R_{\rm WD}= 10^9$ cm and $d=$ 1.6 kpc.  
However, if we use $R_{\rm WD}= 0.003 ~R_\sun$ 
and $d=$ 0.6 kpc, the estimated value is 
$I(\mbox{H}\beta) = 3.8 \times 10^{-11}$ ergs cm$^{-1}$ s$^{-1}$,
being consistent with the observed intensity.
\par
     X-rays were also observed at the quiescent phase \citep{ori93},
but the flux is too low to be compatible with the TNR model.
One possible explanation is an absorption by the massive cool wind 
\citep[e.g.,][]{sho96} from the RG component as suggested by
\citet{anu99}.  On the other hand,
soft X-rays were observable just after the outburst, because
the ejecta swept away the cool wind around the RS Oph system.
The cool red giant wind fills again its Roche lobe and 
can cover the hot component in a time scale of 
$\Delta t = a / v \sim 300 R_\sun / 10 \mbox{~km~s}^{-1}
= 300$ days, which is roughly consistent with the recovery
of the quiescent $V$ luminosity as shown in Figure 1 of
\citet{eva88}.
\par
     \citet{hje86} estimated the distance 
to RS Oph of $d= 1.6$ kpc from H{\sc ~I} absorption line measurements,
using the hydrogen column density 
of $N_{\rm H}= (2.4 \pm 0.6) \times 10^{21}$ cm$^{-2}$ 
and the relation of $N_{\rm H}/ (T_{\rm s}~d) = 1.59 \times 10^{19}$
cm$^{-2}$ K$^{-1}$ kpc$^{-1}$ together with $T_{\rm s}=100$ K.
If a large part of this hydrogen column density stems not from 
the Galactic absorption but from the local one belonging to RS Oph,
the distance is overestimated.  For example, in the direction 
of the recurrent nova U Sco, \citet{kah99}
reported a hydrogen column density of (3---4)$\times 10^{21}$ cm$^{-2}$,
which is much higher than the Galactic hydrogen column density
of $1.4 \times 10^{21}$ cm$^{-2}$.
%Here, U Sco is at least 2 kpc above the Galactic plane ($b=22\arcdeg$).
If RS Oph is in the same case as in U Sco, the distance of 1.6 kpc is 
just an upper limit and we possibly derive a much shorter distance.

\subsection{The Ejected Envelope Mass 
and Possibility of SN Ia Explosion}
     The envelope mass at the optical maximum is estimated to be
$\Delta M= 2.2 \times 10^{-6} M_\sun$, which indicates a mass 
accretion rate of $1.2 \times 10^{-7} M_\sun$ yr$^{-1}$
during the quiescent phase between the 1968 and 1985 outbursts.
In our calculation, about 90\% of the envelope mass 
($2.0 \times 10^{-6} M_\sun$) 
has been blown off in the optically 
thick wind, and the residual 10\% ($0.2 \times 10^{-6} M_\sun$) 
has been left and added to the helium layer of the WD.  
The residual mass itself depends on both the hydrogen content $X$ and
the WD mass.  It is roughly ranging from 10\% ($X=0.70$) to 15\%($X=0.50$)
including ambiguities of the WD mass ($M_{\rm WD}=1.377\pm 0.01 M_\sun$).
Therefore, the net mass increasing rate of the WD 
is $\dot M_{\rm He} = 1.2 \times 10^{-8} M_\sun$ yr$^{-1}$.
\par
     Our calculated ejecta (wind) mass seems to be inconsistent with
observation.  \citet{boh89} estimated the (fully ionized hydrogen) 
shell mass from their observation 201 days after maximum.  
The obtained value is 
$M_{\rm shell} \sim 1.5 \times 10^{-5} (d / \mbox{1~kpc})^2 M_\sun$.
Then, they obtained the ejecta mass form their equation (8), i.e.,
$M_{\rm ejecta} \sim 3(V_{\rm s}/V_{\rm e})^2 M_{\rm shell}$, where 
$V_{\rm s}$ is the shell velocity after the shock has broken out of 
the circumstellar envelope and $V_{\rm e}$ is the velocity 
of the ejecta before it collides with the circumstellar matter.
Assuming $V_{\rm s}= 200$ km s$^{-1}$ and $V_{\rm e}= 2000$ km s$^{-1}$,
Bohigas et al. estimated 
$M_{\rm ejecta} \sim 4.5 \times 10^{-7} (d / \mbox{1~kpc})^2 M_\sun$.
This ejecta mass is not consistent with our numerical model
($\Delta M_{\rm wind} = 2.0 \times 10^{-6} M_\sun$)
if we use $d= 0.6$ kpc.
\par
     This inconsistency comes from the inappropriate pair of velocities
because the estimated mass depends sensitively on the
ratio $V_{\rm s} / V_{\rm e}$.
If we adopt an appropriate pair of velocities, our wind model
is consistent with the observational shell mass.
Starting from the set of $d= 0.6$ kpc and 
$M_{\rm ejecta} \equiv \Delta M_{\rm wind} \sim 2.0 \times 10^{-6} M_\sun$,
we obtain
$M_{\rm shell} \sim 5.4 \times 10^{-6} M_\sun$ and
$V_{\rm s}/V_{\rm e} \sim 0.35$.  
Adopting $V_{\rm e} \sim 4000$ km s$^{-1}$ from the observation
\citep[e.g.,][]{ros87i, sni87a, sni87b} 
gives $V_{\rm s} \sim 1400$ km s$^{-1}$.
\par
     This velocity of 1400 km s$^{-1}$ is consistent with
the radio map obtained 77 days after maximum \citep{tay89}.
The VLBI map shows an elongated
structure of angular dimensions of $\sim 0 \farcs 22$
(long axis) and $\sim 0 \farcs 08$ (short axis), 
which is consistent with the distance of 0.6 kpc
and the fastest expansion velocity of 1400 km s$^{-1}$ after shock,
i.e.,
$ 0 \farcs 092 ~(d / \mbox{1~kpc})^{-1} 
\cdot (V_{\rm s} / \mbox{1000~km~s}^{-1}) \sim 0 \farcs 22$.
Thus, we have obtained a consistent set of physical quantities,
$M_{\rm shell} \sim 5.4 \times 10^{-6} M_\sun$,
$M_{\rm ejecta} \sim 2.0 \times 10^{-6} M_\sun$,
and $V_{\rm s}= 1400$ km s$^{-1}$ for the distance of $d=0.6$ kpc.
\par
     Thus, we obtain the cool wind mass loss rate of 
$4 \times 10^{-7} M_\sun$ yr$^{-1}$ ($= M_{\rm shell}/18$~yr), 
indicating that
one fourth of the cool wind has been captured by the WD.
Observationally, a very high mass loss rate of the cool wind,
$\sim 10^{-5} M_\sun$ yr$^{-1}$, was suggested 
by \citet{sho96} based on the UV data. 
However, \citet{dob96} argued against such 
a high mass loss rate of the M-giant cool wind.  If the wind mass
loss rate approaches such a high rate of $10^{-5} M_\sun$ yr$^{-1}$,
a very large 10$\mu$m excess should be observed.  But, the giant of
RS Oph has only a modest 10---20$\mu$m excess.
They estimated a few times $10^{-7} M_\sun$ yr$^{-1}$ 
\citep*[e.g.,][]{ken88}.  This mass loss rate of the cool wind
is very consistent with the present estimation.
\par
      Finally, we examine whether or not 
RS Oph will explode as a Type Ia supernova in quite a near
future.  One of the conditions for SN Ia explosions is changed
as follows:
\begin{equation}
\dot M_{\rm acc} > \dot M_{\rm He-det} \sim 
4 \times 10^{-8} M_\sun \mbox{~yr}^{-1},
\label{low_accretion_condition_for_SNIa_metal_poor}
\end{equation}
for low metallicity stars \citep{nom82, nom91} instead of
equation (\ref{low_accretion_condition_for_SNIa}).
Our estimated mass accretion rate of
$\dot M_{\rm acc}= 1.2 \times 10^{-7} M_\sun$ yr$^{-1}$
meets the above condition.  
Thus, we may conclude that RS Oph is an immediate progenitor of
an SN Ia even if the donor is a metal-poor star.

\subsection{Absence of Secondary Maximum}
     Among the four recurrent novae with a red giant companion,
only T CrB shows a secondary maximum of $\Delta V \sim 2$ mag brighter 
than the quiescent level 120---190 days after the optical maximum.
We discuss in more detail on the difference between T CrB and the others.
The essential difference exists in the fact that the red giant component
in T CrB fills its inner critical Roche lobe but the others do not
(see the following sections for V745 Sco and V3890 Sgr).
\par
     In the T CrB system, the mass transfer continues 
from the red giant component to the white dwarf even in the outburst 
(hot) wind phase.  Moreover, the strong irradiation increases 
the mass transfer rate from the red giant component; thus increasing
the radius of the optically thick region of the accretion disk.
The unstable warping grows shortly after
the hot wind stops because the mass transfer rate satisfies 
the warping condition (\ref{radiation_condition_T_CrB}).
\par
     On the other hand, the RS Oph system cannot maintain
the mass transfer because the cool red giant wind has once been
evacuated by the outburst (hot) wind/ejecta from the white dwarf.
During the day 0---80 (hot wind phase), 
the growth of warping was suppressed 
by the hot wind because the hot wind (matter) momentum is much 
larger than the photon momentum as shown by \citet{kat94h}.
During the day 80---120 (steady burning phase),
the warping instability may grow to some extent.
However, the disk cannot contribute much to the $V$ light,  
since the optically thick region of the accretion disk
has become as small as or smaller than $\sim 1 R_\sun$. 
\par
     The mass transfer to the white dwarf 
resumes $\sim 400$ days after the optical maximum, that is,
after the hot wind stops ($\sim 80$ days)
and then the cool red giant wind fills again the binary system
($\Delta t = a/v \sim 300~R_\sun / 10$ km s$^{-1}$ = 300 days)
as already discussed earlier.  
At that time, the luminosity of the white dwarf has once decreased
as low as $\sim 100$---$200~L_\sun$ 
and too low (two orders of magnitude lower) 
to excite the warping instability as easily understood from
the condition for RS Oph \citep{sou97}, i.e.,
\begin{eqnarray}
{{\dot M_{\rm acc}} \over {3 \times 10^{-8}~M_\sun \mbox{~yr}^{-1}}}
& \lesssim & 
\left( {{R_{\rm disk}} \over {R_\sun}} \right)^{1/2}
\left( {{L_{\rm bol}} \over {2 \times 10^{38} 
\mbox{~ergs~s}^{-1}}} \right) \cr
&\times & \left( {{R_{\rm WD}} \over {0.003 ~R_\sun}} \right)^{1/2}
\left( {{M_{\rm WD}} \over {1.377 ~M_\sun}}\right)^{-1/2}.
\label{radiation_condition_RS_Oph}
\end{eqnarray}
In the cool wind accretion phase, our estimated accretion rate 
of the WD in RS Oph is 
$\dot M_{\rm acc} \sim 1.2 \times 10^{-7} M_\sun$ yr$^{-1}$
(therefore, $L_{\rm bol} \sim 5 \times 10^{36}$ ergs s$^{-1}$
for the accretion luminosity), 
which also does not meet the above condition even if 
$R_{\rm disk}= 10 ~R_\sun$, so that 
the radiation-induced instability does not occur in RS Oph.  
This is the reason why RS Oph does not show a second peak.
\par
     The other two, V745 Sco and V3890 Sgr, are similar systems
to RS Oph, so that a secondary maximum was not observed, 
as will be shown in the following sections.

\subsection{Summary of RS Oph Outburst}
     The main features of the RS Oph outbursts can be well understood
with a thermonuclear runaway model on a very massive white dwarf
close to the Chandrasekhar mass limit.
We have summarized the following important points: \\
(1) It has been suggested that the M-giant in RS Oph is metal-poor.
Assuming an upper limit of white dwarf mass, i.e.,
$M_{\rm WD}= 1.377 ~M_\sun$, we have obtain a lower limit of 
metallicity, i.e., $Z \ge 0.004$ from the light curve fitting
of the early phase ($\sim 0$---4 days).
In other words, the early visual light curve 
can be well reproduced only by a bloated white dwarf photosphere
of $M_{\rm WD}= 1.377 \pm 0.01~M_\sun$ for a low metallicity 
of $Z=0.004$.  On the other hand, $M_{\rm WD}= 1.35\pm 0.01 ~M_\sun$ 
is obtained for the solar metallicity of $Z=0.02$. \\
(2) The ensuing phase ($\sim 5$---80 days) visual light
curve cannot be reproduced only by a white dwarf photosphere.
It requires an irradiation effect of the accretion disk,
the size of which is gradually decreasing from $\sim 10 ~R_\sun$ 
to $\sim 1~R_\sun$.  \\
(3) The UV light curve can also be well reproduced with the same
model as that of the visual luminosity.  The early phase UV flux mainly
comes from the bloated white dwarf photosphere  
while the late phase UV flux comes from the irradiated 
disk surface because the main emitting region of the white dwarf
photosphere has shift from UV to soft X-ray in the late stage. \\
(4) The optically thick wind stops on the day $\sim$80.
The WD photosphere quickly shrinks.
The sharp drop in the UV flux around the day 80
can be understood from this quick shrink of the WD photosphere. \\
(5) The distance to RS Oph is estimated to be 0.6 kpc
from the fitting with the dereddened UV flux. 
This distance is consistent with the $V$ light curve fitting. \\
(6) The envelope mass of the bloated white dwarf is
$2.2 \times 10^{-6} M_\sun$, suggesting the mass accretion
rate of $1.2 \times 10^{-7} M_\sun$ yr$^{-1}$. \\
(7) The wind carries away about 90\% of the envelope mass,
so that the residual 10\% can be accumulated to the helium
layer of the white dwarf.  As a result, the white dwarf grows 
at a rate of $1.2 \times 10^{-8} M_\sun$ yr$^{-1}$. \\
(8) RS Oph will certainly explode as a Type Ia supernova
in quite a near future if the white dwarf consists of carbon and oxygen.

\section{V745 SCORPII}
      V745 Scorpii underwent the second recorded outburst in 1989 
(the first recorded outburst in 1937), and
has been recognized as a member of recurrent novae 
\citep[e.g.,][and references therein]{sek90}.
Its light curve is characterized by a very rapid rise 
(in about half a day) and decline ($t_3 \sim 9$ days,
$t_3$ is the time taken to drop 3 mag from maximum).
The general spectral evolution closely resembles that of
RS Oph \citep[e.g.,][]{sek90}.  Although its orbital period, 
the most important system parameter, is still not known,
the secondary cool component has been classified as 
M6---8 III \citep[from optical spectra,][]{sek90, due90, wil91}, 
M4 III \citep[from the infrared spectra and colors,][]{har93},
or M4---6 III \citep[from the absorption line indices,][]{anu99}.  
Therefore, V745 Sco is a twin system to RS Oph, 
a member of the recurrent novae with a red giant secondary.  

\subsection{The Model of the 1989 Outburst}
\begin{figure}
%\epsscale{.9}
%\plotone{vmag1350va1_v745sco_x70i80p560.eps}
\plotone{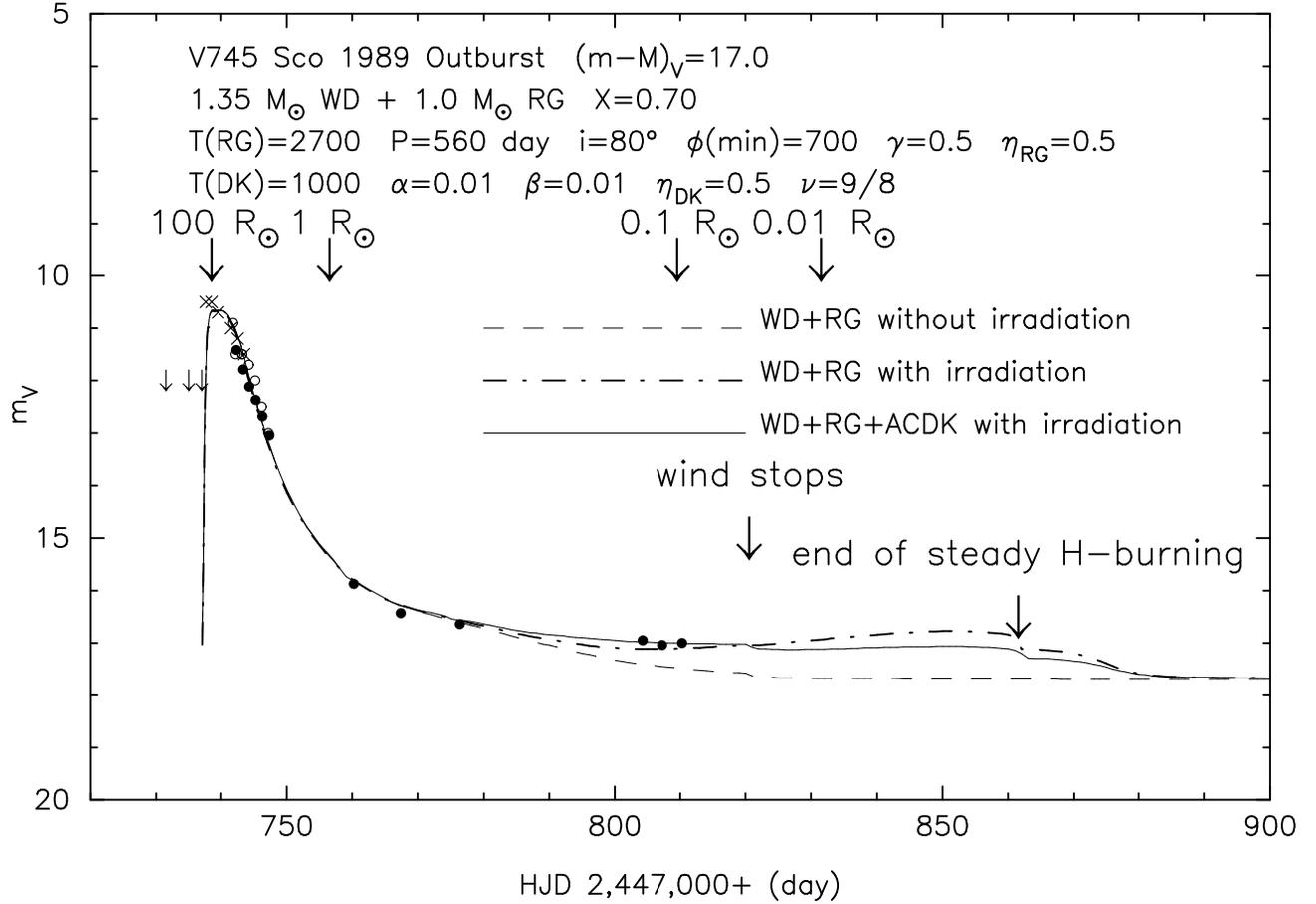}
%\plotfiddle{sn1a96let/evolution1.ps}{5.0cm}{270}{0.4}{0.4}{-170}{220}
\caption{
Model light curves are plotted against time (HJD 2,447,000+) 
together with the observational points of the V745 Sco 1989 outburst.
Here, the inclination angle of $i=80\arcdeg$ is assumed.
The other system parameters are shown in the figure.
Filled circles indicate the observational points with $V$ filter 
\citep[taken from][]{sek90}.  Crosses \citep[taken from][]{lil89a,
lil89b} are shifted by 0.8 mag down because they are systematically 
0.8 mag brighter than the $V$ magnitude.  Small allows (downwards) 
indicate an upper limit before the 1989 outburst 
\citep[taken from][]{hon89}. 
Open circles are visual magnitudes taken from 
\citet{sho89} and \citet{ovb89}. 
Dashed line denotes the $V$ light curve of the WD photosphere and
the nonirradiated RG only.  Dash-dotted line corresponds to
the $V$ light from the WD photosphere and the irradiated RG.
Solid line represents the total $V$ light of the WD photosphere,
the irradiated RG, and the irradiated accretion disk.
The photospheric radius of the WD is indicated above the $V$ light
curve.  Two epochs are indicated by arrows 
with "wind stops" and "end of steady H-burning."
The optically thick wind stops at HJD 2,447,819.
The steady hydrogen shell burning ends at HJD 2,447,860.
Fitting with the observations indicates an apparent distance modulus 
of $(m-M)_V= 17.0$.
\label{vmag1350va1_v745sco_x70i80p560}}
\end{figure}

     We have calculated $V$ light curves for five cases 
of the WD mass ($M_{\rm WD}= 1.3$, 1.35, 1.36, 1.37, and $1.377 M_\sun$)
by assuming the hydrogen content $X=0.70$, 0.50, or 0.35 with 
the solar metallicity ($Z=0.02$). 
By fitting light curves in the first 25 days of the 1989 outburst, 
we have determined the WD mass of $M_{\rm WD}= 1.35 \pm 0.01 M_\sun$
as shown in Figure \ref{vmag1350va1_v745sco_x70i80p560}.
Here, we have shifted the visual magnitude obtained with orange 
filter or 2415 films \citep{lil89a, lil89b} 
by $\sim 0.8$ mag down (crosses, $\times$)
because they are systematically about 0.8 mag brighter 
than the $V$-magnitude obtained by \citet{sek90}. 
The decline rate of the $V$ light in the early phase depends 
very sensitively on the WD mass, but depends hardly on the hydrogen
content $X$, the companion (its mass and radius), or the size 
of the accretion disk, so that we can safely determine
the WD mass from the early light curve \citep[see also][]{kat99}.

\placefigure{vmag1350va1_v745sco_x70i80p560}

\par
   Our fitting indicates an apparent distance modulus 
of $(m-M)_V= 17.0$. 
Thus, the distance to V745 Sco is derived to be $d = 5.0$ kpc
if we adopt the absorption of $A_V= 3.1 \times E(\bv) = 3.5$,
where $E(\bv)=1.1$ \citep{har93, wil94}.
\par
     From our solutions, the envelope mass at the maximum $V$ light
is estimated to be $\Delta M = 4.6 \times 10^{-6} M_\sun$,
indicating a mass accretion rate 
of $\dot M_{\rm acc}= 0.9 \times 10^{-7} M_\sun$ yr$^{-1}$ 
between the 1937 and 1989 outbursts.
About 95\% of the envelope mass has been blown off
by the wind and the residual 5\% 
($2.3 \times 10^{-7} M_\sun$) has been left and added
to the helium layer of the WD.  Therefore, we obtain the net
mass increasing rate of the WD 
as $\dot M_{\rm He}= 0.45 \times 10^{-8} M_\sun$ yr$^{-1}$.

\subsection{Irradiation Effect of Companion M-giant} 
     In the late phase 
($t \sim 25$---200 days after the optical maximum), 
contributions of the irradiated red giant (RG)
and accretion disk (ACDK) to the $V$ light
may become dominant like as in T CrB and RS Oph.  
First, in this subsection, we examine the irradiation effect of 
the M-giant, and then, in the next subsection, we also study 
the irradiation of the ACDK.  
\par
     In order to examine companion's irradiation effect,
we need to know the separation, mass ratio, inclination angle,
and the radius of the red giant, but none of them is known.
The lack of both the orbital information and the visual luminosity
observation in the late phase clouds the accurate determination of 
the outburst model because of ample ambiguity in the system 
parameters.
\par
     The orbit is assumed to be circular.
It is observationally suggested that orbital periods
of S-type symbiotic stars are related to their spectral type by
\begin{equation}
P \sim (1.31)^S \times 190 {\rm ~days}, 
\label{symbiotic_period_spectral}
\end{equation} 
where the power of $S$ means the sub-spectral type of M-giants
\citep[$S=4$ for M4 III, see, e.g.,][]{mur99, har00}.  
From this relation, we assume $P=560$ days for V745 Sco
\citep[M4 III, e.g.,][]{har93} and  
the ephemeris for inferior conjunction of the M-giant in front,
$\phi_{\rm min}=$ HJD $2,448,700.0 + 560 E$.
In what follows, we refer this ephemeris simply 
as $\phi_{\rm min}= 700$ day.
\par
     It has also been suggested that M-giants in S-type symbiotic stars
lies well within the inner critical 
Roche lobe, that is, $\gamma \sim 0.3-0.6$ \citep[e.g.,][]{mur99}.
Therefore, we assume roughly the ratio of 
\begin{equation}
{{R_{\rm RG}} \over {R_2^*}} \equiv \gamma \sim {1 \over 2}.
\label{ratio_of_RG_radius_to_Roche_lobe}
\end{equation} 
In addition, we also assume 50\% efficiency of the irradiation 
($\eta_{\rm RG}=0.5$).  For the companion M-giant, 
\citet{sek90} suggested $m_V= 17$ in quiescence
from the $V$ magnitude just after the outburst.
\citet{anu99} obtained $m_{5500}= 17.7$
in March 1998.   We adopt here $m_V = 17.7$ in quiescence. 
\par
     The light curves are calculated for six cases of the companion mass, 
i.e., $M_{\rm RG}= 0.5$, 0.6, 0.7, 0.8, 1.0, and $1.2 M_\sun$. 
Since we have obtained similar light curves for all of these six masses,
we show here only the results for $M_{\rm RG}= 1.0 M_\sun$.
For this case, the separation is $a= 380.1 R_\sun$, the effective 
radii of the inner critical Roche lobes are $R_1^*= 154.0 R_\sun$
for the WD component, and $R_2^*= 134.3 R_\sun$ for the RG component.
If $\gamma= 0.5$, then we have 
$R_2 = R_{\rm RG} = 0.5 R_2^* \sim 67 R_\sun$.
% AV. DM2/DT=  0.1284E-04  WIND=  0.9533  M1= 1.3500  M2= 1.0000
% AA=380.073  RL1=153.986  RL2=134.262  RDISK=  1.540  FLUPD=  0.010
Then, the nonirradiated photospheric temperature of the RG
is estimated to be $T_{\rm ph, RG} = 2700$ K from
the light curve fitting in quiescence.
\par
\begin{figure}
\epsscale{.9}
%\plotone{vmag1350va1_v745sco_x70imixp560.eps}
\plotone{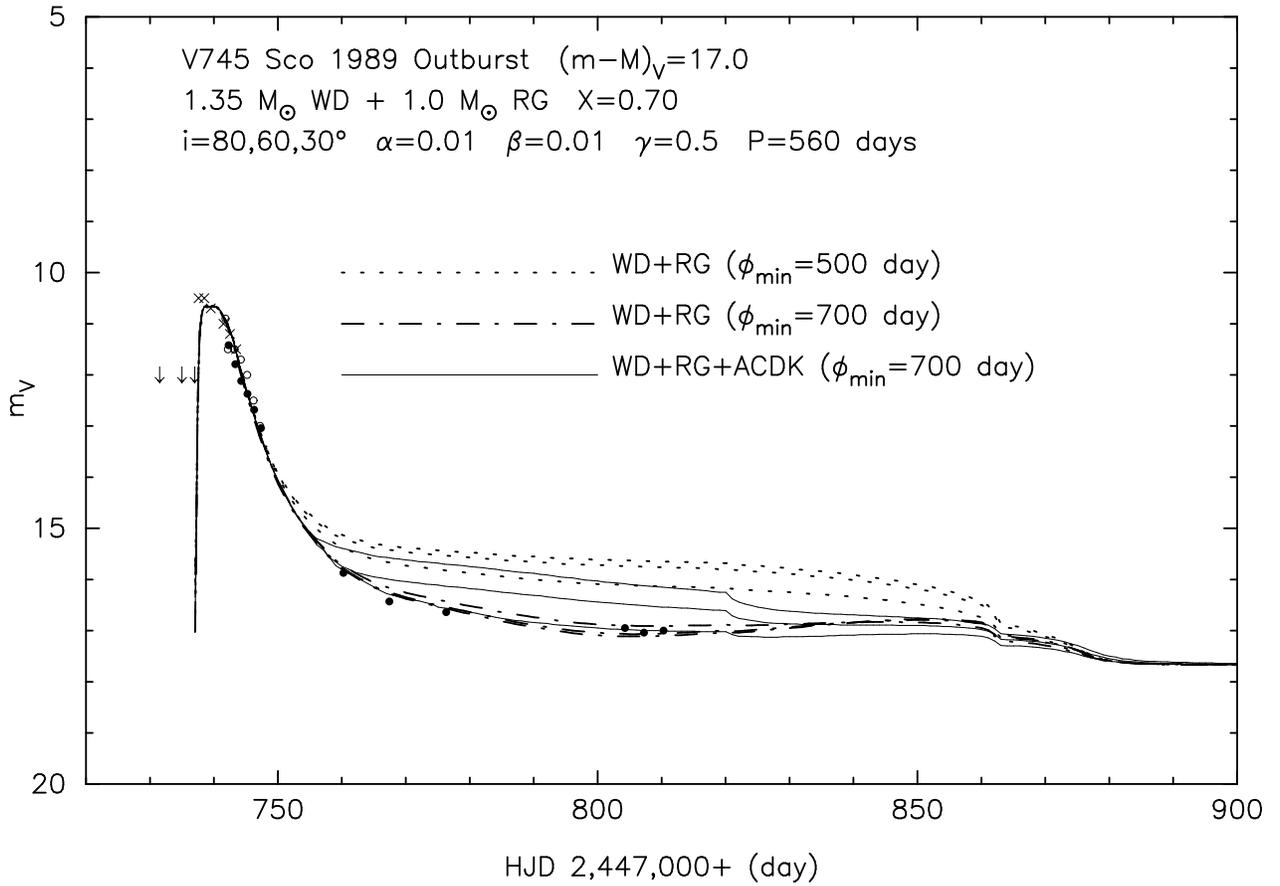}
%\plotfiddle{sn1a96let/dmdt_env1.ps}{5.0cm}{270}{0.4}{0.4}{-170}{220}
\caption{
Same as Fig. 13 but for different inclination angles 
($i=30$, 60, $80\arcdeg$ from top to bottom).
Dot-dashed lines denote the model light curves for the cases of
no accretion disk while solid lines denote the cases with an accretion
disk.  The model light curves with an accretion disk depend largely on
the inclination angle, while we do not find any significant differences 
for the cases of no accretion disk.  
Dotted lines indicate a different ephemeris of 
$\phi_{\rm min} = 500$ day, which seems not to be compatible
with the observation for all three inclination angles.
\label{vmag1350va1_v745sco_x70imixp560}}
\end{figure}

     We have finally adopted the best fit set of 
the inclination angle $i=80\arcdeg$ for the ephemeris of
$\phi_{\rm min}=700$ day as shown 
in Figure \ref{vmag1350va1_v745sco_x70i80p560}.
Dash-dotted curves in Figure \ref{vmag1350va1_v745sco_x70imixp560} show
$V$ light curves for various inclination angles
($i=30\arcdeg$, $60\arcdeg$, $80\arcdeg$)
of the orbit for the ephemeris of $\phi_{\rm min}=700$ day.
Their light curves depends hardly on the inclination angle
because the non-irradiated hemisphere of the M-giant faces 
towards the Earth.  On the other hand, $V$ light is much brighter
(dotted curves) if the ephemeris of $\phi_{\rm min}=500$ day is adopted 
because the irradiated hemisphere of the M-giant faces 
towards the Earth,  which is not compatible with the observation.

\begin{figure}
\epsscale{.9}
%\plotone{vmag1350va1_v745sco_xxmixp560.eps}
\plotone{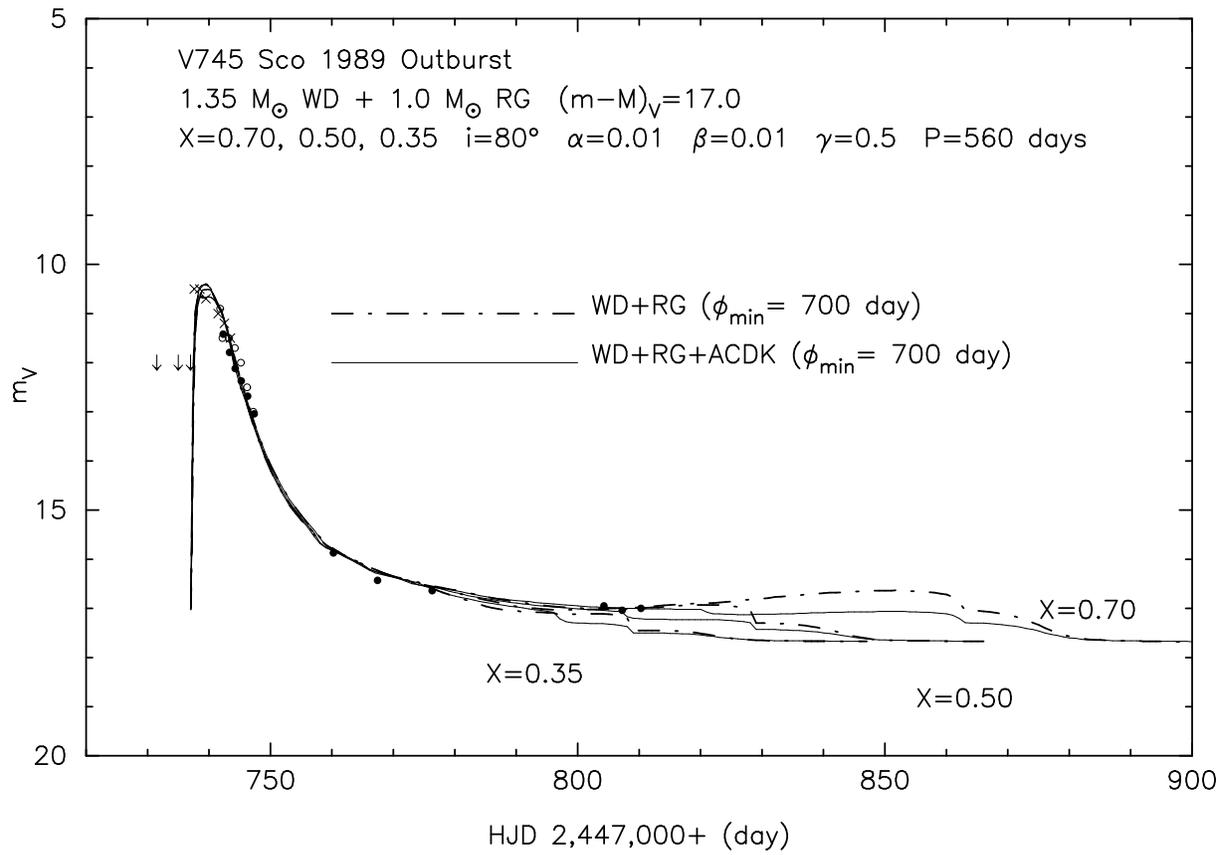}
%\plotfiddle{sn1a96let/evolution1.ps}{5.0cm}{270}{0.4}{0.4}{-170}{220}
\caption{
Same as Fig. 13 but for three different hydrogen content of $X=0.35$, 
0.50, and 0.70.   We do not determine the hydrogen content from
these observational points, but may conclude that it is somewhere
between $X=0.5$ and $0.7$.
\label{vmag1350va1_v745sco_xxmixp560}}
\end{figure}

\placefigure{vmag1350va1_v745sco_x70imixp560}
\placefigure{vmag1350va1_v745sco_xxmixp560}

\par
     Figure \ref{vmag1350va1_v745sco_xxmixp560} shows the dependency
of the hydrogen content, $X$.
The early-phase $V$ light curve depends hardly on the hydrogen content, 
while the late-phase $V$ light curve is weakly identified
by the hydrogen content of the WD envelope.
The optically thick, outburst wind stops 59, 69, 82 days after maximum,
and the steady hydrogen shell burning ends 71, 90, 123 days 
after maximum, for $X= 0.35$, 0.50, and 0.70, respectively.
The $V$ light curve indicates that the hydrogen content is somewhere
between $X=0.5$ and $X=0.7$ (dash-dotted curves).
\par
     We have also calculated light curves for a much longer 
orbital period 
\citep[$P= 1000$ days, for M6 III, e.g.,][]{anu99}, 
and have obtained similar results to those for $P= 560$ days.  
In this case, the binary parameters are
$a= 559.4 R_\sun$, $R_1^* = 226.7 R_\sun$, $R_2^* = 197.6 R_\sun$,
$R_2 = 99 R_\sun$.  We have estimated $T_{\rm RG}= 2500$ K 
for the best fit model in the late phase.  
% AV. DM2/DT=  0.1271E-04  WIND=  0.9624  M1= 1.3500  M2= 1.0000
% AA=559.424  RL1=226.650  RL2=197.619  RDISK=  0.000  FLUPD=  0.000

\subsection{Small Irradiation Effect of Accretion Disk}
     In this subsection, we show that the ACDK should not contribute
so much to the $V$ light.  Since the visual light curve is well 
reproduced by M-giant's irradiation with no accretion disk,
one may imagine that the accretion disk does not contribute much.
Here, we determine the upper limit 
of the ACDK size in relation to the reason why 
a secondary maximum does not occur in V745 Sco. 
\par
     We here again assume 50\% efficiency of the ACDK irradiation
($\eta_{\rm DK}= 0.5$).
The temperature of the unheated surface of the ACDK including
the rim is assumed to be $T_{\rm disk}= 1000$ K.
We have examined another case of $T_{\rm disk}= 0$ K, but
found no significant differences in the $V$ light curves.
The power of $\nu$ is 
assumed to be $\nu=9/8$ from the standard disk model.  
\par
     We have finally obtain allowable sets of the disk parameters 
of $\alpha \lesssim 0.01$ and $\beta \lesssim 0.01$ 
for the orbital inclination
angle of $i \sim 80\arcdeg$ as shown 
in Figure \ref{vmag1350va1_v745sco_x70i80p560}.  
Here, the thin solid line 
denotes the case of $\alpha=0.01$ and $\beta=0.01$, while
the dash-dotted line corresponds to $\alpha=0$ (no accretion disk).
Both can reproduce the $V$ light curve of the 1989 outburst.
\par
     We have examined six cases of the inclination angle of the orbit,
$i= 80\arcdeg$, $70\arcdeg$, $60\arcdeg$, $50\arcdeg$, $40\arcdeg$,
and $30\arcdeg$ and shown three of them 
in Figure \ref{vmag1350va1_v745sco_x70imixp560}.
The $V$ light curves are more luminous 
for a lower inclination angle of the orbit,
since a larger area of the ACDK can be seen from the Earth.
This means that lower inclinations of the orbit
contradicts the existence of a large ACDK.
To summarize, for any case, the ACDK should not  
contribute so much to the $V$ light.
\par
     Figure \ref{vmag1350va1_v745sco_xxmixp560} shows
the dependency of the hydrogen content 
for the case with an accretion disk.
The $V$ light curves suggest that the hydrogen content is not clearly
identified by the observation 
but it may be somewhere between $X=0.5$ and $0.7$ (solid curves).
\par
     Thus, a very small size of the optically thick ACDK, 
i.e., $R_{\rm disk} \lesssim 1 ~R_\sun$, in the late phase  
indicates no secondary maximum of the light curve
as discussed in RS Oph even if the warping instability sets in.

\subsection{The Distance, Envelope Mass of the White Dwarf,
and Possibility of SN Ia Explosion}
     The distance to V745 Sco is estimated 
to be $d= 5.0$ kpc with the absorption of $A_V= 3.5$.
This is roughly consistent with Harrison et al.'s (1993) estimation 
of $d= 4.6$ kpc.  Harrison et al. attributed M4 III to the cool
component from the infrared CO lines. 
Assuming the absolute magnitude of $M_K = -5.5$ (M4 III), 
they obtain the distance modulus of 
$(m-M)_0 = m_K - M_K - A_K = 8.2 - (-5.5) - 0.4 =13.3$, where 
they adopted $m_K = 8.2$ from \citet{sek90} 
and used a relation between the absorptions of $A_K$ and
$A_V$ given by \citet{rie85}, i.e.,
$A_K / A_V = 0.112$.
\par
     Sekiguchi et al.'s (1990) K-band luminosity $m_K= 8.21$ 
(87 days after maximum) is about 0.2 mag brighter than 
Harrison et al.'s (1993) K-band luminosity $m_K= 8.41$ 
(two years after the 1989 outburst).  This is consistent with 
our models as seen in Figure \ref{vmag1350va1_v745sco_x70i80p560}, 
because the cool component had been strongly irradiated 
by the WD photosphere always during Sekiguchi et al.'s observations.
Using $m_K= 8.41$, we obtain a distance modulus of 
$(m-M)_0 = m_K - M_K - A_K =8.41 - (-5.5) - 0.4 = 13.5$, i.e., 
$d=5.0$ kpc, being coincided with our distance estimation.
\par
     At the end of this section, we discuss the possibility of
SN Ia explosion.  
We have already obtained the average mass accretion rate
of $\dot M_{\rm acc}= 0.9 \times 10^{-7} M_\sun$ yr$^{-1}$ 
between the 1937 and 1989 outbursts from 
the envelope mass of $\Delta M = 4.6 \times 10^{-6} M_\sun$. 
This meets condition (\ref{low_accretion_condition_for_SNIa})
for SN Ia explosions \citep{nom91},
For all the three cases of $X=0.70$, 0.50, and 0.35,
about 95\% of the envelope mass has been blown off
in the outburst wind and the residual 5\% 
($2.3 \times 10^{-7} M_\sun$) has been left and added
to the helium layer of the WD.  Therefore, we obtain the net
mass increasing rate of the WD 
of $\dot M_{\rm He}= 0.45 \times 10^{-8} M_\sun$ yr$^{-1}$.  
Thus, we may conclude that V745 Sco is also an immediate progenitor of
Type Ia supernova if the hot component is a carbon-oxygen 
white dwarf.

\section{V3890 SAGITTARII}
      V3890 Sagittarii underwent the second recorded outburst in 1990 
(since the 1962 first recorded outburst), and
has been recognized as a member of the recurrent novae 
\citep{jon90a, lil90}.  
Its light curve is characterized by a very rapid rise 
(within two days) and fast decline \citep[$t_3 \sim 17$ days;
see, e.g., Fig. 1 in][]{gon92}. 
The general spectral evolution closely resembles that of
RS Oph \citep[e.g.,][]{sek90iauc}.  
The cool component has been classified from optical spectra as 
M4 III \citep{sek90iauc}, M8 III \citep{wil91}, 
or M5 III \citep{anu99}.
The infrared spectra and colors indicate a spectral type 
M5 III \citep{har93}.  Thus, V3890 Sgr is
a similar system to RS Oph and V745 Sco, 
a member of the recurrent novae with an M-giant secondary.

\subsection{The Model of the 1990 Outburst}
\begin{figure}
\epsscale{.9}
%\plotone{vmag_mix_mass_v3890sgr.eps}
\plotone{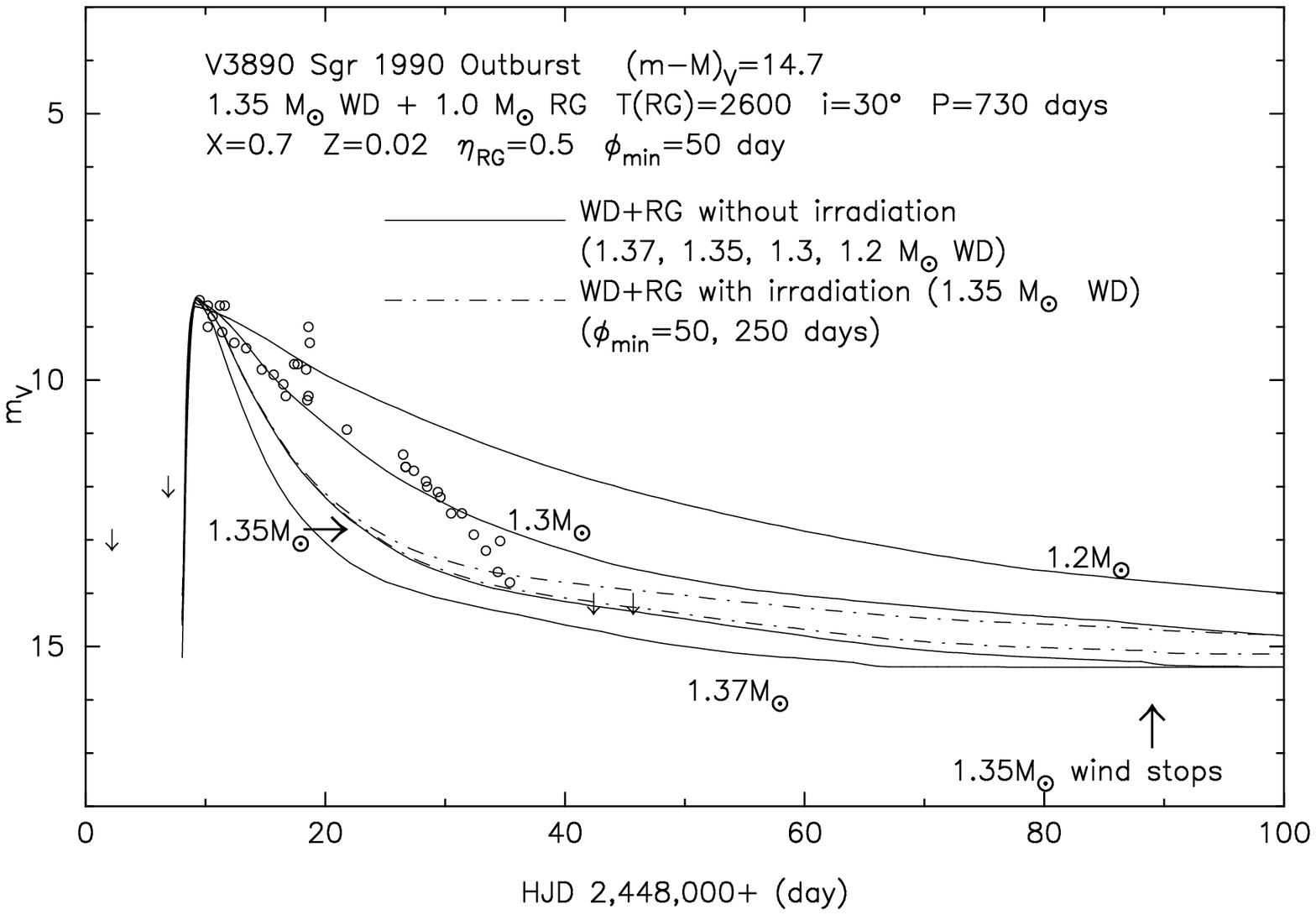}
%\plotfiddle{sn1a96let/evolution1.ps}{5.0cm}{270}{0.4}{0.4}{-170}{220}
\caption{
Model light curves are plotted against time (HJD 2,448,000+) 
together with the observational points of the V3890 Sgr 1990 outburst.
Here, the inclination angle of $i=30\arcdeg$ is assumed.
The other system parameters are shown in the figure.
Open circles indicate the observational points 
taken from IAU Circulars 
\citep{jon90a, jon90b, jon90c, 
pea90a,  pea90b,  pea90c,  pea90d,  pea90e,  pea90f,
per90a, per90b, boa90, lil90, 
smr90a, smr90b, smr90c} and   
Figure 1 of \citet{gon92}, while
small downward arrows show an upper limit.  
Solid lines denote the total $V$ light of the WD photosphere and
the nonirradiated M-giant secondary.  
Dot-dashed lines depict the total $V$ light of the WD photosphere and
the irradiated M-giant secondary with two different $\phi_{\rm min}$,
i.e., $\phi_{\rm min}=50$ day (lower) and $\phi_{\rm min}= 250$ day
(upper) only for $M_{\rm WD}=1.35 ~M_\sun$.  
Fitting with the observations suggests an apparent distance modulus 
of $(m-M)_V= 14.7$.  
Here, we assume that the WD photosphere reached its maximum
expansion on HJD 2,448,009.
\label{vmag_mix_mass_v3890sgr}}
\end{figure}

     We have calculated the total $V$ light of the WD photosphere
and the non-irradiated RG photosphere 
for the WD mass of $M_{\rm WD}= 1.2$, 
1.3, 1.34, 1.35, 1.36, 1.37, and $1.377 M_\sun$
by assuming the hydrogen content $X=0.70$
and the solar metallicity of $Z=0.02$. 
Four of them (1.2, 1.3, 1.35, and 1.37 $M_\sun$) are 
plotted in Figure \ref{vmag_mix_mass_v3890sgr} (solid curves).
Here, we assume that the optical maximum was reached on HJD 2,2448,009.
This is because the irradiation of the M-giant cannot contribute
to the early phase $V$ light.
\par
     The observed $V$ light curve in the first 20 days can be roughly
reproduced by the model of $1.3 M_\sun$ WD, but it deviates
greatly in the ensuing phase (20---40 days after maximum).
The envelope mass of the $1.3 M_\sun$ WD 
at the optical maximum is too large
to be compatible with the TNR model.  The envelope mass of
$\Delta M \sim 5 \times 10^{-6} M_\sun$ suggests 
$\dot M_{\rm acc} \sim 2 \times 10^{-7} M_\sun$ yr$^{-1}$,
which is just below the lower limit of
steady hydrogen shell burning for $X=0.7$, i.e., $\dot M_{\rm std}$ 
in equation (\ref{steady_hydrogen_burning}) and also 
in Figure \ref{accmap_z02}.
\par
     Only a bloated WD photosphere cannot reproduce 
the $V$ light curve even in the early phase of outburst.
Moreover, our simple picture of the WD photosphere never
shows a sharp spike on the light curve as seen 
in Figure \ref{vmag_mix_mass_v3890sgr}.  These features 
strongly suggest that the irradiation of the accretion disk is
important even in the early phase:
if the optically thick region of the accretion disk is as large as 
or larger than $\sim 10~R_\sun$, the decay of the $V$ light
becomes much slower as seen in RS Ohp 
(see, e.g., Fig. \ref{vmag1377va1_rsoph1985fixdisk}).
Therefore, we seek a reasonable set of system parameters based on
the assumption that the V3890 Sgr system consists of the WD photosphere, 
the irradiated M-giant, and the irradiated accretion disk.

\placefigure{vmag_mix_mass_v3890sgr}

\subsection{Large Size of Accretion Disk in Early Phase}
     In order to obtain a consistent set of system parameters,
we have to adopt the WD mass more massive than 
$M_{\rm WD} \gtrsim 1.35 M_\sun$,
and a size of the irradiated ACDK larger than
$R_{\rm disk} \gtrsim 10~R_\sun$ at the initial phase of outburst.
The condition of $M_{\rm WD} \gtrsim 1.35 M_\sun$ is required from
the consistencies with 
the TNR model, i.e., the mass accretion rate is smaller than
$\dot M_{\rm acc} \lesssim 1 \times 10^{-7} M_\sun$ yr$^{-1}$, 
and the rapid decline of the $V$ light curve 20---40 days after maximum
as shown in Figure \ref{vmag_mix_mass_v3890sgr}.
The $V$ light curve of a $1.34 M_\sun$ WD model lies above 
the observational points on the day 26 and the upper limits
on the day 33 and 37, being not consistent with the observation.
On the other hand, $V$ light curve of a $1.35 M_\sun$ 
or $1.37 M_\sun$ WD is less bright during the day 3---40. 
The lack of the luminosity during these days
can be supplied by the irradiation of a large size 
accretion disk.
\par
      Figure \ref{vmag1350va1_fixdk_v3890sgr} shows
$V$ light curves of
fixed-size disk models for the WD masses of 1.35 $M_\sun$ 
(solid curves) and 1.37 $M_\sun$ (dashed curves).  
The disk size parameters are $\alpha=0.7$, 0.1, and 0.01. 
It is clear that the fixed-size disk model cannot reproduce 
the $V$ light curve in the later stage.  
This is simply because the large disk contributes too much 
in the later phase.
\par
     Here, we summarize our parameters:
the orbital period of $P= 730$ days
from equation (\ref{symbiotic_period_spectral}) and 
the spectral type of M5 III \citep{anu99} for the cool component,
the ephemeris for the M-giant in front is 
$\phi_{\rm min} =$ HJD 2,448,050.0 $+ 730 \times E$, 
$\gamma= 0.5$ from equation (\ref{ratio_of_RG_radius_to_Roche_lobe}),
50\% efficiency of the irradiation ($\eta_{\rm RG}=0.5$),
the M-giant mass of $M_{\rm RG}= 1.0~M_\sun$. 
In this case, the separation is $a= 453.5 R_\sun$, the effective radii 
of the inner critical Roche lobes are $R_1^*= 183.8 R_\sun$ 
for the WD component and $R_2^*= 160.2 R_\sun$ for the RG component.  
Then, we have $R_{\rm RG} = 0.5 R_2^* \sim 80 R_\sun$.
The nonirradiated photospheric temperature of the RG
is estimated to be $T_{\rm ph, RG} = 2600$ K from fitting with
the quiescent $V$ magnitude of $m_V \sim 16$ \citep{anu99}.
% AV. DM2/DT=  0.8278E-05  WIND=  0.9445  M1= 1.3500  M2= 1.0000
% AA=453.548  RL1=183.754  RL2=160.218  RDISK=  0.000  FLUPD=  0.000
For the accretion disk, we assume 50\% efficiency 
of the irradiation ($\eta_{\rm DK}= 0.5$),
the unheated surface temperature of $T_{\rm disk}= 1000$ K.

\subsection{Evaporating Accretion Disk}
     The optically thick strong wind blows
during the outburst so that the disk surface
is certainly blown off in the wind due to drag and ablation.  
Moreover, the mass transfer from the M-giant is certainly inhibited
during the strong wind phase because the cool red giant wind 
has been evacuated by the hot wind from the WD as discussed in RS Oph.
In such a situation, it is very likely that
the edge of the disk is gradually dissipating with time.
We here simulate such an effect by shrinking the disk size 
exponentially, i.e.,
\begin{equation}
\alpha = \alpha_0 
\left( {{\alpha_1} \over {\alpha_0}}\right)^{(t-t_0)/30},
\end{equation} 
where we adopt $\alpha_0=0.7$ and $\alpha_1=0.01$, $t_0$ is the time of
the optical maximum, and the time $t$ is in units of day.
We set the minimum size of the ACDK at $\alpha= 0.001$ 
in our calculations.  
\par
\begin{figure}
\epsscale{.9}
%\plotone{vmag1350va1_fixdk_v3890sgr.eps}
\plotone{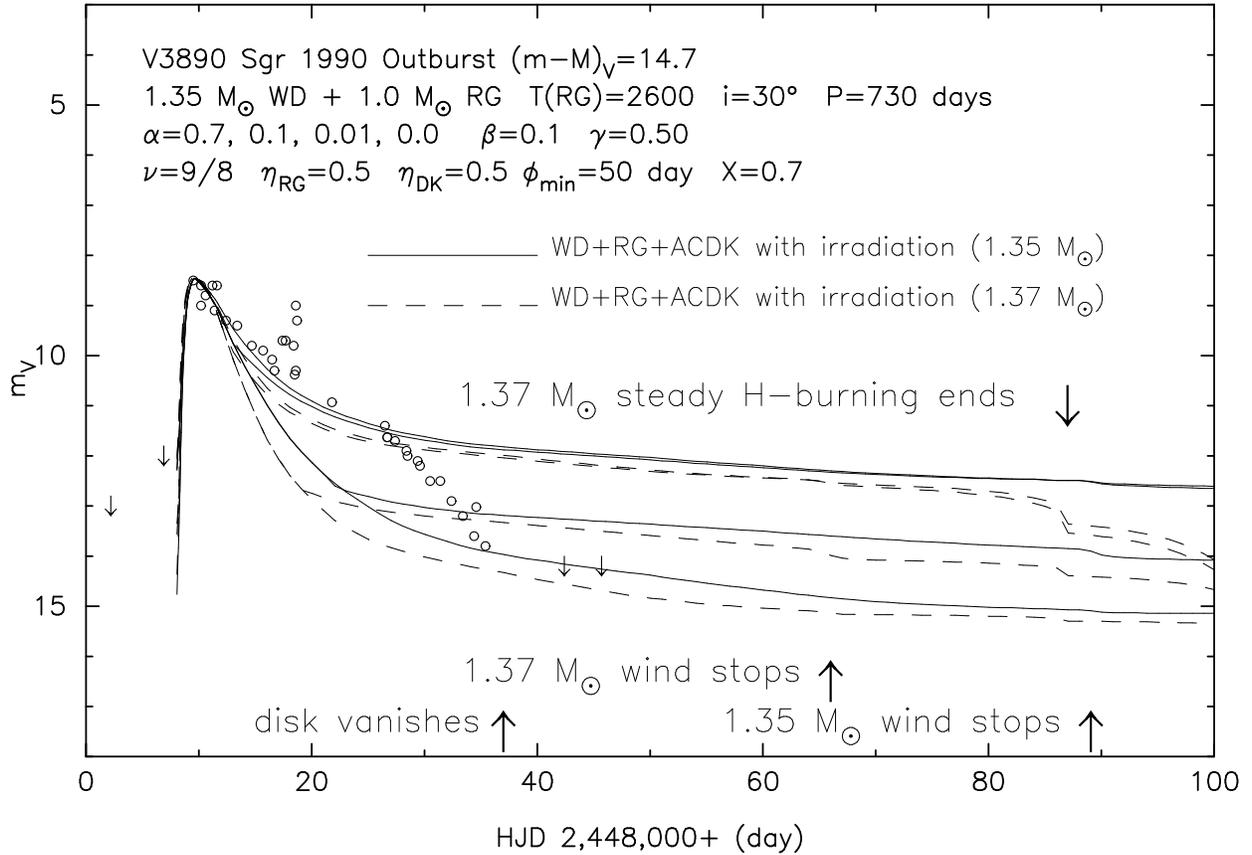}
%\plotfiddle{sn1a96let/evolution1.ps}{5.0cm}{270}{0.4}{0.4}{-170}{220}
\caption{
Same as Fig. 16 but for two white dwarf masses, i.e.,
1.35 $M_\sun$ (solid lines) and 1.37 $M_\sun$ (dashed lines).
Both solid and dashed lines represent the total $V$ light 
of the WD photosphere, the irradiated M-giant, 
and the irradiated accretion disk, 
the size of which is fixed during the outburst as defined 
by equations (\ref{accretion-disk-size}) and (\ref{flaring-up-disk}).
The optically thick wind blows during the outburst, from the beginning 
of the outburst (HJD 2,448,008) to 80 days after maximum 
(HJD 2,448,089), and then the steady hydrogen shell-burning ends 
at 122 days after maximum (HJD 2,448,131)
for the 1.35 $M_\sun$ WD and the hydrogen content of $X=0.7$,
while 57 days and 78 days, respectively, 
for the 1.37 $M_\sun$ WD and the hydrogen content of $X=0.7$.
The slow decline just after the optical maximum can be reproduced
by the relatively large disk size around the 1.35 $M_\sun$ WD
but not by the 1.37 $M_\sun$ WD.
\label{vmag1350va1_fixdk_v3890sgr}}
\end{figure}
\begin{figure}
\epsscale{.9}
%\plotone{vmag1350va1_comp_v3890sgr.eps}
\plotone{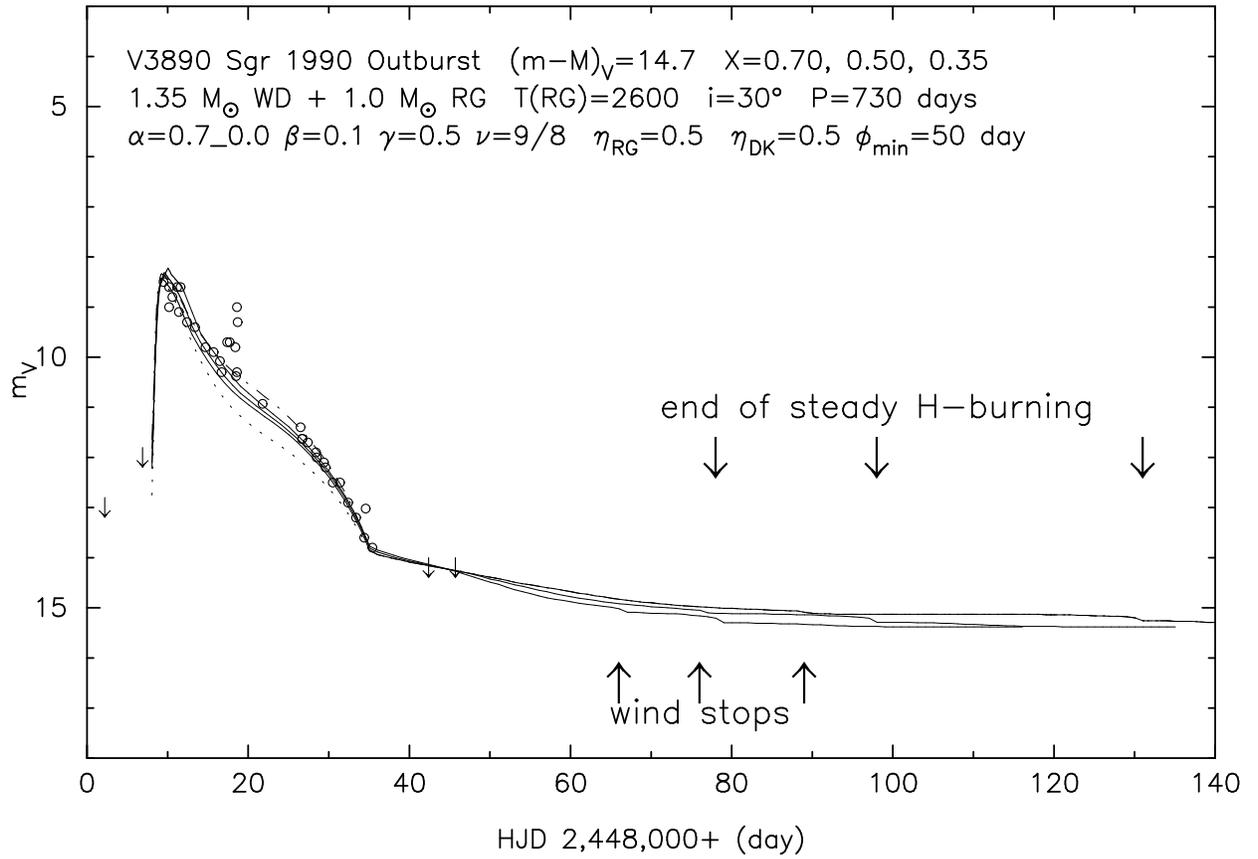}
%\plotfiddle{sn1a96let/evolution1.ps}{5.0cm}{270}{0.4}{0.4}{-170}{220}
\caption{
Same as Fig. 17 but for the case of evaporating accretion disk 
with three different hydrogen content of $X=0.35$, 0.50, and 0.70
({\it solid lines}).   
We do not pose any constraints on the hydrogen content from these 
observational points, because of a lack of data points in the
later phase.
For comparison, two other cases with different irradiation efficiencies, 
$\eta_{\rm DK}= 1.0$ ({\it dot-dashed}) and 0.25  ({\it dotted}), 
are added only for $X=0.70$.
\label{vmag1350va1_comp_v3890sgr}}
\end{figure}

     Thus, we have obtained the best fit model
for the WD mass of $M_{\rm WD}= 1.35 \pm 0.01 ~M_\sun$ 
as shown in Figures \ref{vmag1350va1_fixdk_v3890sgr} 
and \ref{vmag1350va1_comp_v3890sgr}.  
The envelope mass at the maximum $V$ light is estimated to be
$\Delta M = 3.1 \times 10^{-6} M_\sun$,
indicating a mass transfer rate 
of $\dot M_{\rm acc}= 1.1 \times 10^{-7} M_\sun$ yr$^{-1}$ 
between the 1962 and 1990 outbursts.
The efficiency of the ACDK irradiation has been examined 
for other two cases of $\eta_{\rm DK}= 1.0$ (dot-dashed line) 
and 0.25 (dotted line) only for $X=0.70$ and $M_{\rm WD}= 1.35~M_\sun$ 
as shown in Figure \ref{vmag1350va1_comp_v3890sgr}.  
\par
     We have examined the dependency of the $V$ light curve 
on the orbital inclination angle.  The luminosity of the ACDK
depends largely on the the orbital inclination angle 
mainly because a larger area of the ACDK can be seen from the Earth
for a smaller inclination angle.  Therefore, the inclination angle 
should be as low as $i \lesssim 30\arcdeg$ 
to reproduce the $V$ light curve
during $t \sim 5$---20 days after maximum.

\placefigure{vmag1350va1_fixdk_v3890sgr}

\par
     We have also calculated $V$-light curves for other four cases of 
the companion mass, i.e., $M_{\rm RG}=$0.6, 0.7, 0.8, and $1.2~M_\sun$,  
but obtained similar light curves to $M_{\rm RG}= 1.0~M_\sun$.
We have checked the other case of $T_{\rm disk}= 0$ K, but
found no significant differences in the light curves.
The power of $\nu$ is fixed to be $\nu=9/8$, 
but we cannot find any significant 
differences in the light curves even if we adopt $\nu=2$. 
Finally, we have study the effect of hydrogen content.
Either the early- or late-phase $V$ light curve depends hardly 
on the hydrogen content $X$
as shown in Figure \ref{vmag1350va1_comp_v3890sgr}.  
The optically thick, outburst wind 
stops 56, 68, and 80 days after maximum,
and the steady hydrogen shell burning ends 68, 89, and 122 days 
after maximum, for $X= 0.35$, 0.50, and 0.70, respectively.

\placefigure{vmag1350va1_comp_v3890sgr}

     \citet{anu94} obtained optical spectra 
18---20 days after maximum.  
They pointed out that these line features closely resembles 
the spectrum of RS Oph 60 days after maximum.
Anupama \& Sethi derived the Zanstra temperature
$T_* = 300,000$ K of the ionizing central source from the helium and
hydrogen line (H$\beta$ and \ion{He}{2} 4686 \AA) ratios
and its radius $R_* = 0.06 R_\odot$   
from H$\beta$ line flux.  For RS Oph, our model gives us
$T_{\rm ph}= 300,000$ K and $R_{\rm ph}= 0.08~R_\odot$ about 70 days
after maximum (see the previous section), being roughly consistent
with the Zanstra temperature derived by Anupama \& Sethi.
For V3890 Sgr, however, our model suggests 
$T_{\rm ph}= 83,000$ K and $R_{\rm ph}= 1.0~R_\odot$ 
on the day 18, being not consistent.
This may indicate the importance of the contribution 
from the shocked layer photons. 
\par
     \citet{muk90} obtained optical spectrum on May 14
(17 days after maximum).  Comparing them with the earlier 
spectra observed by \citet*{wag90} and \citet{buc90},
Mukai et al. pointed out that V3890 Sgr had 
progressed to a higher ionization state on that day.
In our modeling of the light curve, the accretion disk plays a key
role in these days.  It is very likely that the conspicuous change 
in the spectrum $\sim 20$ days after maximum is caused by the evaporation
of the accretion disk.  As seen in Figure \ref{vmag1350va1_fixdk_v3890sgr},
the disk size decreases to $\lesssim 1~R_\odot$ 
on the day $\sim 20$.
Then, the mean temperature of the disk surface increases to
$\gtrsim 50,000$ K.  The same thing happens in RS Oph
around the day $\sim$60---70.
\par
     We thus suggest a very small size of the optically 
thick ACDK or no disk, that is,
$R_{\rm disk} \lesssim 0.1 ~R_\sun$, in the late phase 
(after HJD 2,448,036, indicated by "disk vanishes" 
in Figure \ref{vmag1350va1_fixdk_v3890sgr}).  
Therefore, no secondary maximum of the light curve is expected
as well as in RS Oph and V745 Sco.

\subsection{The Distance, Envelope Mass of the White Dwarf,
and Possibility of SN Ia Explosion}
   Fitting indicates an apparent distance modulus of $(m-M)_V= 14.7$. 
Thus, the distance to V3890 Sgr is derived to be $d = 4.2$ kpc
if we adopt the absorption of $A_V= 3.1 \times E(\bv) = 1.6$,
where $E(\bv)=0.5$ \citep{har93},
because the distance modulus is 
$(m-M)_0= (m-M)_V - A_V = 14.7 - 1.6 = 13.1$.
This is roughly consistent with Harrison et al.'s (1993) estimation 
of $d= 5.2$ kpc.  Harrison et al. determined 
the infrared spectral type of the cool companion as M5 III. 
Assuming $M_K= -5.5$ (M5 III), 
%
%   from Koornef
%
%        typical M III                Harrison et al. 1993
%
%        MK[M5III] = -7.0 mag         M_K[M5III]= -5.5 mag
%        MK[M4III] = -6.1 mag         M_K[M4III]= -5.5 mag
%        MK[M3III] = -5.3 mag         M_K[M4III]= - mag
%        MK[M0III] = -4.1 mag         M_K[M4III]= - mag
%        MK[K5III] = -3.9 mag         M_K[M4III]= - mag
%        MK[K0III] = -1.9 mag         M_K[M4III]= - mag
%
%   from Bessell and Brett 1988, PASP, 100, 1134-1151
%
%        MK[M6III] = -7.7 mag         M_K[M5III]= -5.5 mag
%        MK[M5III] = -6.7 mag         M_K[M5III]= -5.5 mag
%        MK[M4III] = -5.8 mag         M_K[M4III]= -5.5 mag
%        MK[M3III] = -5.2 mag         M_K[M4III]= - mag
%        MK[M0III] = -4.2 mag         M_K[M4III]= - mag
%        MK[K5III] = -3.9 mag         M_K[M4III]= - mag
%        MK[K0III] = -1.9 mag         M_K[M4III]= - mag
%
%
they obtained the distance modulus of 
$(m-M)_0 = m_K - M_K - A_K = 8.24 - (-5.5) - 0.17 =13.6$
by using \citet{rie85} relation of $A_K / A_V = 0.112$.
It should be noted here that there are a large divergence
in the observed color excesses: \citet{gon92} determined 
$E(\bv)= 1.1$ from UV lines; \citet{wil94} obtained
$E(\bv)= 0.76$ from H I and He II line ratios;
\citet{anu94} obtained $E(\bv)=1.1$ from the \ion{He}{1} line ratios,
and $E(\bv)=1.2$ from the H$\alpha/$H$\beta$ ratio.
\par
     Finally, we examine the possibility of Type Ia supernova explosion.
The envelope mass at the maximum $V$ light is estimated 
to be $\Delta M = 3.1 \times 10^{-6} M_\sun$ from our solutions,
indicating a mass transfer rate 
of $\dot M_{\rm acc}= 1.1 \times 10^{-7} M_\sun$ yr$^{-1}$ 
between the 1962 and 1990 outbursts.
Moreover, about 90\% of the envelope mass has been blown off
in the outburst wind and the residual 10\% 
($3 \times 10^{-7} M_\sun$) has been left and added
to the helium layer of the WD.  Therefore, we obtain the net
mass increasing rate of 
$\dot M_{\rm He}= 1.1 \times 10^{-8} M_\sun$ yr$^{-1}$.  
This meets condition (\ref{low_accretion_condition_for_SNIa})
of SN Ia explosion \citep[e.g.,][]{nom91}, 
if the WD consists of carbon and oxygen.
Thus, we may conclude that V3890 Sgr is an immediate progenitor of
Type Ia supernova if the hot component is a carbon-oxygen white 
dwarf.  The white dwarf can grow to $1.378 M_\sun$
even if mild helium shell flashes repeatedly occur
after the helium layer grows to the critical mass at which 
helium ignites \citep[see, e.g.,][]{kat99h}.

\section{CONCLUSIONS}
     We have developed a numerical method for calculating 
light curves of recurrent nova outbursts.  Our basic model consists of
a white dwarf with an accretion disk and a red giant companion.
The calculated light curves includes contributions from 
the white dwarf photosphere, the irradiated accretion disk,
and the irradiated red giant.
We summarize our main results on the light curve analysis
of the four recurrent novae with a red giant companion,
T CrB, RS Oph, V745 Sco, and V3890 Sgr.

\subsection{Relevance to Type Ia Supernova Progenitors}
\par \noindent
{\bf 1.}
     The mass of white dwarf (WD) has been estimated, from
the decline of early-phase light curve, to be
$M_{\rm WD}= 1.37 \pm 0.01 ~M_\sun$ for T CrB, 
$1.35 \pm 0.01 ~M_\sun$ with solar metallicity ($Z=0.02$) or
$1.377 \pm 0.01 ~M_\sun$ with low metallicity ($Z=0.004$) for RS Oph,
$1.35 \pm 0.01 ~M_\sun$ for V745 Sco, and
$1.35 \pm 0.01 ~M_\sun$ for V3890 Sgr. 
\par \noindent
{\bf 2.} 
     The apparent distance moduli are obtained, with the light curve 
fitting, to be
$(m-M)_V=10.2$ for T CrB, $(m-M)_V=11.2$ for RS Oph, 
$(m-M)_V=17.0$ for V745 Sco, and $(m-M)_V=14.7$ for V3890 Sgr.
Therefore, the distances are estimated to be
$d= 0.94$ kpc with $A_V=0.35$ for T CrB, 
$d=0.57$ kpc with $A_V=2.3$ for RS Oph, 
$d=5.0$ kpc with $A_V=3.5$ for V745 Sco, and
$d=4.2$ kpc with $A_V=1.6$ for V3890 Sgr.
\par \noindent
{\bf 3.}
     Each envelope mass at the optical maximum is also calculated to be
$\Delta M \sim 3 \times 10^{-6} M_\sun$ (T CrB),
$2 \times 10^{-6} M_\sun$ (RS Oph),
$5 \times 10^{-6} M_\sun$ (V745 Sco),
$3 \times 10^{-6} M_\sun$ (V3890 Sgr), which indicates
an average mass accretion rate of
$\dot M_{\rm acc} \sim 0.4 \times 10^{-7} M_\sun$ yr$^{-1}$ 
(80 yrs, T CrB), 
$1.2 \times 10^{-7} M_\sun$ yr$^{-1}$ (18 yrs, RS Oph), 
$0.9 \times 10^{-7} M_\sun$ yr$^{-1}$ (52 yrs, V745 Sco), 
$1.1 \times 10^{-7} M_\sun$ yr$^{-1}$ (28 yrs, V3890 Sgr) 
during quiescent phase.
\par \noindent
{\bf 4.}
     Although a large part of the envelope mass is blown in the wind,
each white dwarf can retain a substantial part of the envelope mass
after hydrogen burning.  Thus, we have obtained net mass-increasing
rates of the WDs,
$\dot M_{\rm He} \sim 0.10 \times 10^{-7} M_\sun$ yr$^{-1}$ (T CrB), 
$0.12 \times 10^{-7} M_\sun$ yr$^{-1}$ (RS Oph), 
$0.05 \times 10^{-7} M_\sun$ yr$^{-1}$ (V745 Sco), 
$0.11 \times 10^{-7} M_\sun$ yr$^{-1}$ (V3890 Sgr). 
\par \noindent
{\bf 5.}
     These results strongly indicate that the white dwarfs 
in recurrent novae have now grown up to near the Chandrasekhar mass
limit and will soon explode as a Type Ia supernova if the white dwarf
consists of carbon and oxygen.

\subsection{Appearance/Absence of Secondary Maximum}
\par \noindent
{\bf 6.}
          Among the four recurrent novae with a red giant companion,
only T CrB shows a secondary maximum 120---190 days 
after the optical maximum.
This secondary maximum can be well reproduced if 
an irradiated accretion disk of $R_{\rm disk} \sim 6 ~R_\sun$  
around the white dwarf tilts due to a radiation-induced warping
instability.  If we neglect the accretion disk, we cannot 
fully reproduce the secondary peak by an irradiated M-giant model.
\par \noindent
{\bf 7.} 
     The difference between T CrB and the others is in the companion:
it fills its inner critical Roche lobe only in T CrB 
but in the others does not.  Therefore, in T CrB
the mass transfer is mainly driven by the Roche lobe overflow 
while in the other systems matter is fed by cool red giant winds.
In T CrB, the mass transfer continues even in the outburst 
wind phase, which maintains an optically thick accretion disk 
large enough to show a secondary maximum 
after a tilting instability grows up.
\par \noindent
{\bf 8.} 
     The other three systems, RS Oph, V745 Sco, and V3890 Sgr, 
cannot maintain the mass transfer during the outburst 
because the cool red giant wind has once been evacuated 
by the ejecta of the outburst.
The light curves of RS Oph and V3890 Sgr require a relatively 
large size accretion disk 
of $R_{\rm disk} \sim 10~R_\sun$ in the mid-phase 
but a small size of
$R_{\rm disk} \lesssim 1 ~R_\sun$ in the late-phase.
The late-phase UV data of RS Oph can be well reproduced 
with a small accretion disk of $R_{\rm disk} \sim 0.5 R_\sun$.
No or a very small accretion disk is also derived in V745 Sco.
These accretion disks are quickly dissipating 
to become as small as $R_{\rm disk} \lesssim 1 ~R_\sun$,
at least, in the late phase of the outbursts, because no mass is fed. 
Such a small disk cannot contribute much to the $V$ light.
Even if the tilting instability begins to grow after the wind stops, 
it is therefore very unlikely that secondary maxima like in T CrB 
are observed in these three systems.
\par \noindent
{\bf 9.} 
      The post-outburst minima of RS Oph outbursts, which is 1 mag 
(in $V$) below the quiescent level and continues for $\sim 300$ days 
from 100 to 400 days 
after the optical maximum, is understood from the evacuation 
of the cool red giant wind by the ejecta of nova.
After the hot WD wind stops 80 days after the optical maximum, 
the cool red giant wind expands to recover again in $\sim 300$ days.
Then, $\sim 400$ days after maximum, mass accretion from the red giant
wind to the white dwarf resumes.
Supersoft X-rays from the white dwarf, which were observed about 250 
days after maximum, are very consistent with the end of hot WD
wind and the evacuation of the cool red giant wind.


%% Included in this acknowledgments section are examples of the
%% AASTeX hypertext markup commands. Use \url without the optional [HREF]
%% argument when you want to print the url directly in the text. Otherwise,
%% use either \url or \anchor, with the HREF as the first argument and the
%% text to be printed in the second.

\acknowledgments
     This research has been supported in part by the Grant-in-Aid for
Scientific Research (08640321, 09640325, 11640226) 
of the Japanese Ministry of Education, Science, 
Culture, and Sports.

\clearpage
\begin{deluxetable}{llll}
\tabletypesize{\scriptsize}
\tablecaption{Recurrent novae
\label{recurrent_novae}}
\tablewidth{0pt}
\tablehead{
%\multicolumn{1}{c}{$M_{\rm WD}$} &
%\multicolumn{10}{c}{$Z=0.02$} \\
%\multicolumn{1}{c}{$(M_\sun)$} &
%\multicolumn{10}{c}{$X$}
%%%%\colhead{$\dot P/P$} 
%\\
\colhead{Star} & 
\colhead{Subclass} &
\colhead{$P_{\rm orb}$} &
\colhead{Outburst Dates} 
\\
\colhead{} &
\colhead{} &
\colhead{(day)} &
\colhead{} 
%\colhead{$(M_\sun)$} &
%\colhead{$(10^{-6}$yr$^{-1})$} &
%\colhead{$(10^{-6}M_\sun$ yr$^{-1}$)} &
%\colhead{$(M_\sun)$} &
%\colhead{$(10^{-6}$yr$^{-1})$} 
} 
\startdata
%1.377\tablenotemark{a} 
RS Oph & RS Oph & 457 & 1898,1933,1958,1967,1985 \\
T CrB &  & 227.6 & 1866,1946 \\
V745 Sco &  & --- & 1937,1989 \\
V3890 Sgr &  & --- & 1962,1990 \\
U Sco & U Sco & 1.23 & 1863,1906,1936,1979,1987,1999 \\
V394 CrA &  & 0.757 & 1949,1987 \\
LMC RN &  & --- & 1968,1990 \\
CI Aql &  & 0.618 & 1917,2000 \\
T Pyx & T Pyx & 0.076 & 1890,1902,1920,1944,1966 \\
\enddata
%\tablenotetext{a}{
%hydrogen content, and $Z=0.02$ is assumed for 
%heavy elements.
%}
\end{deluxetable}

%\clearpage
\begin{deluxetable}{llcccccccccc}
\tabletypesize{\scriptsize}
\tablecaption{Physical properties of Recurrent novae outbursts
\label{recurrent_novae_outburst}}
\tablewidth{0pt}
\tablehead{
%\multicolumn{1}{c}{$M_{\rm WD}$} &
%\multicolumn{10}{c}{$Z=0.02$} \\
%\multicolumn{1}{c}{$(M_\sun)$} &
%\multicolumn{10}{c}{$X$}
%%%%\colhead{$\dot P/P$} 
%\\
\colhead{Star} & 
\colhead{$P_{\rm orb}$} &
\colhead{$M_{\rm WD}$} &
\colhead{$X$} &
\colhead{$\Delta M_{\rm max}$} &
\colhead{$\tau_{\rm rec}$} &
\colhead{$\dot M_{\rm acc}$} &
\colhead{$\eta_{\rm H}$} &
\colhead{$\dot M_{\rm He}$} &
\colhead{$E(\bv)$} &
\colhead{$d$} &
\colhead{Fate} 
\\
\colhead{} &
\colhead{(day)} &
\colhead{$(M_\sun)$} &
\colhead{} &
\colhead{$(M_\sun)$} &
\colhead{(yr)} &
\colhead{($M_\sun$ yr$^{-1}$)} &
\colhead{} &
\colhead{($M_\sun$ yr$^{-1}$)} &
\colhead{} &
\colhead{(kpc)} &
\colhead{} 
%\colhead{$(M_\sun)$} &
%\colhead{$(10^{-6}$yr$^{-1})$} &
%\colhead{$(10^{-6}M_\sun$ yr$^{-1}$)} &
%\colhead{$(M_\sun)$} &
%\colhead{$(10^{-6}$yr$^{-1})$} 
} 
\startdata
%%%%%1.377\tablenotemark{a} 
T CrB & 227.67 & 1.37 & 0.70 & $3 \times 10^{-6}$ & 80 
& $4.1 \times 10^{-8}$ & 0.10 & $0.4 \times 10^{-8}$ & 0.11 & 0.9 & SN Ia \\
RS Oph\tablenotemark{a} &  460 & 1.35 & 0.70 & $2 \times 10^{-6}$ & 18 
& $1.2 \times 10^{-7}$ & 0.10 & $1.2 \times 10^{-8}$ & 0.73 & 0.6 & SN Ia \\
RS Oph\tablenotemark{b} &  460 & 1.377 & 0.70 & $2 \times 10^{-6}$ & 18 
& $1.2 \times 10^{-7}$ & 0.10 & $1.2 \times 10^{-8}$ & 0.73 & 0.6 & SN Ia \\
V745 Sco & \nodata & 1.35 &  0.70 & $5 \times 10^{-6}$ & 52 
& $9.0 \times 10^{-8}$  & 0.05 & $0.5 \times 10^{-8}$ & 1.1 & 5 & SN Ia \\
V3890 Sgr & \nodata & 1.35 & 0.70 & $3 \times 10^{-6}$ & 28 
& $1.1 \times 10^{-7}$ & 0.10 & $1.1 \times 10^{-8}$ & 0.5 & 4 & SN Ia \\
U Sco\tablenotemark{c} & 1.23056 & 1.37 & 0.05 & $3 \times 10^{-6}$ & 12 
& $2.5 \times 10^{-7}$ & 0.40 & $1.0 \times 10^{-7}$ & 0.20 & 6 & SN Ia \\
V394 CrA\tablenotemark{d} &  0.7577 & 1.37 & 0.05 & $6 \times 10^{-6}$ & 39 
& $1.5 \times 10^{-7}$ & 0.23 & $3.4 \times 10^{-8}$ & 0.26 & 4 & SN Ia \\
\enddata
\tablenotetext{a}{taken from \citet{hac00ka} of solar metallicity 
model for RS Oph}
\tablenotetext{b}{the present low metallicity ($Z=0.004$) model for RS Oph}
\tablenotetext{c}{taken from \citet{hkkm00}}
\tablenotetext{d}{taken from \citet{hac00kb}}
%\tablenotetext{d}{the primary component will explode as 
%a Type Ia supernovae (SN Ia) when it reaches $M_{\rm Ia}= 1.378
%M_\sun$ (Nomoto et al. 1984).}
%\tablenotetext{e}{taken from Selville et al. (1991).}
\end{deluxetable}


%% If you are not including electronic art with your submission, you may
%% mark up your captions using the \figcaption command. See the 
%% No more than seven \figcaption commands are allowed per page, 
%% so if you have more than seven captions, insert a \clearpage 

%% Tables should be submitted one per page, so put a \clearpage before

%% Two options are available to the author for producing tables:  the
%% deluxetable environment provided by the AASTeX package or the LaTeX
%% table environment.  Use of deluxetable is preferred.
%%

%% Three table samples follow, two marked up in the deluxetable environment,
%% one marked up as a LaTeX table.

\clearpage

%\clearpage

%\clearpage

\clearpage

\clearpage
%\epsfverbosetrue

\clearpage
%\epsfverbosetrue

\clearpage
%\epsfverbosetrue

\clearpage
%\epsfverbosetrue

\clearpage
%\epsfverbosetrue

\clearpage
%\epsfverbosetrue

%% The following command ends your manuscript. LaTeX will ignore any text

\end{document}